\documentclass[12pt]{article}
\pdfoutput=1

\usepackage{epsfig}
\usepackage{psfrag}
\usepackage{latexsym}
\usepackage{indentfirst}
\usepackage{fancyhdr}
\usepackage{dsfont}
\usepackage{amssymb}
\usepackage{amsmath}
\usepackage{amsfonts}
\usepackage{pifont}
\usepackage{xcolor}
\usepackage{cite}
\usepackage{hyperref}
\hypersetup{colorlinks=true,linkcolor=red,anchorcolor=black,citecolor=green}
\usepackage{bbold}
\usepackage{color}
\usepackage{colordvi}
\usepackage{fancybox}
\usepackage[footnotesize]{caption2}
\usepackage{graphicx}
\usepackage[center,footnotesize,hang]{subfigure}
\usepackage{bbm}
\usepackage{bm}
\usepackage{diagbox}
\usepackage{url}
\usepackage{multirow}
\usepackage{array}
\usepackage{arydshln}
\usepackage{enumitem}
\usepackage{rotating}
\usepackage{tikz}
\newcommand{\ignore}[1]{}
\newcommand{\PreserveBackslash}[1]{\let\temp=\\#1\let\\=\temp}
\newcolumntype{C}[1]{>{\PreserveBackslash\centering}p{#1}}
\newcolumntype{R}[1]{>{\PreserveBackslash\raggedleft}p{#1}}
\newcolumntype{L}[1]{>{\PreserveBackslash\raggedright}p{#1}}
\allowdisplaybreaks  

\newcommand{\bq}{\begin{eqnarray}}
\newcommand{\nq}{\end{eqnarray}}

\makeatletter
\@addtoreset{equation}{section}
\makeatother

\begin{document}
\title{
\begin{flushright}
\hfill\mbox{{\small\tt USTC-ICTS/PCFT-21-27}} \\[5mm]
\begin{minipage}{0.2\linewidth}
\normalsize
\end{minipage}
\end{flushright}
{\Large \bf
Modular symmetry at level 6 and a new route towards finite modular groups
\\[2mm]}
}
\date{}

\author{
Cai-Chang Li$^{1,2,3}$\footnote{E-mail: {\tt
ccli@nwu.edu.cn}},  \
Xiang-Gan Liu$^{4,5}$\footnote{E-mail: {\tt
hepliuxg@mail.ustc.edu.cn}},  \
Gui-Jun Ding$^{4,5}$\footnote{E-mail: {\tt
dinggj@ustc.edu.cn}} \
\\*[20pt]
\centerline{
\begin{minipage}{\linewidth}
\begin{center}
$^1${\it\small School of Physics, Northwest University, Xi'an 710127, China}\\[2mm]
$^2${\it\small Peng Huanwu Center for Fundamental Theory, Xi'an 710127, China}\\[2mm]
$^3${\it\small Shaanxi Key Laboratory for Theoretical Physics Frontiers, Xi'an 710127, China} \\[2mm]
$^4${\it \small
Interdisciplinary Center for Theoretical Study and  Department of Modern Physics,\\
University of Science and Technology of China, Hefei, Anhui 230026, China} \\[2mm]
$^5${\it \small Peng Huanwu Center for Fundamental Theory, Hefei, Anhui 230026, China}
\end{center}
\end{minipage}}
\\[10mm]}
\maketitle
\thispagestyle{empty}

\centerline{\large\bf Abstract}
\begin{quote}
\indent
We propose to construct the finite modular groups from the quotient of two principal congruence subgroups as $\Gamma(N')/\Gamma(N'')$, and the modular group $SL(2,\mathbb{Z})$ is extended to a principal congruence subgroup $\Gamma(N')$. The original modular invariant theory is reproduced when $N'=1$. We perform a comprehensive study of $\Gamma'_6$ modular symmetry corresponding to $N'=1$ and $N''=6$, five types of models for lepton masses and mixing with $\Gamma'_6$ modular symmetry are discussed and some example models are studied numerically. The case of $N'=2$ and $N''=6$ is considered, the finite modular group is $\Gamma(2)/\Gamma(6)\cong T'$, and a benchmark model is constructed.
\end{quote}
\newpage
\section{Introduction}

The neutrino masses and lepton mixing parameters have been precisely measured by a large number of neutrino oscillation experiments. If neutrinos are Majorana particles, the lepton mixing are described by three mixing angles, one Dirac CP violation phase $\delta_{CP}$ and two Majorana CP violation phases $\alpha_{21}$ and $\alpha_{31}$. The $3\sigma$ ranges of lepton mixing angles and neutrino masses for normal ordering (NO) spectrum are determined to be~\cite{Esteban:2020cvm}
\begin{eqnarray}
\nonumber &&\hskip-0.1in 0.02032\leq\sin^2\theta_{13}\leq0.02410,\quad 0.269\leq\sin^2\theta_{12}\leq0.343,\quad
0.415\leq\sin^2\theta_{23}\leq0.616,\\
\label{eq:3s_ran} &&\hskip-0,1in6.82~\text{eV}^2\leq\Delta m^2_{21}\times10^{5}\leq8.04~\text{eV}^2,\quad
2.435~\text{eV}^2\leq\Delta m^2_{31}\times10^{3}\leq2.598~\text{eV}^2\,,
\end{eqnarray}
where $\theta_{12}$, $\theta_{13}$ and $\theta_{23}$ are three lepton mixing angles in the standard parametrization~\cite{ParticleDataGroup:2020ssz}, and $\Delta m^2_{21}\equiv m^2_2-m^2_1$ and $\Delta m^2_{31}\equiv m^2_3-m^2_1$ are two mass squared differences. Similar results for inverted ordering (IO) spectrum have been obtained~\cite{Esteban:2020cvm}. The Dirac phase $\delta_{CP}$ is less constrained by the current data, and the forthcoming long-baseline experiments are expected to make precise measurement of $\delta_{CP}$. The  Majorana phases of the neutrino mixing matrix can not have any effect in neutrino oscillations, and almost nothing is known about their possible values at present.

Understanding the origin of fermion masses and mixing patterns is a greatest challenge in particle physics. It was found that the non-abelian finite discrete flavor symmetry is particularly suitable to explain the large lepton mixing angles~\cite{Altarelli:2010gt,Ishimori:2010au,King:2013eh,King:2014nza,King:2015aea,King:2017guk,Feruglio:2019ybq}. Certain mixing patterns can be obtained if the discrete flavor group is broken down to different remnant subgroups in the charged lepton and neutrino sectors. In the paradigm of traditional discrete flavor symmetry, the flavor group is spontaneously broken by the vacuum expectation values of a set of flavons. In this approach, additional dynamics of vacuum alignment, field content and auxiliary symmetry are necessary to achieve the desired non-trivial vacuum configuration so that the resulting models are rather complicated and the scalar potential has to be cleverly designed.

However, the origin of finite flavor symmetry is still unknown. Recently modular symmetry which naturally appears in string theory has been suggested as the origin of such a finite flavor symmetry~\cite{Feruglio:2017spp}. The homogeneous (inhomogeneous) finite modular groups $\Gamma'_N$ ($\Gamma_N$) appearing as the quotient group of the two dimensional (projective) special linear group over the principal congruence subgroups is considered in the framework of the modular invariance approach. In this modular approach, the flavon fields other than the modulus $\tau$ are unnecessary and the Yukawa couplings are modular forms which are holomorphic functions of the complex modulus $\tau$. The flavor symmetry will be broken when the modulus $\tau$ gets vacuum expectation value (VEV). Moreover, the modular invariance and supersymmetry completely determine all higher dimensional operators in the superpotential. As the lepton mixing parameters and masses are only fixed by a small number of free parameters in modular invariant lepton models, the resulting models have strong predictive power.

The inhomogeneous finite modular group $\Gamma_N$ is the quotient group of the modular group $PSL(2,\mathbb{Z})\cong\overline{\Gamma}$ over the principal congruence subgroup $\Gamma(N)$. The phenomenologically viable modular invariant models have been widely discussed by using the inhomogeneous finite modular group $\Gamma_N$ in the literature, such as models based on the finite modular groups $\Gamma_2\cong S_3$~\cite{Kobayashi:2018vbk,Kobayashi:2018wkl,Kobayashi:2019rzp,Okada:2019xqk,Novichkov:2021evw}, $\Gamma_3\cong A_4$~\cite{Feruglio:2017spp,Criado:2018thu,Kobayashi:2018vbk,Kobayashi:2018scp,deAnda:2018ecu,Okada:2018yrn,Kobayashi:2018wkl,Novichkov:2018yse, Nomura:2019jxj,Okada:2019uoy,Nomura:2019yft,Ding:2019zxk,Okada:2019mjf,Nomura:2019lnr,Kobayashi:2019xvz,Asaka:2019vev, Gui-JunDing:2019wap,Zhang:2019ngf,Nomura:2019xsb,Wang:2019xbo,Kobayashi:2019gtp,King:2020qaj,Ding:2020yen,Okada:2020rjb,Nomura:2020opk,Asaka:2020tmo, Okada:2020brs,Yao:2020qyy,Feruglio:2021dte,Chen:2021zty,Okada:2021qdf,Novichkov:2021evw}, $\Gamma_4\cong S_4$~\cite{Penedo:2018nmg,Novichkov:2018ovf,deMedeirosVarzielas:2019cyj,Kobayashi:2019mna,King:2019vhv,Criado:2019tzk,Wang:2019ovr,Gui-JunDing:2019wap,Wang:2020dbp,King:2021fhl,Ding:2021zbg,Qu:2021jdy}, $\Gamma_5\cong A_5$~\cite{Novichkov:2018nkm,Ding:2019xna,Criado:2019tzk}
and $\Gamma_7\cong PSL(2, Z_7)$~\cite{Ding:2020msi} have been studied. The modular forms of integral weights will be decomposed into irreducible representations of the homogeneous finite modular group $\Gamma'_N$ which is the double covering of $\Gamma_N$~\cite{Liu:2019khw}. Modular invariant models based on the integral weight modular forms have been built and the corresponding finite modular groups $\Gamma'_3\cong T'$~\cite{Liu:2019khw,Lu:2019vgm}, $\Gamma'_4\cong S'_4$~\cite{Novichkov:2020eep,Liu:2020akv} and $\Gamma'_5\cong A'_5$~\cite{Wang:2020lxk,Yao:2020zml} have been studied in the modular invariance approach. Recently the modular weight $k$ has been extended to a rational number, and the corresponding finite modular group will be extended to its metaplectic covers~\cite{Liu:2020msy,Yao:2020zml}. It is remarkable that the modular symmetry also has the merit of conventional abelian flavor symmetry group, the structure of the modular form can produce texture zeros of fermion mass matrices exactly~\cite{Zhang:2019ngf,Lu:2019vgm}, the modular weights can play the role of Froggatt-Nielsen charges~\cite{Criado:2019tzk,King:2020qaj}, and the hierarchical charged lepton masses can arise solely due to the proximity of the modulus $\tau$ to a residual symmetry conserved point~\cite{Okada:2020ukr,Feruglio:2021dte,Novichkov:2021evw}. Furthermore, the modular invariance approach has been extended to the quark sector to explain the quark mixing and quark masses. Then a unified description of leptons and quarks can be achieved from a common finite discrete modular group~\cite{Okada:2019uoy,Lu:2019vgm,Okada:2020rjb,Liu:2020akv,Yao:2020zml,Yao:2020qyy}. In order to further improve the predictive power of the modular invariance approach, the setup of combining modular symmetry with the generalized CP (gCP) symmetry has been discussed~\cite{Novichkov:2019sqv,Ding:2021iqp}. The consistency of modular and gCP symmetries requires that the complex modulus $\tau$ transforms as $\tau\stackrel{\mathcal{CP}}{\longrightarrow}-\tau^{*}$ up to modular transformations~\cite{Acharya:1995ag,Dent:2001cc,Giedt:2002ns,Baur:2019kwi,Novichkov:2019sqv,Ding:2021iqp}. In this context, the VEV of $\tau$ is the unique source of both modular and gCP symmetries. In the basis where the representation matrices of both modular generators $S$ and $T$ are symmetric, the gCP transformation would reduce to canonical CP and all the coupling constants would be real. Moreover, the possibility of coexistence of multiple moduli has been investigated~\cite{deMedeirosVarzielas:2019cyj,Ding:2020zxw}.

In the present work, we generalized the modular invariance approach~\cite{Feruglio:2017spp} by replacing the modular group $SL(2,\mathbb{Z})$ with the principal congruence subgroup $\Gamma(N')$, the original modular invariant theory is the special case of $N'=1$. The finite modular group is the quotient of two principal congruence subgroups $\Gamma(N')/\Gamma(N'')$, and the known homogeneous finite modular groups can be reproduced. As examples, we study two cases of $\Gamma(N')=\Gamma(1)$ and $\Gamma(N')=\Gamma(2)$ with  $\Gamma(N'')=\Gamma(6)$ to construct models of lepton masses and mixing. The modular symmetry is extended to combine with the gCP symmetry, and invariance under gCP requires all coupling constants to be real in our working basis. We find that the modular forms of weight $k=1$ and level $N=6$ can be obtained from the modular forms of $\Gamma(3)$ with weight 1 and they can be expressed in terms of the products of Dedekind eta functions. A systematical analysis of the modular invariant lepton models with $\Gamma^\prime_6$ modular symmetry is performed. The finite modular group $\Gamma^\prime_6$ has abundant two-dimensional representations which provides new model building possibilities. The left-handed lepton fields, the right-handed charged lepton and the right-handed neutrinos can be assigned to a triplet, doublet plus singlet or three singlets of $\Gamma'_6$. We are concerned with the phenomenologically viable models with small number of free parameters. We have discussed five types of lepton models classified according to the representation assignments of the lepton fields under $\Gamma'_6$, and some examples models are presented. The neutrino masses are generated by the effective Weinberg operator in type IV model and by the seesaw mechanism in other types of models. Three right-handed neutrinos are introduced in type I, II, III models while there are two right-handed neutrinos in type V models so that the lightest neutrino is massless. Furthermore, we consider the case of $\Gamma(N')=\Gamma(2)$ and $\Gamma(N'')=\Gamma(6)$, the finite modular group is $\Gamma(2)/\Gamma(6)=\Gamma'_3\cong T'$. The modular forms of level 6 can be arranged into $T'$ multiplets, and a benchmark model is constructed.

The outline of this paper is as follows. The concept of modular symmetry is recapitulated and a new route towards finite modular groups is given in section~\ref{sec:new_rou}. In section~\ref{Sec:MS_MF}, we give the formalism of the generalized modular invariant supersymmetry theory and modular forms of level 6 are presented. In section~\ref{Sec:model_G6p}, lepton models with $\Gamma'_6$ modular symmetry are studied, and we have discussed five types of models differing in the representation assignments of the lepton fields under $\Gamma^\prime_{6}$. As an example, a modular invariant model based on $\Gamma(2)$ modular symmetry with finite modular group $T^\prime$ is given in section~\ref{Sec:model_Tp}. We conclude the paper in section~\ref{sec:Con}. The group theory of the $\Gamma^\prime_6$ modular group is presented and the Clebsch-Gordan (CG) coefficients in our working basis are reported in appendix~\ref{app:group-theory}. The linearly independent higher integral weight modular multiplets of level 6 with finite modular group $\Gamma^\prime_6$ are listed in appendix~\ref{app:hig_wei}. We present the decomposition of level 6 modular forms with respect to the $T^\prime$ modular symmetry in appendix~\ref{app:hig_wei_Tp},.

\section{\label{sec:new_rou}A new route towards finite modular groups}

The finite modular groups and their cover groups have been widely studied as the flavor symmetry groups. They usually arise from the quotient of the
group $SL(2,\mathbb{Z})$ or $PSL(2,\mathbb{Z})$  modulo its normal subgroups $\Gamma(N)$, where $SL(2,\mathbb{Z})\equiv\Gamma$ called the full modular group is defined as
\begin{equation}
SL(2,\mathbb{Z})=\left\{\begin{pmatrix}
a &~ b \\ c &~ d
\end{pmatrix}  \Big| ad- bc=1\,,\quad  a,b,c,d \in \mathbb{Z}\,\right\}\,.
\end{equation}
The group $SL(2,\mathbb{Z})$ has infinite elements and it can be generated by two generators $S$ and $T$ with
\begin{equation}
	S=\begin{pmatrix}
		0 &~ 1 \\ -1 &~ 0
	\end{pmatrix}\,, \qquad T=\begin{pmatrix}
		1 &~ 1 \\ 0 &~ 1
	\end{pmatrix}\,,
\end{equation}
which fulfill the relations
\begin{equation}
	S^4=(ST)^3=\mathbb{1}\,, \qquad S^2 T=T S^2\,.
\end{equation}
The infinite normal subgroups $\Gamma(N)$ which are called principal congruence subgroup of level $N$ are defined as
\begin{equation}
	\Gamma(N)=\left\{ \gamma \in SL(2,\mathbb{Z}) ~~\Big |~~ \gamma \equiv \begin{pmatrix}
		1 &~ 0 \\ 0 &~ 1
	\end{pmatrix} \mod N \right\}\,.
\end{equation}
Note that $\Gamma(1) = SL(2,\mathbb{Z})$ and $T^N\in \Gamma(N)$. The homogeneous finite modular group $\Gamma'_N$ is the quotient group $\Gamma/\Gamma(N)$~\footnote{The quotient group $\Gamma_N=\overline{\Gamma}/\overline{\Gamma}(N)$ is the
inhomogeneous finite modular group of level $N$, where $\overline{\Gamma}$ and $\overline{\Gamma}(N)$ are the projective groups $\overline{\Gamma}=\Gamma/\{\pm 1\}$ and $\overline{\Gamma}(N)=\Gamma(N)/\{\pm 1\}$ respectively. Obviously, $\Gamma_N$ is isomorphic to $\Gamma'_N/\{\pm 1\}$ for $N>2$. Hence $\Gamma'_N$ is the double cover group of $\Gamma_N$. }:
\begin{equation}
\Gamma'_N\equiv \Gamma/\Gamma(N): \quad S^4=(ST)^3=T^N=1, \quad  S^2T=TS^2\,, \quad N<6\,.
\end{equation}
Note that additional relations are necessary to render the group $\Gamma'_N$ finite for $N\geq6$~\cite{deAdelhartToorop:2011re}. In the present work, we are concerned with the modular group $\Gamma'_6$ which satisfies the defining relations
\begin{equation}
  S^{4}=T^{6}=(ST)^{3}=ST^2ST^3ST^4ST^3=1, \qquad S^2T=TS^2\,.
\end{equation}

The above construction of finite modular group is not unique. The quotient of two congruence subgroups of $SL(2,\mathbb{Z})$ can also give finite groups.
Let $N=p^{n_1}_1 \dots p^{n_k}_k$ be the prime factorization of the positive integer $N$, where $n_1,\dots , n_k \in \mathbb{N}$ and $p_1, \dots , p_k$ are distinct primes. The prime factorization of all positive integers less than 21 is listed in table~\ref{tab:primeFactor}. It is well known that Chinese remainder theorem gives the isomorphism of rings~\cite{Serre-1973A}:
\begin{equation}
\mathbb{Z}/N\mathbb{Z} \cong \left(\mathbb{Z}/p^{n_1}_1\mathbb{Z}\right) \times \dots \times \left(\mathbb{Z}/p^{n_k}_k \mathbb{Z}\right)\,.
\end{equation}
Then this gives isomorphism~\cite{miyake2006modular}:
\begin{equation}
\label{eq:factor-hom-FMG}SL(2,\mathbb{Z}/N\mathbb{Z}) \cong SL(2,\mathbb{Z}/p^{n_1}_1\mathbb{Z}) \times \dots \times SL(2,\mathbb{Z}/p^{n_k}_k\mathbb{Z})\,.
\end{equation}
\begin{table}[t!]
\centering
\resizebox{1.0\textwidth}{!}{
\begin{tabular}{|c||c|c|c|c|c|c|c|c|c|c|c|c|c|c|c|c|c|c|c|c|c|}
\hline\hline
N & 1 & 2 & 3 & 4 & 5 & 6 & 7 & 8 & 9 & 10 & 11 & 12 & 13 & 14 & 15 & 16 & 17 & 18 & 19 & 20\\ \hline
 & 1 & 2 & 3 & $2^2$ & 5 & $2\cdot 3$ & 7 & $2^3$ & $3^2$ & $2\cdot 5$& 11 & $2^2\cdot 3$ & 13 & $2\cdot 7$ & $3\cdot 5$ & $2^4$ & 17 & $2\cdot 3^2$ & $19$ & $2^2\cdot 5$ \\ \hline \hline
\end{tabular} }
\caption{\label{tab:primeFactor}The prime factorization of positive integer $0<N<21$.}
\end{table}
Using isomorphism $SL(2,\mathbb{Z}/N\mathbb{Z}) \cong \Gamma(1)/\Gamma(N)$, we have
\begin{equation}
\Gamma(1)/\Gamma(N)\cong (\Gamma(1)/\Gamma(p^{n_1}_1)) \times \dots \times (\Gamma(1)/\Gamma(p^{n_k}_k)) \,,
\end{equation}
which implies
\begin{equation}
\Gamma(p^{n_i}_i)/\Gamma(N) \cong \prod_{j \neq i} \Gamma(1)/\Gamma(p^{n_j}_j) = \prod_{j \neq i} \Gamma'_{p^{n_j}_j}\,.
\end{equation}
As concrete examples, we have~\footnote{Note that $S_3 \cong \Gamma_2 = \overline{\Gamma}(1)/\overline{\Gamma}(2) = \Gamma(1)/\Gamma(2) =\Gamma'_2$.}
\begin{align}
\nonumber
&\Gamma'_2 = \Gamma(1)/\Gamma(2)  \cong \Gamma(3)/\Gamma(6) \cong \Gamma(5)/\Gamma(10)\,, \\
\nonumber
&\Gamma'_3 = \Gamma(1)/\Gamma(3) \cong \Gamma(2)/\Gamma(6) \cong \Gamma(4)/\Gamma(12)  \,,\\
\nonumber
&\Gamma'_4 = \Gamma(1)/\Gamma(4) \cong \Gamma(3)/\Gamma(12) \cong
\Gamma(5)/\Gamma(20) \,,\\
\label{eq:finite-modGrp}&\Gamma'_5 = \Gamma(1)/\Gamma(5) \cong \Gamma(2)/\Gamma(10) \cong
\Gamma(3)/\Gamma(15)  \,.
\end{align}
This implies that quotient of the principal congruence subgroups can also give rise to the homogeneous finite modular groups $\Gamma'_N$. If one considers the more general congruent subgroups of $SL(2,\mathbb{Z})$, other finite modular groups can be constructed by the quotient procedure. Besides the principal congruence subgroups, there are two another important congruence subgroups~\cite{diamond2005first}
\begin{align}
\label{eq:congruenceG0G1}
&\Gamma_0(N)=\Big\{\begin{pmatrix}
a &~ b \\ c &~ d
\end{pmatrix} \in SL(2,\mathbb{Z}) ~\Big|~ \begin{pmatrix}
a &~ b \\ c &~ d
\end{pmatrix} \equiv \begin{pmatrix}
* &~ * \\ 0 &~ *
\end{pmatrix}\mod N \Big\} \,,\\
&\Gamma_1(N)=\Big\{\begin{pmatrix}
a &~ b \\ c &~ d
\end{pmatrix} \in SL(2,\mathbb{Z}) ~\Big|~ \begin{pmatrix}
a &~ b \\ c &~ d
\end{pmatrix} \equiv \begin{pmatrix}
1 &~ * \\ 0 &~ 1
\end{pmatrix}\mod N \Big\}\,,
\end{align}
where * means any positive integer small than $N$. Obviously we have relation $\Gamma(N) \subset \Gamma_1(N) \subset \Gamma_0(N) \subset \Gamma$. In fact, it is known that $\Gamma(N)$ and $\Gamma_1(N)$ are the normal sungroups of $\Gamma_1(N)$ and $\Gamma_0(N)$ respectively, and~\cite{diamond2005first}
\begin{equation}
\Gamma_1(N)/\Gamma(N) \cong Z_N , \qquad \Gamma_0(N)/\Gamma_1(N) \cong Z^*_N \,,
\end{equation}
where $Z_N$ are cyclic groups of order $N$, also known as additive group of integers modulo $N$. And $Z^*_N$ is known as the multiplicative group of integers modulo $N$, and its order is equal to $\phi(N)\equiv N\prod_{p|N}\left(1-\frac{1}{p}\right)$ which is Euler's totient function. $Z^*_N$ is a finite abelian group and it reduces to a cyclic group for prime $N$. Since both $Z_N$ and $Z^*_N$ are abelian groups and they only have one-dimensional irreducible representations, we don't consider the cases of $Z_N$ and $Z^*_N$ as flavor symmetry in the following.

We denote the two principal congruence subgroups as $\Gamma(N')$ and $\Gamma(N'')$ in the following, $N''$ is divisible by $N'$ so that $\Gamma(N'')$ is a normal subgroup of  $\Gamma(N')$. $\Gamma(N')$ acts on the upper half plane $\mathcal{H}=\{\tau\in\mathbb{C}| \Im{\tau}>0\}$ by the linear fractional transformation
\begin{equation}
\gamma \tau = \dfrac{a\tau+b}{c\tau+d}\,,\qquad \gamma=\begin{pmatrix}
a &~ b \\c &~ d
\end{pmatrix}\in \Gamma(N')\,.
\end{equation}
As a consequence, the fundamental domain becomes $\mathcal{D}'=\mathcal{H}/\Gamma(N')$ in which no two points are related by the modular transformations of $\Gamma(N')$. The modular forms of the modular subgroup $\Gamma(N'')$ with weight $k$ are holomorphic functions $f(\tau)$ satisfying the property
\begin{equation}
f_i(h\tau)=(c\tau+d)^k f_i(\tau),\qquad h\in \Gamma(N'')\,.
\end{equation}
The modular forms $f_i(\tau)$ span a finite dimensional linear space, and they are usual modular forms of level $N''$. It is possible to choose a basis such that the transformations of modular forms $f_i(\tau)$ under $\Gamma(N')$ can be described by~\cite{Feruglio:2017spp,Liu:2019khw}
\begin{equation}
f_i(\gamma\tau)=(c\tau+d)^{k}\rho_{ij}(\gamma)f_j(\tau)\,, \qquad \forall\, \gamma=\begin{pmatrix}
a &~ b \\ c &~ d
\end{pmatrix}\in  \Gamma(N')\,,
\end{equation}
where $\rho(\gamma)$ is the irreducible representation of quotient group $\Gamma(N')/\Gamma(N'')$.

\section{\label{Sec:MS_MF}Modular symmetry and modular forms of level 6}


We generalize the modular invariant supersymmetry theory~\cite{Feruglio:2017spp} by replacing $\Gamma\cong SL(2,\mathbb{Z})$ with the principal congruence subgroup $\Gamma(N')$, the original modular invariant theory is the special case of $N'=1$. Notice that this generalization is already contained in the more general automorphic supersymmetric framework~\cite{Ding:2020zxw}. The field content consists of a set of chiral matter superfields $\Phi_I$ and a modulus superfield $\tau$, their modular transformation under $\Gamma(N')$ are given by:
\begin{equation}
\label{eq:modularTrs_Phi}
\begin{cases}
\tau\to \gamma\tau=\dfrac{a\tau+b}{c\tau+d}\,,\\
\Phi_I\to (c\tau+d)^{-k_I}\rho_I(\gamma)\Phi_I\,,
\end{cases}
\qquad \gamma=\begin{pmatrix}
a &~ b \\ c &~ d
\end{pmatrix} \in \Gamma(N')\,,
\end{equation}
where $k_I$ is called the modular weight of matter field $\Phi_I$, and $\rho_I(\gamma)$ is the unitary representation of quotient group $\Gamma(N')/\Gamma(N'')$.
The K\"ahler potential is still taken to be the minimal~\cite{Feruglio:2017spp}
\begin{equation}
	\mathcal{K}(\Phi_I,\bar{\Phi}_I; \tau,\bar{\tau}) =-h\Lambda^2 \log(-i\tau+i\bar\tau)+ \sum_I (-i\tau+i\bar\tau)^{-k_I} |\Phi_I|^2~~~,
\end{equation}
which leads to the kinetic terms for matter fields and the modulus field $\tau$ after the modular symmetry breaking induced by non-vanishing VEV of $\tau$. The superpotential $\mathcal{W}(\Phi_I,\tau)$ can be expanded in power series of the involved supermultiplets $\Phi_I$,
\begin{equation}
	\mathcal{W}(\Phi_I,\tau) =\sum_n Y_{I_1...I_n}(\tau)~ \Phi_{I_1}... \Phi_{I_n}\,,
\end{equation}
where $Y_{I_1...I_n}(\tau)$ is a modular multiplet of weight $k_Y$ and level $N''$ as introduced in the previous section, and it fulfills
\begin{equation}
Y_{I_1...I_n}(\tau)\stackrel{\gamma}{\longrightarrow}Y_{I_1...I_n}(\gamma\tau)=(c\tau+d)^{k_Y}\rho_{Y}(\gamma)Y_{I_1...I_n}(\tau),~~~\gamma=\begin{pmatrix}
a &~ b \\ c &~ d
\end{pmatrix} \in \Gamma(N')\,.
\end{equation}
Modular invariance entails that each term of the $\mathcal{W}(\Phi_I,\tau)$ should satisfy the following conditions:
\begin{equation}
	k_Y=k_{I_1}+...+k_{I_n},~\quad~ \rho_Y\otimes\rho_{I_1}\otimes\ldots\otimes\rho_{I_n}\ni\bm{1}\,.
\end{equation}
In this work, we will study the normal subgroup $\Gamma(N'')=\Gamma(6)$ as an example, and consider two cases of $\Gamma(N')=\Gamma(1)$ and $\Gamma(N')=\Gamma(2)$ as the full modular symmetries to construct the corresponding lepton models. Accordingly the finite modular groups are $\Gamma(1)/\Gamma(6)=\Gamma'_6$ and $\Gamma(2)/\Gamma(6)=\Gamma'_3\cong T'$ respectively.

\subsection{\label{Sec:MF_N=6}Modular forms of level $N=6$}

The modular forms of weight $k$ and level $N$ form a linear space $\mathcal{M}_{k}(\Gamma(N))$, and the general dimension formula of $\mathcal{M}_{k}(\Gamma(N))$ is~\cite{diamond2005first,schultz2015notes}
\begin{eqnarray}
\label{eq:dimNLa}\texttt{dim}\mathcal{M}_{k}(\Gamma(N))=\dfrac{(k-1)N+6}{24}N^2 \prod_{p|N}\left(1-\dfrac{1}{p^2}\right),
\end{eqnarray}
for $12>N>2,\,k\geq 1$ or $N>2,\,k\geq 2$, where the product is over the prime divisors $p$ of $N$. For the concerned level $N=6$, the above dimensional formula is
\begin{equation}
\texttt{dim}\mathcal{M}_k(\Gamma(6))=6k , ~~~k \geq 1 \,.
\end{equation}
Therefore there are six linear independent modular forms of weight $k=1$ at level 6.

The basis (only $q$-series form) of weight $k=1$ and level $N=6$ modular forms space can be extracted by using the SageMath algebra system~\cite{SageMath:2018} as follows:
\begin{align}
\nonumber
&a_1(\tau) = 1 + 6 q^2 + 6 q^6 + 6 q^8 + 12 q^{14} + 6 q^{18} + 6 q^{24} + 12 q^{26}+\dots \,,\\
\nonumber
&a_2(\tau) = q^{1/6} (1 + 2 q + 2 q^2 + 2 q^3 + q^4 + 2 q^5 + 2 q^6 + 2 q^7 +\dots )\,, \\
\nonumber
&a_3(\tau) = q^{1/3} (1 + q + 2 q^2 + 2 q^4 + q^5 + 2 q^6 + q^8 + 2 q^9 +\dots) \,, \\
\nonumber
&a_4(\tau) =q^{1/2} (1 + q + 2 q^3 + q^4 + 2 q^6 + 2 q^9 + 2 q^{10} + q^{12}+ \dots) \,, \\
\nonumber
&a_5(\tau) =q^{2/3} (1 + q^2 + 2 q^4 + 2 q^8 + q^{10} + 2 q^{12} + q^{16} + 2 q^{18}  + \dots) \,, \\
&a_6(\tau) = q(1 - q + q^2 + q^3 - q^5 + 2 q^6 - q^7 + q^8 + \dots) \,,
\end{align}
with $q=e^{2\pi i \tau}$. Analytically the integral weight modular forms of level $N=6$ can be constructed from the product of the Dedekind eta function defined by
\begin{equation}
\eta(\tau)=q^{1/24}\prod_{n=1}^\infty \left(1-q^n \right),\qquad q= e^{ 2 \pi i\tau}\,.
\end{equation}
The eta function transforms under the actions of $S$ and $T$ as follows~\cite{diamond2005first}
\begin{equation}\label{eq:eta_tra}
 \eta(\tau)\stackrel{S}{\longrightarrow}\eta(-1/\tau)=\sqrt{-i \tau}~\eta(\tau),\qquad
\eta(\tau)\stackrel{T}{\longrightarrow}\eta(\tau+1)=e^{i \pi/12}\eta(\tau)\,.
\end{equation}
Notice that $\Gamma(6) \subset \Gamma(3)$, therefore we have $\mathcal{M}_1(\Gamma(3)) \subset \mathcal{M}_1(\Gamma(6))$. By performing the lift $f(\tau) \to f(2\tau)$ where $f(\tau)$ is modular form of $\Gamma(3)$, we can obtain some modular forms of $\Gamma(6)$ from those of $\Gamma(3)$~\footnote{
For any positive divisor $d$ of $N/M$, the transformation $f(\tau)\to f(d\tau)$ maps the modular form space $\mathcal{M}_k(\Gamma(M))$ into the modular form space $\mathcal{M}_k(\Gamma(N))$~\cite{diamond2005first}.}. To be more specific, the modular forms space of level $N=3$ and weight $k=1$ is~\cite{schultz2015notes,Liu:2019khw}
\begin{equation}
\mathcal{M}_{1}(\Gamma(3)) = \Big\{\frac{\eta^3(3\tau)}{\eta(\tau)},\quad  \frac{\eta^3(\tau/3)}{\eta(\tau)} \Big\}\,,
\end{equation}
from which we can obtain four weight 1 modular forms of $\Gamma(6)$:
\begin{equation}
\frac{\eta^3(3\tau)}{\eta(\tau)},~~ \frac{\eta^3(\tau/3)}{\eta(\tau)}, ~~ \frac{\eta^3(6\tau)}{\eta(2\tau)}, ~~\frac{\eta^3(2\tau/3)}{\eta(2\tau)}\,.
\end{equation}
The modular form space $\mathcal{M}_k(\Gamma(N))$ must be closed up to the automorphy factor $(c\tau+d)^k$ under the action of full modular group $\Gamma$. Considering the modular generator $S$, we can obtain the other two weight 1 modular forms of $\Gamma(6)$:
\begin{eqnarray}
\nonumber&&\frac{\eta^3(6\tau)}{\eta(2\tau)} \stackrel{S}{\longrightarrow}\frac{\eta^3(-6/\tau)}{\eta(-2/\tau)}=-\tau\frac{i}{6\sqrt{3}}\frac{\eta^3(\tau/6)}{\eta(\tau/2)}\,,\\
&&\frac{\eta^3(2\tau/3)}{\eta(2\tau)}\stackrel{S}{\longrightarrow}\frac{\eta^3(-2/(3\tau))}{\eta(-2/\tau)}=-\tau\frac{3\sqrt{3}\,i}{2}\frac{\eta^3(3\tau/2)}{\eta(\tau/2)}\,.
\end{eqnarray}
As a consequence, the six independent basis vectors of the linear space $\mathcal{M}_1(\Gamma(6))$ can be chosen to be
\begin{eqnarray}
&&\nonumber
e_1(\tau)= \frac{\eta^3(3\tau)}{\eta(\tau)}, \qquad ~ e_2(\tau)= \frac{\eta^3(\tau/3)}{\eta(\tau)},\qquad~~ e_3(\tau)=\frac{\eta^3(6\tau)}{\eta(2\tau)}, \\
&&e_4(\tau)= \frac{\eta^3(\tau/6)}{\eta(\tau/2)} ,\qquad  e_5(\tau)=\frac{\eta^3(2\tau/3)}{\eta(2\tau)} ,\qquad e_6(\tau)=\frac{\eta^3(3\tau/2)}{\eta(\tau/2)}\,.
\end{eqnarray}
From the expressions of $q-$expansion, we find that the above basis vectors $e_{1,2,3,4,5,6}(\tau)$ are linear combinations of $a_{1,2,3,4,5,6}(\tau)$ given by SageMath,
\begin{eqnarray}
\nonumber &&e_1(\tau)=a_3(\tau)\,,   \qquad~ \quad \qquad \quad e_2(\tau)=a_{1}(\tau)-3 a_{3}(\tau)+6 a_{6}(\tau)\,, \\
\nonumber &&e_3(\tau)=a_5(\tau)\,,  \qquad \quad \qquad \quad~ e_4(\tau)=a_{1}(\tau)-3 a_{2}(\tau)+6 a_{4}(\tau)-3 a_{5}(\tau)\,, \\
&&e_5(\tau)=a_{1}(\tau)-3 a_{5}(\tau)\,,   \qquad~ e_6(\tau)=a_{2}(\tau)+a_{5}(\tau)\,.
\end{eqnarray}
As one can check, under the actions of the generators $S$ and $T$, the modular forms $e_i(\tau)$ transform as follows:
\begin{align}
&\begin{pmatrix}
e_1(\tau) \\ e_2(\tau) \\ e_3(\tau) \\ e_4(\tau) \\ e_5(\tau) \\ e_6(\tau)
\end{pmatrix} \stackrel{S}{\longrightarrow}  \begin{pmatrix}
e_1(-\frac{1}{\tau}) \\ e_2(-\frac{1}{\tau}) \\ e_3(-\frac{1}{\tau}) \\ e_4(-\frac{1}{\tau}) \\ e_5(-\frac{1}{\tau}) \\ e_6(-\frac{1}{\tau})
\end{pmatrix}=
(-i\tau) \left(
\begin{array}{cccccc}
 0 & \dfrac{1}{3 \sqrt{3}} & 0 & 0 & 0 & 0 \\
 3 \sqrt{3} & 0 & 0 & 0 & 0 & 0 \\
 0 & 0 & 0 & \dfrac{1}{6 \sqrt{3}} & 0 & 0 \\
 0 & 0 & 6 \sqrt{3} & 0 & 0 & 0 \\
 0 & 0 & 0 & 0 & 0 & \dfrac{3 \sqrt{3}}{2} \\
 0 & 0 & 0 & 0 & \dfrac{2}{3 \sqrt{3}} & 0 \\
\end{array}
\right) \begin{pmatrix}
e_1(\tau) \\ e_2(\tau) \\ e_3(\tau) \\ e_4(\tau) \\ e_5(\tau) \\ e_6(\tau)
\end{pmatrix} \,,\\
&\begin{pmatrix}
e_1(\tau) \\ e_2(\tau) \\ e_3(\tau) \\ e_4(\tau) \\ e_5(\tau) \\ e_6(\tau)
\end{pmatrix} \stackrel{T}{\longrightarrow}  \begin{pmatrix}
e_1(\tau+1) \\ e_2(\tau+1) \\ e_3(\tau+1) \\ e_4(\tau+1) \\ e_5(\tau+1) \\ e_6(\tau+1)
\end{pmatrix}=
\left(
\begin{array}{cccccc}
 \omega  & 0 & 0 & 0 & 0 & 0 \\
 3 (1-\omega ) & 1 & 0 & 0 & 0 & 0 \\
 0 & 0 & \omega ^2 & 0 & 0 & 0 \\
 0 & 0 & 6 \left(1-\omega ^2\right) & -1 & 2 & -3 \left(1-\omega ^2\right) \\
 0 & 0 & 3 \left(1-\omega ^2\right) & 0 & 1 & 0 \\
 0 & 0 & 2 \omega ^2 & 0 & 0 & -\omega ^2 \\
\end{array}
\right)\begin{pmatrix}
e_1(\tau) \\ e_2(\tau) \\ e_3(\tau) \\ e_4(\tau) \\ e_5(\tau) \\ e_6(\tau)
\end{pmatrix} \,,
\end{align}
with $\omega=e^{i2\pi/3}$. These six linearly independent modular forms $e_i(\tau)$ can be arranged into two irreducible representations of $\Gamma'_6 \cong S_{3}\times T^{\prime}$: a doublet $\bm{2^0_2}$ and a quartet $\bm{4_1}$,
\begin{align}
	\label{eq:TprimexS3_Multiplets}
\nonumber	&Y_{\bm{2^0_2}}^{(1)}(\tau) \equiv \begin{pmatrix}
		Y_1 \\ Y_2
	\end{pmatrix} = \begin{pmatrix}
		3e_1(\tau) +  e_2(\tau)  \\[0.1in]
		3\sqrt{2}e_1(\tau)
	\end{pmatrix}\,, \\[0.1in]
	&Y_{\bm{4_1}}^{(1)}(\tau) \equiv \begin{pmatrix}
		Y_3 \\ Y_4 \\ Y_5 \\ Y_6
	\end{pmatrix} = \begin{pmatrix}
		3\sqrt{2}e_3(\tau) \\[0.1in]
		-3e_3(\tau)-e_5(\tau) \\[0.1in]
		\sqrt{6}(e_3(\tau) -  e_6(\tau))\\[0.1in]
		-\sqrt{3}e_3(\tau)+\frac{1}{\sqrt{3}}e_4(\tau)-\frac{1}{\sqrt{3}}e_5(\tau)+\sqrt{3}e_6(\tau)
	\end{pmatrix}\,.
\end{align}
The $q$-expansions of $Y_i(\tau)$ read as
\begin{align}
\nonumber
&Y_1(\tau)=1 + 6 q + 6 q^3 + 6 q^4 + 12 q^7 + 6 q^9 + 6 q^{12} + 12 q^{13} +\dots \,,\\
\nonumber
&Y_2(\tau)=3\sqrt{2}q^{1/3}( 1 + q + 2 q^2 + 2 q^4 + q^5 + 2 q^6 + q^8 + 2 q^9 +\dots) \,,\\
\nonumber
&Y_3(\tau)=3\sqrt{2} q^{2/3} (1 + q^2 + 2 q^4 + 2 q^8 + q^{10} + 2 q^{12} + q^{16} + 2 q^{18}+\dots)\,,\\
\nonumber
&Y_4(\tau)= -(1 + 6 q^2 + 6 q^6 + 6 q^8 + 12 q^{14} + 6 q^{18} + 6 q^{24} + 12 q^{26} +\dots )\,,\\
\nonumber
&Y_5(\tau)=-\sqrt{6}q^{1/6}(1 + 2 q + 2 q^2 + 2 q^3 + q^4 + 2 q^5 + 2 q^6 + 2 q^7 +\dots )\,,\\
\label{eq:Yi-expansion}&Y_6(\tau)= 2\sqrt{3} q^{1/2} (1 + q + 2 q^3 + q^4 + 2 q^6 + 2 q^9 + 2 q^{10} + q^{12}+\dots) \,.
\end{align}
The $Y_i(\tau)$ generate the ring of the modular forms of level 6, and the modular forms of higher weights can be built from the products of $Y_i(\tau)$, as shown in appendix~\ref{app:hig_wei}.  From Eq.~\eqref{eq:Yi-expansion}, we see that all the Fourier coefficients of $ Y_i$ are all real, thus we have the identity $Y_i(-\tau^*)=Y^*_i(\tau)$. Furthermore, the CG coefficients of the $\Gamma'_6$ group are real in the working basis, hence all the integral weight modular forms of level 6 satisfy $Y^{(k)}_{\mathbf{r}}(-\tau^{*})=Y^{(k)*}_{\mathbf{r}}(\tau)$.

\subsection{\label{Sec:gCP}Generalized CP consistent with $\Gamma^\prime_{6}$ modular symmetry}

In order to make modular invariant models more predictive, we shall assume the theory is invariant under both modular symmetry and gCP symmetry. It is well known that the modular symmetry $SL(2,\mathbb{Z})$ can consistently combine the gCP symmetry~\cite{Acharya:1995ag,Dent:2001cc,Giedt:2002ns,Baur:2019kwi,Novichkov:2019sqv,Ding:2021iqp}, and the modulus $\tau$ transforms under CP as
\begin{equation}\label{eq:tau_CP}
	\tau \stackrel{\mathcal{CP}}{\longrightarrow} -\tau^* \, .
\end{equation}
If the gCP symmetry is imposed, the modular symmetry $\Gamma\cong SL(2,\mathbb{Z})$ is extended to $\Gamma^{*}\cong GL(2,\mathbb{Z})$ with
\begin{equation}
\Gamma^{*}= \Big\{ \tau \stackrel{S}{\longrightarrow} -1/\tau, ~~ \tau \stackrel{T}{\longrightarrow} \tau + 1, ~~  \tau \stackrel{\mathcal{CP}}{\longrightarrow} -\tau^{*} \Big\}
\end{equation}
The action of $\Gamma^{*}$ on the complex modulus $\tau$
is given by
\begin{equation}
	\begin{pmatrix} a &~ b \\ c &~ d \end{pmatrix}
	\in \Gamma^*:\quad
	\begin{cases}
		\tau \to \cfrac{a\tau + b}{c\tau + d} &\quad \text{for} \quad ad - bc = 1\,, \\
		\tau \to \cfrac{a\tau^{*} + b}{c\tau^{*} + d}
		&\quad \text{for} \quad ad - bc = -1\,.
	\end{cases}
\end{equation}
Under the action of gCP transformation, the matter fields and the modular form multiplets transform as
\begin{equation}
	\psi(x)\stackrel{\mathcal{CP}}{\longrightarrow} X_{\mathbf{r}}\overline{\psi}(t, -{\bf x}), \qquad Y(\tau)\stackrel{\mathcal{CP}}{\longrightarrow} Y(-\tau^{*})=X_{\mathbf{r}}Y^{*}(\tau)\,,
\end{equation}
where $X_{\mathbf{r}}$ is the gCP transformation matrix in the irreducible representation $\mathbf{r}$. The consistency between the modular symmetry and gCP symmetry requires the following condition has to be fulfilled~\cite{Novichkov:2019sqv,Ding:2021iqp}
\begin{equation}
\label{eq:consistency-cond}X_{\mathbf{r}}\rho^{*}_{\mathbf{r}}(\gamma)X^{-1}_{\mathbf{r}}=\chi^{-k}(\gamma)\rho_{\mathbf{r}}(u(\gamma)), \qquad \gamma\in\Gamma\,,
\end{equation}
where $\chi(\gamma)$ called character is a homomorphism of $SL(2,\mathbb{Z})$ to $\left\{+1, -1\right\}$ with $\chi(S)=\chi(T)=1$ or $\chi(S)=\chi(T)=-1$, and $u(\gamma)$ is an outer automorphism of the modular group,
\begin{equation}
\gamma=\begin{pmatrix}
a  &  b \\
c  &  d
\end{pmatrix}\longmapsto u(\gamma)=\chi(\gamma)\begin{pmatrix}
a  &  -b \\
-c  &  d
\end{pmatrix}\,.
\end{equation}
The gCP transformation with $\chi(S)=\chi(T)=-1$ will leads to vanishing mixing angles~\cite{Novichkov:2020eep,Qu:2021jdy}, consequently we consider the case with trivial character $\chi(S)=\chi(T)=1$. Then the consistency condition in Eq.~\eqref{eq:consistency-cond} for the generators $\gamma=S, T$ takes the following form,
\begin{equation} X_{\mathbf{r}}\rho^{*}_{\mathbf{r}}(S)X^{-1}_{\mathbf{r}}=\rho_{\mathbf{r}}(S^{-1}),\qquad X_{\mathbf{r}}\rho^{*}_{\mathbf{r}}(T)X^{-1}_{\mathbf{r}}=\rho_{\mathbf{r}}(T^{-1})\,.
\end{equation}
The representation matrices of generators $S$ and $T$ are unitary and symmetric in our working basis, as shown in table~\ref{tab:rep}. Therefore the gCP transformation matrix $X_{\mathbf{r}}$ is fixed to be unity matrix up to an overall phase, i.e.
\begin{equation}
X_{\mathbf{r}}=\mathbb{1}_{\mathbf{r}}\,.
\end{equation}
Moreover, the CG coefficients given in appendix~\ref{app:group-theory} are real, hence gCP invariance requires all coupling constants to be real in the following models.

\subsection{\label{sub:MF_Tp}Decomposing modular forms of level $N=6$ into the representations of $\Gamma(2)/\Gamma(6) \cong T'$}
The $\Gamma(6)$ is a normal subgroup of $\Gamma(2)$, Eq.~\eqref{eq:factor-hom-FMG} gives
\begin{equation}
\Gamma(2)/\Gamma(6)=\Gamma'_3\cong T' = \langle \tilde{a}, \tilde{b} ~|~ \tilde{a}^4 = (\tilde{a} \tilde{b})^3 = \tilde{b}^3=1, ~\tilde{a}^2 \tilde{b} = \tilde{b} \tilde{a}^2  \rangle
\end{equation}
where $\tilde{a}=TST^4S^3T$ and $\tilde{b}=T^4$ are the two generators of principal congruence subgroup $\Gamma(2)$. Here $T^\prime$ is the double covering of the tetrahedron group $A_4$. The representation matrices of $\tilde{a}$ and $\tilde{b}$ in different irreducible representations of $T'$ are given in table~\ref{tab:rep2}.
From the general properties of modular forms~\cite{Liu:2019khw}, we know  that the modular forms of $\Gamma(6)$ can be decomposed into the irreducible multiplets of the finite group $\Gamma(2)/\Gamma(6) \cong T'$. The six weight 1 modular forms $Y_i(\tau)$ of level 6 in Eq.~\eqref{eq:Yi-expansion} can be arranged into the three doublets of $T'$
\begin{equation}
	Y_{\bm{2^0_2}}^{\prime(1)}(\tau) = \begin{pmatrix}
		Y_1 \\ Y_2
	\end{pmatrix} , \qquad
	Y_{\bm{2^0_1}i}^{\prime(1)}(\tau) = \begin{pmatrix}
		Y_3 \\ Y_4
	\end{pmatrix} , \qquad
	Y_{\bm{2^0_1}ii}^{\prime(1)}(\tau) = \begin{pmatrix}
		Y_5 \\ Y_6
	\end{pmatrix} \,,
\end{equation}
where $Y_{\bm{2^0_1}i}^{\prime(1)}$ and $Y_{\bm{2^0_1}ii}^{\prime(1)}$ denote the two weight 1 modular forms transforming in the representation $\bm{2^0_1}$ of $T'$. It is worth mentioning that the multiplet $Y_{\bm{2^0_2}}^{\prime(1)}(\tau)$ is the same as $Y_{\bm{2^0_2}}^{(1)}(\tau)$ in Eq.~\eqref{eq:TprimexS3_Multiplets}. However the quartet $Y_{\bm{4_1}}^{(1)}(\tau)$ in Eq.~\eqref{eq:TprimexS3_Multiplets} breaks down to two doublets $Y_{\bm{2^0_1}i}^{\prime(1)}(\tau) $ and $Y_{\bm{2^0_1}ii}^{\prime(1)}(\tau)$. Furthermore $Y_{\bm{2^0_2}}^{\prime(1)}(\tau)$ corresponds to the unique weight 1 modular form $Y^{(1)}_{\bm{2}}(\tau)$ of level 3 given in~\cite{Liu:2019khw}. The different expressions of $Y_{\bm{2^0_2}}^{\prime(1)}(\tau)$ and $Y^{(1)}_{\bm{2}}(\tau)$ arise from the different representation bases used in the present work and~\cite{Liu:2019khw}. As shown in section~\ref{Sec:gCP}, all couplings would be real if we impose gCP as a symmetry on the theory.

\subsection{Decomposing modular forms of level $N=6$ into representations of $\Gamma(3)/\Gamma(6) \cong S_3$}

Similar to previous case, $\Gamma(6)$ is a normal subgroup of $\Gamma(3)$, and we have
\begin{equation}
\Gamma(3)/\Gamma(6) \cong S_3 = \langle \tilde{c}, \tilde{d} ~|~ \tilde{c}^2 = (\tilde{c} \tilde{d})^3 = \tilde{d}^2=1 \rangle\,,
\end{equation}
where $\tilde{c}=S^3T^3S$, $\tilde{d}=T^3$ are the generators of $\Gamma(3)$. We summarize the irreducible representation matrices of $\tilde{c}$ and $\tilde{d}$ in table~\ref{tab:rep2}. Analogously the six weight 1 modular forms $Y_i(\tau)$ can be arranged into two singlets and two doublets of $S_3$ as follows
\begin{equation}
Y_{\bm{1^0_0}i}^{\prime\prime(1)}(\tau) = Y_1 , \qquad
Y_{\bm{1^0_0}ii}^{\prime\prime(1)}(\tau) =Y_2, \qquad
Y_{\bm{2_0}i}^{\prime\prime(1)}(\tau) = \begin{pmatrix}
Y_3 \\ Y_5
\end{pmatrix} , \qquad
Y_{\bm{2_0}ii}^{\prime\prime(1)}(\tau) = \begin{pmatrix}
Y_4 \\ Y_6
\end{pmatrix} \,.
\end{equation}
Similarly the decomposition of higher weight modular forms under $S_3$ can also be performed, we will not enumerate them here. It is interesting to note that the modular weight can be generic nonnegative integer while only even weight modular forms are allowed in the original modular invariant theory with $\Gamma_2\cong S_3$ modular symmetry~\cite{Feruglio:2017spp,Liu:2019khw}

\section{\label{Sec:model_G6p}Model building based on $\Gamma^\prime_{6}$ modular symmetry}

In this section we shall construct some typical lepton models based on the $\Gamma^\prime_6$ modular symmetry, and no flavon other than complex modulus $\tau$ will be introduced. The modular symmetry is broken by the VEV of $\tau$. If gCP symmetry is imposed on the theory, the VEV $\langle\tau\rangle$ also results in CP violation. Without loss of generality, we can limit the value of $\langle\tau\rangle$ in the fundamental domain $\mathcal{D}_{\Gamma}$ of $SL(2,\mathbb{Z})$:
\begin{equation}
\label{eq:fundomain_Gamma}
\mathcal{D}_{\Gamma}=\left\{\tau ~\big|~ |\Re{\tau}|\leq1/2\,,~\Im{\tau}>0\,,~ |\tau|\geq1 \right\}\,.
\end{equation}
Other values of $\langle\tau\rangle$ are related by modular symmetry to those in $\mathcal{D}_{\Gamma}$.
In the present work, the construction of modular models is guided by the principle of minimality and simplicity. The neutrinos are assumed to be Majorana particles in these models, and the two Higgs doublets fields $H_{u,d}$ are assumed to be trivial singlet $\bm{1^{0}_0}$ under  $\Gamma^\prime_6$ with vanishing modular weights. The modular group $\Gamma'_6$ is isomorphic to $S_3\times T'$, the direct products of the irreducible representations of $S_3$ and $T'$ give all irreducible representations of $\Gamma'_6$. Therefore all models based on $S_3$ and $T'$ modular symmetry can be reproduced from $\Gamma'_6$. Besides the singlet, doublet and triplet representations, $\Gamma^\prime_{6}$ has the four-dimensional representations $\bm{4_0}$, $\bm{4_1}$, $\bm{4_2}$ and six-dimensional representations $\mathbf{6}$ which don't appear in both $S_{3}$ and $T^\prime$ modular symmetries, and there are level 6 modular forms in the above high dimensional representations. Hence the modular group $\Gamma'_6$ opens up new model building possibilities not available in these two finite modular groups. The lepton fields are assigned to be three singlets, singlet plus doublet or a triplet of $\Gamma'_6$ multiplets. We consider five possible representation assignments in the present work, as summarized in table~\ref{tab:five_mod}. More free parameters are needed to explain the experimental data for other assignments, consequently we don't show them in this work. In the following, we present the five types of models with $\Gamma'_6$ modular symmetry together with come typical examples.

\begin{table}[t!]
\centering
\resizebox{1.0\textwidth}{!}{
\begin{tabular}{|c||c|c|c|c|c|c|c|c|c|c|c|c|c|c|c|c|c|c|c|c|c|}
\hline\hline
\diagbox{Fields}{Models} & Type I & Type II & Type III & Type IV & Type V \\ \hline
$L$	& triplet & triplet & triplet & singlet+doublet & singlet+doublet   \\ \hline
		$E^{c}$	& three singlets & three singlets & singlet+doublet & singlet+doublet & triplet   \\ \hline
		$N^{c}$	& triplet & singlet+doublet & triplet & --- & doublet   \\ \hline \hline
	\end{tabular} }
\caption{\label{tab:five_mod}The representation assignments of lepton matter fields $L$, $E^{c}$ and $N^{c}$ under the finite modular group $\Gamma^\prime_{6}$ in the five types of models.}
\end{table}

\subsection{\label{sec:typ_1_model} Type I models }

In the first type of models, the three generations of left-handed (LH) lepton doublets $L$ are assigned to a $\Gamma^\prime_6$ triplet $\bm{3^r}$ with the modular weight $k_L$, and the three right-handed (RH) charged leptons $E^{c}_{1,2,3}$ are $\Gamma^\prime_6$ singlets transforming as $\bm{1^{s}_k}$ with weights $k_{E^c_{1,2,3}}$. In these modes, the neutrino masses are described by the type I seesaw mechanism with 3RHN.

\subsubsection{\label{subsec:ch1}Charged lepton sector  }

In this kind of models, the modular form which couples to $E^{c}_{i}$ and $L$ should be in the representation $\bm{3^{t}}$ of $\Gamma^\prime_6$ with $t\equiv -(r+s)~(\text{mod}~2)$. Thus the most general superpotential for the charged lepton masses can be written as :
\begin{equation}\label{eq:We1}
\mathcal{W}_e\,= \alpha (E^c_1 L f_{1}(Y))_{\bm{1^0_0}}H_d +\beta( E^c_2 L f_{2}(Y))_{\bm{1^0_0}}H_d +\gamma (E^c_3L f_{3}(Y))_{\bm{1^0_0}}H_d\,,
\end{equation}
where $f_{i}(Y)~(i=1,2,3)$ stand for the level 6 modular forms which transform as $\bm{3^{t}}$ given in table~\ref{tab:MF_summary}. The three terms in Eq.~\eqref{eq:We1} have similar form, and they contribute to three rows of the charged lepton mass matrix respectively. Guided by the principle of minimality and simplicity, we shall consider the lower weight modular forms in the following. As an example, let us discuss all possible forms of the first row of the charged lepton mass matrix. Without loss of generality, we assume $L\sim\bm{3^0}$ and $E^c_1\sim\bm{1^r_i}$. For the assignment $E^c_1\sim\bm{1^0_i}$, the modular form function $f_{1}(Y)$  is $Y^{(2)}_{\bm{3^0}}$ for $k_L+k_{E^c_1}=2$ and $Y^{(4)}_{\bm{3^0}}$ for $k_L+k_{E^c_1}=4$. Then we can read out the first row of the charged lepton mass matrix as follows
\begin{equation}\label{eq:row_ch1}
R_{w,i}=\alpha(Y^{(w)}_{\bm{3^0},1},Y^{(w)}_{\bm{3^0},3},Y^{(w)}_{\bm{3^0},2})P^i_{3}, \qquad w=2,4, \quad i=0,1,2\,,
\end{equation}
where $w=2$ for $k_L+k_{E^c_1}=2$ and $w=4$ for $k_L+k_{E^c_1}=4$, and the permutation matrix $P_{3}$ is shown in Eq.~\eqref{eq:sim_mart}. Here we use $Y^{(w)}_{\bm{3^0},1}$, $Y^{(w)}_{\bm{3^0},2}$ and $Y^{(w)}_{\bm{3^0},3}$ to represent the three entries of modular multiple $Y^{(w)}_{\bm{3^0}}$, and this convention will be used throughout the paper. For the assignment $E^c_1\sim\bm{1^1_i}$, the possible values of the modular weights $k_L+k_{E^c_1}$ are $4$ and $6$, and accordingly the modular function $f_{1}(Y)$ is $Y^{(4)}_{\bm{3^1}}$ and $Y^{(6)}_{\bm{3^1}}$, respectively. For these two cases, the first row of the charged lepton mass matrix reads as
\begin{equation}
 R^\prime_{w,i}=\alpha(Y^{(w)}_{\bm{3^1},1},Y^{(w)}_{\bm{3^1},3},Y^{(w)}_{\bm{3^1},2})P^i_{3}, \qquad w=4,6, \quad i=0,1,2\,.
\end{equation}
From appendix~\ref{app:hig_wei} we know
\begin{equation}
		Y^{(4)}_{\bm{3^{1}}}=\left(Y^{(2)}_{\bm{1^{1}_2}}Y^{(2)}_{\bm{3^{0}}}\right)_{\bm{3^{1}}}  =
		Y^{(2)}_{\bm{1^{1}_2}}\left(
		\begin{array}{c}
			Y^{(2)}_{\bm{3^0},2} \\
			Y^{(2)}_{\bm{3^0},3} \\
			Y^{(2)}_{\bm{3^0},1}
		\end{array}
		\right)\,, \quad
		Y^{(6)}_{\bm{3^{1}}}=2\left(Y^{(2)}_{\bm{1^{1}_2}}Y^{(4)}_{\bm{3^{0}}}\right)_{\bm{3^{1}}}=
		2Y^{(2)}_{\bm{1^{1}_2}}\left(
		\begin{array}{c}
			Y^{(4)}_{\bm{3^0},2} \\
			Y^{(4)}_{\bm{3^0},3} \\
			Y^{(4)}_{\bm{3^0},1}
		\end{array}
		\right) \,.
\end{equation}
Therefore one can easily check the following identity is satisfied
\begin{equation}
 R^\prime_{4,i}=Y^{(2)}_{\bm{1^{1}_2}}R_{2,i}P^2_{3},~~~~R^\prime_{6, i}=2Y^{(2)}_{\bm{1^{1}_2}}R_{4, i}P^2_{3}\,.
\end{equation}
The factor $Y^{(2)}_{\bm{1^{1}_2}}$ can be absorbed into the coupling $\alpha$, thus $R_{w, i}$ and $R^\prime_{w, i}$ give the same set of possible forms of the first row of the charged lepton mass matrix. Hence it is sufficient to consider the assignments of $L\sim\bm{3^0}$ and $E^c_1\sim\bm{1^0_i}$ with $k_L+k_{E^c_1}=2$ and $k_L+k_{E^c_1}=4$, and
the first row of the charged lepton mass matrix can take the 6 forms shown in Eq.~\eqref{eq:row_ch1}. Analogously the other two rows of the charged lepton mass matrix can also take these six forms. Notice that any two rows of a charged lepton mass matrix should not be proportional to each other, otherwise at least one charged lepton would be massless.
In the end, we find there are total 20 possible charged lepton mass matrices (up to row permutations). They can be written as

\begin{table}[t!]
\centering
\renewcommand{\tabcolsep}{0.1mm}
\begin{tabular}{c}
\begin{tabular}{|c|c|c|c||c|c|c|c|}  \hline\hline
   Models & $m_{e}$ & $(\rho_{E_{1}^{c}},\rho_{E_{2}^{c}},\rho_{E_{3}^{c}})$ & $(k_{c_1},k_{c_2},k_{c_3})$ &    Models & $m_{e}$ & $(\rho_{E_{1}^{c}},\rho_{E_{2}^{c}},\rho_{E_{3}^{c}})$ & $(k_{c_1},k_{c_2},k_{c_3})$ \\  \hline
  $C_{1}$ & $m_{e1}(2,0;2,1;2,2)$ & $(\bm{1^0_0},\bm{1^0_1},\bm{1^0_2})$ & $(2,2,2)$ & $C_{2}$ & $m_{e1}(2,0;2,1;4,0)$ & $(\bm{1^0_0},\bm{1^0_1},\bm{1^0_0})$ & $(2,2,4)$ \\ \hline

  $C_{3}$ & $m_{e1}(2,0;2,1;4,1)$ & $(\bm{1^0_0},\bm{1^0_1},\bm{1^0_1})$ & $(2,2,4)$ & $C_{4}$ & $m_{e1}(2,0;2,1;4,2)$ & $(\bm{1^0_0},\bm{1^0_1},\bm{1^0_2})$ & $(2,2,4)$ \\ \hline

  $C_{5}$ & $m_{e1}(2,0;2,2;4,0)$ & $(\bm{1^0_0},\bm{1^0_2},\bm{1^0_0})$ & $(2,2,4)$ & $C_{6}$ & $m_{e1}(2,0;2,2;4,1)$ & $(\bm{1^0_0},\bm{1^0_2},\bm{1^0_1})$ & $(2,2,4)$ \\ \hline

  $C_{7}$ & $m_{e1}(2,0;2,2;4,2)$ & $(\bm{1^0_0},\bm{1^0_2},\bm{1^0_2})$ & $(2,2,4)$ & $C_{8}$ & $m_{e1}(2,0;4,0;4,1)$ & $(\bm{1^0_0},\bm{1^0_0},\bm{1^0_1})$ & $(2,4,4)$ \\ \hline

  $C_{9}$ & $m_{e1}(2,0;4,0;4,2)$ & $(\bm{1^0_0},\bm{1^0_0},\bm{1^0_2})$ & $(2,4,4)$ & $C_{10}$ & $m_{e1}(2,0;4,1;4,2)$ & $(\bm{1^0_0},\bm{1^0_1},\bm{1^0_2})$ & $(2,4,4)$ \\ \hline

  $C_{11}$ & $m_{e1}(2,1;2,2;4,0)$ & $(\bm{1^0_1},\bm{1^0_2},\bm{1^0_0})$ & $(2,2,4)$ & $C_{12}$ & $m_{e1}(2,1;2,2;4,1)$ & $(\bm{1^0_1},\bm{1^0_2},\bm{1^0_1})$ & $(2,2,4)$ \\ \hline

  $C_{13}$ & $m_{e1}(2,1;2,2;4,2)$ & $(\bm{1^0_1},\bm{1^0_2},\bm{1^0_2})$ & $(2,2,4)$ & $C_{14}$ & $m_{e1}(2,1;4,0;4,1)$ & $(\bm{1^0_1},\bm{1^0_0},\bm{1^0_1})$ & $(2,4,4)$ \\ \hline

  $C_{15}$ & $m_{e1}(2,1;4,0;4,2)$ & $(\bm{1^0_1},\bm{1^0_0},\bm{1^0_2})$ & $(2,4,1)$ & $C_{16}$ & $m_{e1}(2,1;4,1;4,2)$ & $(\bm{1^0_1},\bm{1^0_1},\bm{1^0_2})$ & $(2,4,4)$ \\ \hline

  $C_{17}$ & $m_{e1}(2,2;4,0;4,1)$ & $(\bm{1^0_2},\bm{1^0_0},\bm{1^0_1})$ & $(2,4,4)$ & $C_{18}$ & $m_{e1}(2,2;4,0;4,2)$ & $(\bm{1^0_2},\bm{1^0_0},\bm{1^0_2})$ & $(2,4,4)$ \\ \hline

  $C_{19}$ & $m_{e1}(2,2;4,1;4,2)$ & $(\bm{1^0_2},\bm{1^0_1},\bm{1^0_2})$ & $(2,4,4)$ & $C_{20}$ & $m_{e1}(4,0;4,1;4,2)$ & $(\bm{1^0_0},\bm{1^0_1},\bm{1^0_2})$ & $(4,4,4)$ \\   \hline\hline
\end{tabular}
\end{tabular}
\caption{\label{tab:ch_mm} The 20 possible charged lepton mass matrices and the corresponding representations and modular weights of charged leptons for the type I models, where $k_{c_i}\equiv k_L+k_{E_{i}^{c}}$ denote the sum of the modular weight of the charged lepton fields.}
\end{table}
\begin{equation}\label{eq:MCh1}
m_{e1}(w_1,i;w_2,j;w_3,k)=\begin{pmatrix}
\alpha R_{w_1,i} \\
\beta R_{w_2,j} \\
\gamma R_{w_3,k}
\end{pmatrix}\,,\quad \text{with} \quad R_{w_1,i}\neq R_{w_2,j}\neq R_{w_3,k} .
\end{equation}

\subsubsection{\label{subsec:nu1}Neutrino sector}

In this kind of models, the neutrino masses are generated through the type I seesaw mechanism. The three right-handed neutrinos are embedded into a triplet $\bm{3^r}$ of $\Gamma^\prime_{6}$. Hence the most general superpotential for the neutrino masses can be generally written as
\begin{equation}
\mathcal{W}_{N}=h(N^cLY_DH_u)_{\bm{1^0_0}}+\frac{g}{2}\Lambda(N^cN^cY_{N})_{\bm{1^0_0}}\,,
\end{equation}
where $Y_D$ and $Y_{N}$ denote the modular form multiplets of $\Gamma^\prime_{6}$. In the case of $k_{N^c}=0$ and $k_{N^c}=1$, the corresponding modular multiplets $Y_{N}$ are constants and $Y^{(2)}_{\bm{3^0}}$, respectively. Then we can read off the mass matrix for the Majorana neutrinos $N^c$,
\begin{eqnarray}
\label{eq:mN1} \hskip-0.4in&& m_{N}=
\frac{g\Lambda}{2}\begin{pmatrix}
1  &~  0  &~ 0 \\
0 &~  0 &~ 1  \\
0 &~ 1  &~ 0
\end{pmatrix},\quad \text{for}~N^c\sim\bm{3^0}, \bm{3^1} \quad \text{and} \quad k_{N^c}=0 \,,\\
\label{eq:mN2}\hskip-0.4in&& m'_{N}=\frac{g\Lambda}{2}\begin{pmatrix}
 2 Y^{(2)}_{\bm{3^0},1} &~ -Y^{(2)}_{\bm{3^0},3} &~ -Y^{(2)}_{\bm{3^0},2} \\
 -Y^{(2)}_{\bm{3^0},3} &~ 2  Y^{(2)}_{\bm{3^0},2} &~ -Y^{(2)}_{\bm{3^0},1} \\
 -Y^{(2)}_{\bm{3^0},2} &~ -Y^{(2)}_{\bm{3^0},1} &~ 2  Y^{(2)}_{\bm{3^0},3} \\
\end{pmatrix}, \quad \text{for}~N^c\sim\bm{3^0}, \bm{3^1} \quad \text{and} \quad k_{N^c}=1\,.
\end{eqnarray}
We proceed to discuss the possible form of the Dirac neutrino mass matrix. If the left-handed lepton $L$ and right-handed neutrinos $N^c$ transform as $L\sim\bm{3^0}$, $N^c\sim\bm{3^0}$ or $L\sim\bm{3^1}$, $N^c\sim\bm{3^1}$, modular invariance requires $Y_D$ should transform as $\bm{1^{0}_{0}}$, $\bm{1^{0}_{1}}$, $\bm{1^{0}_{2}}$ or $\bm{3^0}$ under $\Gamma'_6$. The lowest non-vanishing weight is $k_L+k_{N^c}=2$, and the Yukawa coupling is given by
\begin{equation}
\mathcal{W}_{\nu_D}=h_1\left(\left(N^cL\right)_{\bm{3^0_1}}Y^{(2)}_{\bm{3^0}}\right)_{\bm{1^0_0}}H_u
+h_2\left(\left(N^cL\right)_{\bm{3^0_2}}Y^{(2)}_{\bm{3^0}}\right)_{\bm{1^0_0}}H_u\,.
\end{equation}
Then the Dirac neutrino mass matrix takes the following form
\begin{equation}\label{eq:MD_N1L1}
m_{D}=\begin{pmatrix}
 2 h_{1} Y^{(2)}_{\bm{3^0},1} &~ Y^{(2)}_{\bm{3^0},3} (h_{2}-h_{1}) &~ -(h_{1}+h_{2})Y^{(2)}_{\bm{3^0},2} \\
 -(h_{1}+h_{2})Y^{(2)}_{\bm{3^0},3}  &~ 2 h_{1} Y^{(2)}_{\bm{3^0},2} &~ (h_{2}-h_{1})Y^{(2)}_{\bm{3^0},1}  \\
 (h_{2}-h_{1})Y^{(2)}_{\bm{3^0},2}  &~ -(h_{1}+h_{2})Y^{(2)}_{\bm{3^0},1} &~ 2 h_{1} Y^{(2)}_{\bm{3^0},3} \\
\end{pmatrix}v_u\,.
\end{equation}
where the phase of $h_1$ can be absorbed by field redefinition, while the relative phase of $h_1$ and $h_2$ can not be eliminated. For the second possible assignment $L\sim\bm{3^0}$, $N^c\sim\bm{3^1}$ or $L\sim\bm{3^1}$, $N^c\sim\bm{3^0}$, the resulting Dirac neutrino mass matrix can be obtained from Eq.~\eqref{eq:MD_N1L1} by permutation of columns. After taking into account the charged lepton sector, no new textures of lepton mass matrices are obtained.

Notice that the group $\Gamma'_6$ is isomorphic to $S_3\times T'$, where $T'$ is the double cover of $A_4$, and the representations $\bm{1^0_k}$ ($k=0, 1, 2$) and $\bm{3^0}$ are the direct products of the $A_4$ representations with the trivial unit representation of $S_3$. As a result, the $\Gamma'_6$ modular symmetry can not be distinguished from $A_4$ in the representations $\bm{1^0_k}$ and $\bm{3^0}$. As we have shown, in this type of models it is sufficient to consider the assignments of $L\sim\bm{3^0}$, $N^c\sim\bm{3^0}$ and $E^c_{1,2,3}\sim\bm{1^0_k}$. Hence the type I models coincide with the $A_4$ modular models which have been systematically analyzed in~\cite{Yao:2020qyy}. We have checked that the numerical results for mixing parameters are the same as those of~\cite{Yao:2020qyy}.

\subsection{\label{sec:typ_2_model} Type II models }

In this type of models, the charged lepton sector is the same as that of type I models discussed in section~\ref{subsec:ch1}. The light neutrino masses arise from the type I seesaw mechanism with 3RHN, we assume that one right-handed neutrino $N^c_1$ transforms as singlet of $\Gamma^\prime_{6}$ and the other two $N^c_{d}=(N^c_{2},N^c_{3})$ form a doublet of $\Gamma^\prime_{6}$. As an example, we consider the transformation  assignment $N^c_{1}\sim\bm{1^0_0}$, $N^c_{d}\sim\bm{2_0}$ with the modular weights $k_{N^{c}_1}=k_{N^c_{d}}=0$, the modular weight of the left-handed lepton doublets $L$ is taken to be $k_L=2$. Then the most general superpotential for neutrino masses is
\begin{equation}
	\mathcal{W}_{\nu}=h_{1}N^c_{1}(LY^{(2)}_{\bm{3^0}})_{\bm{1^0_0}}H_u
	+h_{2}((N^c_{d}L)_{\bm{6}}Y^{(2)}_{\bm{6}})_{\bm{1^0_0}}H_u+\frac{g_1\Lambda}{2} N^c_{1}N^c_{1}+\frac{g_2\Lambda}{2} (N^c_{d}N^c_{d})_{\bm{1^0_0}}\,,
\end{equation}
Then one can read out the Dirac and Majorana mass matrices as follows
\begin{equation}
	m_{N}=
	\frac{\Lambda}{2}\begin{pmatrix}
		g_{1} &~ 0 &~ 0 \\
		0 &~ g_{2} &~ 0 \\
		0 &~ 0 &~ g_{2} \\
	\end{pmatrix}, \qquad
	m_{D}=\begin{pmatrix}
		h_{1} Y^{(2)}_{\bm{3^0},1} &~ h_{1} Y^{(2)}_{\bm{3^0},3} &~ h_{1} Y^{(2)}_{\bm{3^0},2} \\
		h_{2} Y^{(2)}_{\bm{6},1} &~ h_{2} Y^{(2)}_{\bm{6},3} &~ h_{2} Y^{(2)}_{\bm{6},2} \\
		h_{2} Y^{(2)}_{\bm{6},4} &~ h_{2} Y^{(2)}_{\bm{6},6} &~ h_{2} Y^{(2)}_{\bm{6},5} \\
	\end{pmatrix}v_u\,.
\end{equation}
Then the light neutrino mass matrix $m_{\nu}$ only depend on the combinations $h^2_{1}/g_{1}$ and $h^2_{2}/g_{2}$ besides the complex modulus $\tau$. The phase of $h^2_{1}/g_{1}$ is unphysical and consequently $h^2_{1}/g_{1}$ can be taken real, while the phase of $h^2_{2}/g_{2}$ can not be removed by a field redefinition. If gCP invariance is imposed, both $h^2_{1}/g_{1}$ and $h^2_{2}/g_{2}$ are required to be real, i.e. their phases are $0$ or $\pi$.

\subsubsection{Numerical results}

The charged lepton mass matrices can take the twenty forms $C_{i}$ ($i=1,\cdots, 20$) given in table~\ref{tab:ch_mm}, and they all have three real coupling constants $\alpha$, $\beta$ and $\gamma$. Thus all the six lepton masses, three lepton mixing angles and three CP violation phases depend on six dimensionless parameters
\begin{equation}
\tau , \quad \arg{(g_{1}h^2_{2}/(g_{2}h^2_{1}))}\,, ~~
\beta/\alpha,~~ \gamma/\alpha, ~~\, |g_{1}h^2_{2}/(g_{2}h^2_{1})|\,,
\end{equation}
and two overall scales $\alpha v_{d}$ and $h^2_1v^2_u/(g_{1}\Lambda)$. If gCP symmetry is considered, the parameter $\arg{(g_{1}h^2_{2}/(g_{2}h^2_{1}))}$ can only take two values $0$ and $\pi$. We shall perform a numerical analysis for each possible model  without gCP and with gCP. In order to quantitatively assess whether a model can accommodate the experimental data on lepton masses and mixing parameters, we perform a conventional $\chi^2$ analysis, and the $\chi^2$ function is defined as

\begin{table}[t!]
\begin{center}
	\renewcommand{\tabcolsep}{0.5mm}
	\centering
	\renewcommand{\arraystretch}{1.04}
	\begin{tabular}{|c|c|c|c|c|c||c|c|c|c|c|} \hline\hline
		&  \multicolumn{5}{|c||}{
charged lepton sector without gCP }
		& \multicolumn{5}{|c|}{
charged lepton sector with gCP
}             \\ \cline{2-11}
		& $C_3$ & $C_6$ & $C_{14}$ & $C_{17}$ & $C_{20}$ & $C_{3}$ & $C_6$ & $C_{14}$ & $C_{17}$ & $C_{20}$ \\ \hline		
		$\Re\langle\tau\rangle$ & $-0.279$ & $-0.244$ & $-0.281$ & 0.281 & 0.281 & 0.241 & $-0.276$ & $-0.277$ & $-0.277$ & $-0.277$ \\
		$\Im\langle\tau\rangle$ &
			1.03 & 0.904 & 1.03 & 1.03 & 1.03 & 0.903 & 1.03 & 1.03 & 1.03 & 1.03  \\
		$\beta/\alpha$ &
			0.00368 & 0.00512 & 274 & 195 & 14.9 & 0.00373 & 0.00513 & 276 & 199 & 14.8 \\
		$\gamma/\alpha$ &
			15.0 & 13.2 & 4090 & 2920 & 0.00512 & 13.3 & 15.2 & 4093 & 2956 & 0.00513 \\
		$|g_{1}h^2_{2}/(g_{2}h^2_{1})|$ &
			0.857 & 0.874 & 0.858 & 0.875 & 0.874 & 0.854 & 0.854 & 0.855 & 0.855 & 0.855 \\
		$\arg{(g_{1}h^2_{2}/(g_{2}h^2_{1}))}/\pi$ & 	0.940 & 0.184 & 1.94 & 1.18 & 1.18 & 1 & 0 & 0 & 1 & 0 \\
		$\alpha v_{d}$/MeV &
			87.5 & 76.7 & 0.321 & 0.450 & 88.0 & 75.8 & 86.7 & 0.323 & 0.447 & 88.9 \\
		$(h^2_1v^2_u/(g_{1}\Lambda))/$eV  &
			15.4 & 12.3 & 15.4 & 16.0 & 16.0 & 11.7 & 15.4 & 15.4 & 15.3 & 15.4  \\ \hline
		
		$m_e/m_{\mu }$ & 0.0048 & 0.0048 & 0.0048 & 0.0048 & 0.0048 & 0.0048 & 0.0048 & 0.0048 & 0.0048 & 0.0048 \\
		$m_{\mu }/m_{\tau }$ &	0.0565 & 0.0565 & 0.0565 & 0.0565 & 0.0565 & 0.0565 & 0.0565 & 0.0565 & 0.0565 & 0.0565  \\
		$\Delta m_{\text{atm}}^2/\Delta m_{\text{sol}}^2$ &
			0.0303 & 0.0303 & 0.0303 & 0.0303 & 0.0303 & 0.0303 & 0.0303 & 0.0303 & 0.0303 & 0.0303  \\ \hline

		$\sin^2\theta_{13}$  & 0.02152 & 0.02152 & 0.02152 & 0.02152 & 0.02152 & 0.02152 & 0.02152 & 0.02152 & 0.02152 & 0.02152 \\
		$\sin^2\theta_{12}$  &  0.297 & 0.297 & 0.297 & 0.297 & 0.297 & 0.297 & 0.297 & 0.297 & 0.297 & 0.297  \\
		$\sin^2 \theta_{23}$  & 0.573 & 0.573 & 0.573 & 0.573 & 0.573 & 0.572 & 0.572 & 0.572 & 0.572 & 0.572  \\
		$\delta_{CP}/\pi$    & 1.095 & 1.095 & 1.095 & 1.095 & 1.094 & 1.187 & 1.187 & 1.185 & 1.185 & 1.185 \\
		$\alpha_{21}/\pi$   & 	1.44 & 0.543 & 1.44 & 0.547 & 0.547 & 1.44 & 1.44 & 1.44 & 1.44 & 1.44  \\
		$\alpha_{31}/\pi$  & 0.260 & 1.97 & 0.259 & 1.97 & 1.97 & 0.374 & 0.373 & 0.371 & 0.371 & 0.371  \\ \hline

		$m_1$/meV  & 13.1 & 14.6 & 13.2 & 14.7 & 14.6 & 12.9 & 12.9 & 12.9 & 12.9 & 12.9 \\
		$m_2$/meV  & 15.7 & 16.9 & 15.8 & 17.0 & 17.0 & 15.5 & 15.5 & 15.5 & 15.5 & 15.5 \\
		$m_3$/meV & 51.1 & 51.5 & 51.1 & 51.5 & 51.5 & 51.0 & 51.0 & 51.0 & 51.0 & 51.0 \\
		$\sum m_{i}$/meV  &	79.9 & 83.0 & 80.1 & 83.1 & 83.1 & 79.3 & 79.3 & 79.5 & 79.5 & 79.5 \\
		$m_{\beta\beta}$/meV & 10.2 & 11.0 & 10.2 & 11.0 & 11.0 & 10.2 & 10.2 & 10.2 & 10.2 & 10.2 \\ \hline
		$\chi^2_{min}$ & 			2.66 & 2.67 & 2.66 & 2.67 & 2.67 & 3.04 & 3.06 & 3.03 & 3.05 & 3.03 \\ \hline \hline
		
\end{tabular} 
\caption{\label{tab:bf_N1100N2320}The best fit results for lepton masses and  mixing parameters in the five phenomenologically viable models of type II models, where both scenarios with gCP and without gCP symmetry are considered. }
\end{center}
\end{table}

\begin{equation}\label{eq:chisq}
	\chi^2 = \sum_{i=1}^7 \left( \frac{P_i-\mu_i}{\sigma_i}\right)^2\,,
\end{equation}
where $\sigma_{i}$, $\mu_{i}$ and $P_i$ refer to the $1\sigma$ deviations, the global best fit values and the theoretical predictions for the seven dimensionless observable quantities as functions of free parameters:
\begin{equation}\label{eq:obs_qua}
m_{e}/m_{\mu}, \quad m_{\mu}/m_{\tau},\quad	\sin^2\theta_{12},\quad \sin^2\theta_{13},\quad \sin^2\theta_{23}, \quad \delta_{CP},\quad \Delta m^2_{21}/\Delta m^2_{31}\,.
\end{equation}
The overall scales $\alpha v_{d}$ and $  h^2_1v^2_u/(g_{1}\Lambda)$ are fixed by the electron mass and the neutrino mass squared splitting $\Delta m^2_{21}=m^2_2-m^2_1$ respectively. For normal ordering neutrino masses, the best fit values and the $1\sigma$ errors of the observables in Eq.~\eqref{eq:obs_qua} are~\cite{ParticleDataGroup:2020ssz,Esteban:2020cvm}
\begin{eqnarray}
\nonumber &&	\frac{m_e}{m_{\tau}}\Big|_{\text{bf}+1\sigma}=0.0048^{+0.0002}_{-0.0002}\,, \quad \frac{m_{\mu}}{m_{\tau}}\Big|_{\text{bf}+1\sigma}0.0565^{+0.0045}_{-0.0045}\,,\quad \sin^2\theta_{12}|_{\text{bf}+1\sigma}=0.304^{+0.012}_{-0.012}\,,\\
\nonumber &&\sin^2\theta_{13}|_{\text{bf}+1\sigma}=0.02219^{+0.00062}_{-0.00063}\,, \qquad  \sin^2\theta_{23}|_{\text{bf}+1\sigma}=0.573^{+0.016}_{-0.020}\,,\\
&& \frac{\delta_{CP}}{\pi}\Big|_{\text{bf}+1\sigma} =1.094^{+0.15}_{-0.13}\,,\qquad  \frac{\Delta m^2_{21}}{\Delta m^2_{31}}\Big|_{\text{bf}+1\sigma}=0.0294^{+0.00089}_{-0.00089}\,.
\end{eqnarray}

We find that five of the twenty models are phenomenologically viable for both models without gCP and with gCP. The predictions for lepton mixing parameters, neutrino masses and the best fit values of the input parameters for the five phenomenologically viable models are summarized in table~\ref{tab:bf_N1100N2320}.
Here the parameter $m_{\beta\beta}$ refers to the effective Majorana mass in neutrinoless double beta decay and $\sum m_{i}$ is the sum of the light neutrino masses. At present, the most stringent bounds are
$m_{\beta\beta}<(61-165)$ meV from the KamLAND-Zen experiment~\cite{KamLAND-Zen:2016pfg} and $\sum m_i<0.12$ eV  from the Planck Collaboration~\cite{Planck:2018vyg}. We see that the minimal values of the $\chi^2$ are quite small, the experimentally measured lepton masses and mixing angles can be accommodated very well, and the values of $m_{\beta\beta}$ and $\sum_i m_i$ are much below the present experimental bounds~\cite{Planck:2018vyg,KamLAND-Zen:2016pfg}.

\begin{figure}[t!]
	\centering
	\begin{tabular}{c}
		\includegraphics[width=0.95\linewidth]{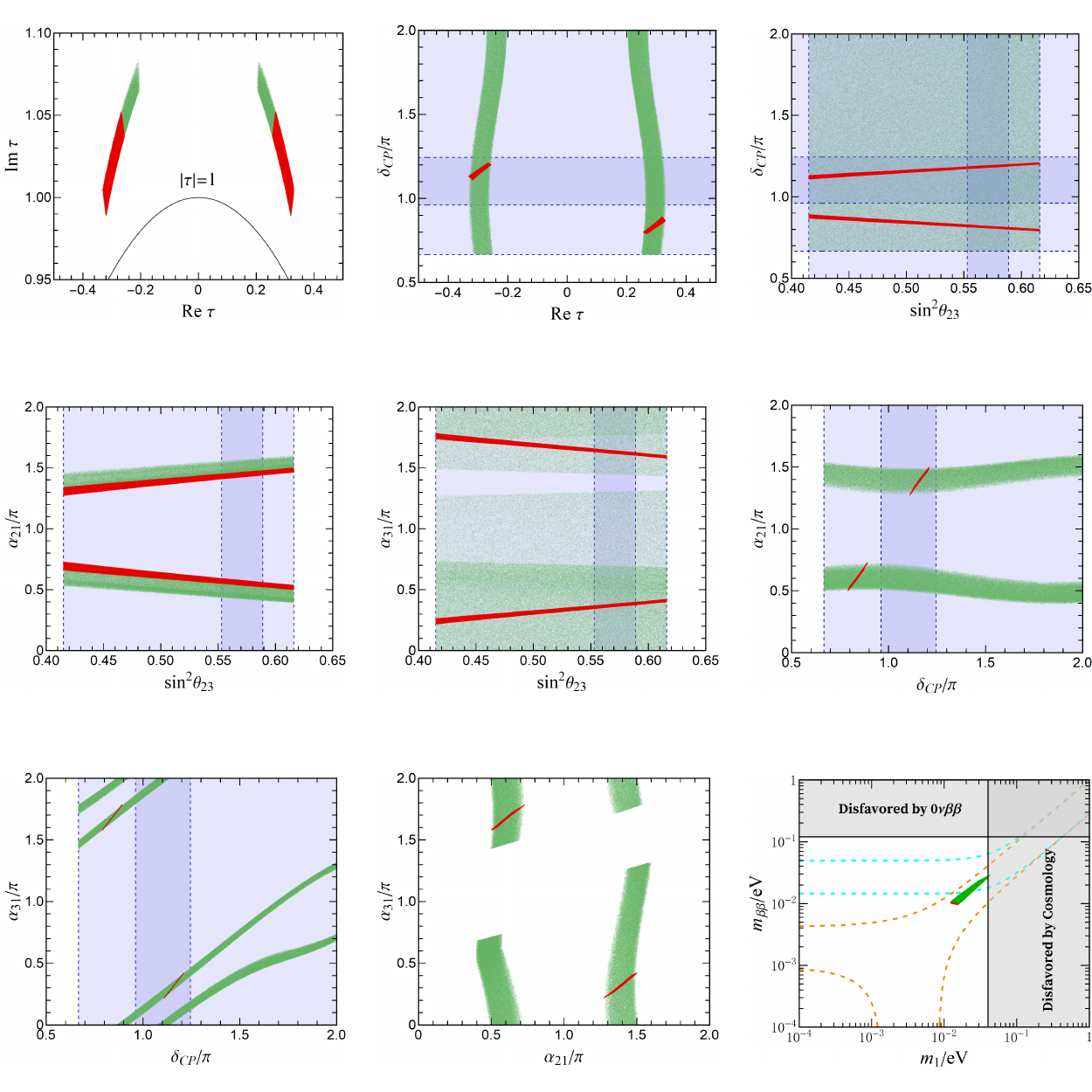}
	\end{tabular}
\caption{\label{fig:M2} The experimentally allowed values of the complex modulus $\tau$ and the correlations between the neutrino mixing parameters in the example model of type II. The green and red  regions denote points compatible with experimental data for models without and with gCP respectively. The light blue bounds represent the $1\sigma$ and $3\sigma$ ranges respectively~\cite{Esteban:2020cvm}.  The  orange (cyan) dashed lines in the last panel represent the most general  allowed regions for NO (IO) neutrino masses respectively, where the neutrino oscillation parameter are varied within their $3\sigma$ ranges. The present upper limit $m_{\beta\beta}<(61-165)$ meV from KamLAND-Zen~\cite{KamLAND-Zen:2016pfg} is shown by horizontal grey band. The vertical grey exclusion band denotes the current bound from the cosmological data of $\sum_im_i<0.120$eV at 95\% confidence level given by the Planck collaboration~\cite{Aghanim:2018eyx}.
}
\end{figure}

In order to show the viability and predictions of this type of models, we shall give detailed numerical results of the model in which the charged lepton sector is $C_6$. We comprehensively scan the parameter space and require all the observables lie in their experimentally preferred $3\sigma$ regions, some interesting correlations among the input parameters and observables are obtained, as shown in figure~\ref{fig:M2}. Here the results without gCP and with gCP are displayed in green and red respectively. From figure~\ref{fig:M2}, we find that the atmospheric mixing $\theta_{23}$ and Dirac CP phase $\delta_{CP}$ can take all possible values in their $3\sigma$ regions if gCP is not considered. However, all couplings are real and $\Re\tau$ is the unique source of CP violation if gCP invariance is required in the model. As a consequence, the three CP violation phases are  constrained to lie in narrow ranges in figure~\ref{fig:M2}.

\subsection{\label{sec:typ_3_model} Type III models }

In this type of models, the left-handed lepton fields $L$ are assigned to a triplet of $\Gamma^\prime_6$, and the right-handed charged leptons are the direct sum of singlet and doublet. Without loss of generality, we assume that $E^c_1$ with modular weight $k_{E^c_{1}}$ transforms as singlet $\bm{1^{r}_k}$ under $\Gamma^\prime_6$, and the right-handed charged leptons $E^c_{d}=(E^c_2,E^c_3)$ with modular weight $k_{E^c_{d}}$ transform as a doublet of $\Gamma^\prime_6$. The most general superpotential for the charged lepton masses is of the following form:
\begin{equation}\label{eq:We3}
	\mathcal{W}_e\,= \alpha (E^c_1 L f_{1}(Y))_{\bm{1^0_0}}H_d +\beta( E^c_{d} L f_{d}(Y))_{\bm{1^0_0}}H_d\,,
\end{equation}
where $f_{1}(Y)$ and $f_{d}(Y)$ are functions of modular forms of level 6 given in appendix~\ref{app:hig_wei}. The charged lepton mass matrix is denoted as
\begin{equation}
m_e=\begin{pmatrix}
v_1  \\
v_2  \\
v_3
\end{pmatrix}\,,
\end{equation}
where $v_i~(i=1,2,3)$ are three rows of charged lepton mass matrix $m_e$. From the analysis in section~\ref{subsec:ch1}, we find that the first row $v_1$ can take the six possible forms in Eq.~\eqref{eq:row_ch1}.

Now let us consider the possible forms of the second row $v_{2}$ and the third row $v_{3}$. For the representation assignment $L\sim\bm{3^r}$ and $E^c_{d}\sim\bm{2_i}$, modular invariance requires the presence of modular form in the representation $\mathbf{6}$. From table~\ref{tab:MF_summary}, we know that the lowest weight modular multiplet transforming as $\mathbf{6}$ under $\Gamma'_6$ is of weight 2, and accordingly the charged lepton mass term is
\begin{equation}
	\beta\left( \left(E^c_{d} L\right)_{\bm{6}} Y_{\bm{6}}^{(2)}\right)_{\bm{1^0_0}}H_d\,,
\end{equation}
which leads to the second and third rows of the charged lepton mass matrix to be
\begin{equation}
	\begin{pmatrix}
		v_2  \\
		v_3 \\
	\end{pmatrix}=\beta P^{r}_{2}\begin{pmatrix}
		Y^{(2)}_{\bm{6},1} &~   Y^{(2)}_{\bm{6},3} &~  Y^{(2)}_{\bm{6},2} \\
		Y^{(2)}_{\bm{6},4} &~   Y^{(2)}_{\bm{6},6} &~   Y^{(2)}_{\bm{6},5} \\
	\end{pmatrix}P^{i}_{3}v_d\,,
\qquad r=0,1, \quad i=0,1,2\,,
\end{equation}
where matrices $P_{2}$ and $P_{3}$ are shown in Eq.~\eqref{eq:sim_mart}. As the general form of the three rows of the charged lepton mass matrix have been obtained, then one can obtain the charged lepton mass matrix $m_{e}$ for all possible models. As the hermitian combination $m^\dagger_em_e$ will take the same form for $r=0$ and $r=1$. Then the effect of $P^{r}_{2}$ in above equation can be absorbed by right-handed charged leptons.

If the lepton fields $L$ and $E^c_{d}$ transform as $L\sim\bm{3^s}$ and $E^c_{d}\sim\bm{2^r_i}$ respectively, from the multiplication rules $\bm{2^{r}_{i}}\otimes\bm{3^{s}}=\bm{2^{t}_{0}}\oplus
\bm{2^{t}_{1}}\oplus\bm{2^{t}_{2}}$ and  $\bm{2^{r}_{i}}\otimes\bm{2^{s}_{j}}
=\bm{1^{t}_{m}}\oplus\bm{3^{t}}$, we find that the doublet modular forms $Y^{(k_L+k_{E^c_{d}})}_{\bm{2^t_i}}$ are necessary to make the second term in Eq.~\eqref{eq:We3} invariant  under the action of $\Gamma^\prime_{6}$. From appendix~\ref{app:hig_wei}, we find that the lowest weight $k_L+k_{E^c_{d}}$ with non-vanishing $Y^{(k_L+k_{E^c_{d}})}_{\bm{2^t_j}}$ is $Y^{(1)}_{\bm{2^0_2}}$. Without loss of generality, we take $E^c_{d}\sim\bm{2^0_i}$ and $L\sim\bm{3^0}$ with $k_L+k_{E^c_{d}}=1$. The second term in Eq.~\eqref{eq:We3} can be written as
\begin{equation}
	\beta\left( \left(E^c_{d} L\right)_{\bm{2^0_1}} Y^{(1)}_{\bm{2^0_2}}\right)_{\bm{1^0_0}}H_d\,.
\end{equation}
Then one can straightforwardly obtain
\begin{equation}
	\begin{pmatrix}
		v_2  \\
		v_3 \\
	\end{pmatrix}=\beta \begin{pmatrix}
		0 &~  Y_{2} &~ -\sqrt{2}  Y_{1} \\
		\sqrt{2}   Y_{2} &~  Y_{1} &~ 0 \\
	\end{pmatrix}P^{i}_{3}v_d\,.
\end{equation}
Guided by the principle of minimality and simplicity, the modular form $Y^{(k_L+k_{E^c_{d}})}_{\bm{2^t_j}}$ can also take to be $Y^{(3)}_{\bm{2^1_1}}$. Then we can take $E^c_{d}\sim\bm{2^1_i}$ and $L\sim\bm{3^0}$ with $k_L+k_{E^c_{d}}=3$, and the charged lepton Yukawa coupling reads as
\begin{equation}
	\beta\left( \left(E^c_{d} L\right)_{\bm{2^1_2}} Y^{(3)}_{\bm{2^1_1}}\right)_{\bm{1^0_0}}H_d\,,
\end{equation}
which gives rise to
\begin{equation}
	\begin{pmatrix}
		v_2  \\
		v_3 \\
	\end{pmatrix}=\beta \begin{pmatrix}
		\sqrt{2}   Y^{(3)}_{\bm{2^1_1},1} &~ 0 &~   -Y^{(3)}_{\bm{2^1_1},2} \\
		0 &~ -\sqrt{2}  Y^{(3)}_{\bm{2^1_1},2} &~   -Y^{(3)}_{\bm{2^1_1},1} \\
	\end{pmatrix}P^{i}_{3}v_d\,.
\end{equation}
We give an example model in the following. Both the left-handed lepton $L$ and the right-handed neutrinos $N^c$ transform as $\bm{3^0}$ under $\Gamma'_6$, and the right-handed charged leptons $E^{c}_{1}$ and $E^c_{d}$ transform as $\bm{1^0_0}$ and $\bm{2_2}$, respectively. The modular weights take to be $k_L=k_{E^c_1}=k_{E^c_{d}}=k_{N^c}=1$. Then the charged lepton  mass matrix is given by
	\begin{equation}
		m_{e}=
		\begin{pmatrix}
			\alpha  Y^{(2)}_{\bm{3^0},1} &~ \alpha  Y^{(2)}_{\bm{3^0},3} &~ \alpha  Y^{(2)}_{\bm{3^0},2} \\
			\beta  Y^{(2)}_{\bm{6},2} &~ \beta  Y^{(2)}_{\bm{6},1} &~ \beta  Y^{(2)}_{\bm{6},3} \\
			\beta  Y^{(2)}_{\bm{6},5} &~ \beta  Y^{(2)}_{\bm{6},4} &~ \beta  Y^{(2)}_{\bm{6},6} \\
		\end{pmatrix}
	\end{equation}
The neutrino masses originate from the type I seesaw mechanism, see section~\ref{subsec:nu1} for details. The Majorana mass matrix of right-handed neutrinos and the Dirac neutrino mass matrix are given by Eq.~\eqref{eq:mN2} and Eq.~\eqref{eq:MD_N1L1} respectively. For this model, we find the minimum of $\chi^2$ is $\chi^2_{\text{min}}=20.975$, and the best fit values of the free parameters are
\begin{eqnarray}
\nonumber &&	\Re\langle\tau\rangle=0.498, ~~~~ \Im\langle\tau\rangle=0.867, ~~~~ \beta/\alpha=0.0713,~~~~ |h_2/h_1|=1.085, \\
&&\arg{(h_2/h_1)}=1.628\pi, \qquad  h^2_1v^2_u/(g\Lambda)=110.499\,\text{meV}\,.
	\end{eqnarray}
At the best fit point, the neutrino masses and mixing parameters are determined to be
\begin{eqnarray}
\nonumber &&\sin^2\theta_{13}=0.02219, \quad \sin^2\theta_{12}=0.304, \quad \sin^2\theta_{23}=0.499, \quad  \delta_{CP}=1.499\pi\,, \\
		\nonumber&&\alpha_{21}=1.9990\pi,\quad \alpha_{31}=0.9995\pi, \quad  m_1=444.081\,\text{meV} ,\quad m_2=444.164\,\text{meV}\,, \\
		&&  m_3=446.164\,\text{meV}, \quad
		m_{\beta\beta}=444.163\,\text{meV}\,.
\end{eqnarray}
Although the predictions for neutrino mixing angels are compatible with experimental data, the light neutrino masses are quasi-degenerate and the effective Majorana mass $m_{\beta\beta}$ is larger than the latest upper bound. We have considered other type II models, and we find the neutrino mass spectrum tends to be degenerate and they are usually disfavored by the data of neutrinoless double beta decay.

\subsection{\label{sec:typ_4_model} Type IV models }

\begin{figure}[t!]
	\centering
	\begin{tabular}{c}
		\includegraphics[width=0.95\linewidth]{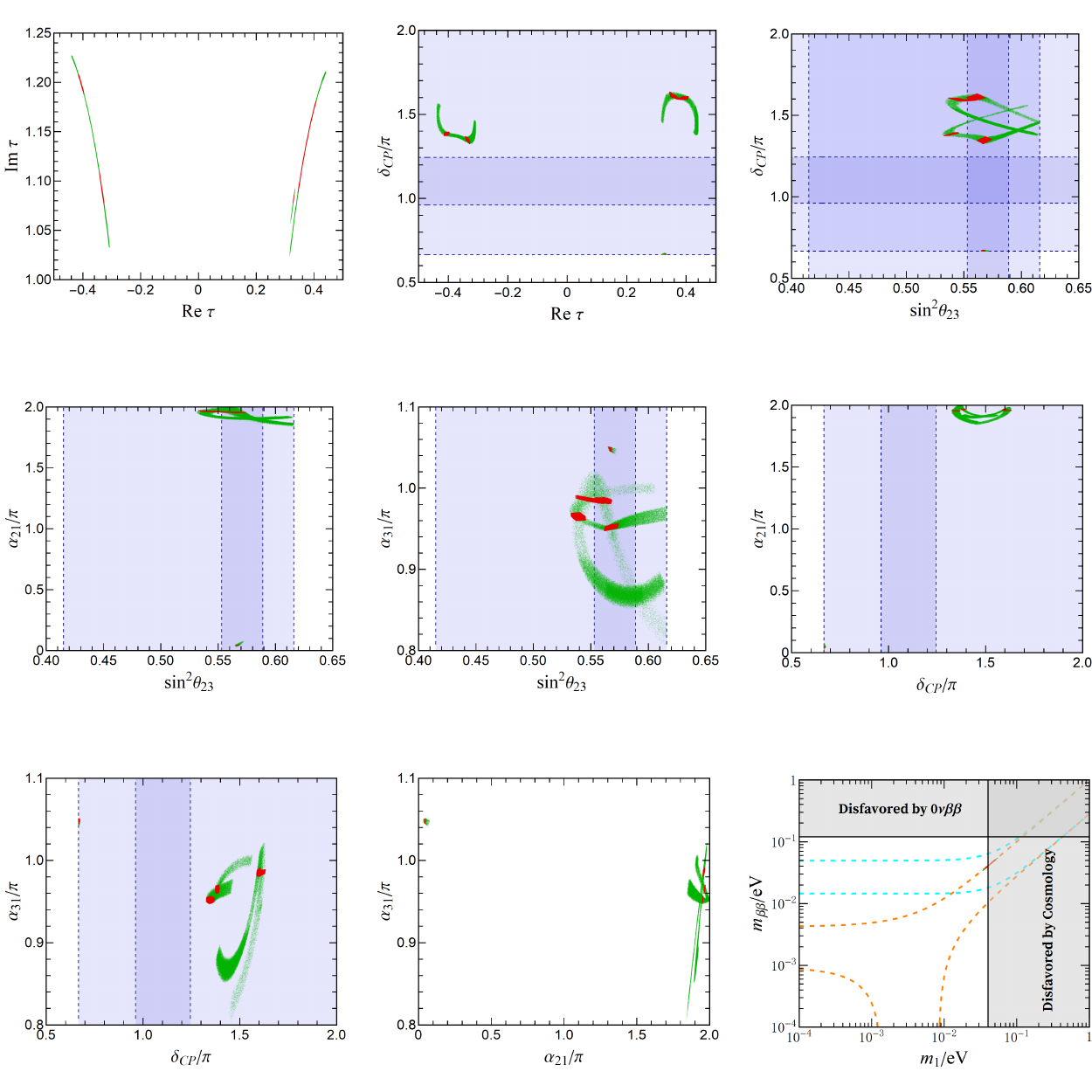}
	\end{tabular}
	\caption{\label{fig:M4} The correlations between the neutrino parameters predicted in the example model of type IV. We adopt the same convention as figure~\ref{fig:M2}.}
\end{figure}

In this type of models, both left-handed leptons $L$ and right-handed charged leptons $E^c$ are assumed to transform as singlet plus doublet under $\Gamma^\prime_{6}$. Since the $\Gamma'_6$ modular group has six one-dimensional representations $\bm{1^{r}_k}$ and nine two-dimensional representation
$\bm{2_k}$, $\bm{2^{r}_k}$ with $r=0, 1$ and $k=0, 1, 2$, there are a lot of possible representation assignment for the lepton fields, and we find many of them can explain the experimental data. We will not shall all these models one by one. In order to see how well this type of models agree with the experimental data on mixing parameters and lepton masses, we will present an example model. We assume that neutrinos masses are described by the  Weinberg operator, the transformation properties and modular weights of the lepton fields are given by
\begin{eqnarray}
\nonumber &&L_{1}\sim\bm{1^0_2}, ~~~~ L_{d}=(L_2,L_3)^T\sim \bm{2^1_1}, ~~~~  E^c_{1}\sim\bm{1^0_1}, ~~~~ E^c_{d}=(E^c_2, E^c_3)^{T}\sim\bm{2_2}, \\
&& k_{L_{1}}=3, ~~~~k_{L_{d}}=2, ~~~~ k_{E^c_1}=-3, ~~~~ k_{E^c_{d}}=3\,.
\end{eqnarray}
Then the superpotential for the lepton mass is determined to be
\begin{eqnarray}
	\nonumber \mathcal{W}&=&\alpha  E^c_1 L_{1}H_d+\beta L_1\left(Y^{(6)}_{\bm{2_2}}E^c_{d}\right)_{\bm{1^0_1}}H_d+\gamma \left(Y^{(5)}_{\bm{4_0}}L_{d}E^c_{d}\right)_{\bm{1^0_0}}H_d +\frac{g_1}{2\Lambda}L_{1}\left(Y^{(5)}_{\bm{2^1_0}}L_{d}\right)_{\bm{1^0_1}}H_uH_u\\
	&& +\frac{g_2}{2\Lambda}\left(Y^{(4)}_{\bm{3^0}}L_{d}L_{d}\right)_{\bm{1^0_0}}H_uH_u
	+\frac{g_3}{2\Lambda}Y^{(4)}_{\bm{1^0_1}}\left(L_{d}L_{d}\right)_{\bm{1^0_2}}H_uH_u\,.
\end{eqnarray}
Note that the term proportional to $g_3$ gives null contributions because the contraction $\bm{2^{1}_{1}}\otimes\bm{2^{1}_{1}}\rightarrow {\bm{1^{0}_{2}}}$ is antisymmetric. In comparison with previous types of models, the modular form $Y^{(5)}_{\bm{4_0}}$ in the four-dimensional representation of $\Gamma'_6$ is involved in this model. We can straightforwardly read off the lepton mass matrices as follow,
\begin{eqnarray}
\nonumber m_{e}&=&\begin{pmatrix}
		\alpha  & ~0 & ~0 \\
		\beta  Y^{(6)}_{\bm{2_2},1} &~ \gamma  Y^{(5)}_{\bm{4_0},4} &~ -\gamma  Y^{(5)}_{\bm{4_0},3} \\
		\beta  Y^{(6)}_{\bm{2_2},2} &~ -\gamma  Y^{(5)}_{\bm{4_0},2}&~ \gamma  Y^{(5)}_{\bm{4_0},1} \\
	\end{pmatrix}v_d\,, \\
m_{\nu}&=&
	\frac{v^2_u}{2\Lambda}\begin{pmatrix}
		0 & ~-g_{1}Y^{(5)}_{\bm{2^1_0},2} &~ g_{1}Y^{(5)}_{\bm{2^1_0},1} \\
		-g_{1}Y^{(5)}_{\bm{2^1_0},2} &~ -2 \sqrt{2} g_{2}Y^{(4)}_{\bm{3^{0}},3} &~ 2 g_{2}Y^{(4)}_{\bm{3^{0}},2} \\
		g_{1}Y^{(5)}_{\bm{2^1_0},1} &~ 2 g_{2}Y^{(4)}_{\bm{3^{0}},2} & ~2 \sqrt{2} g_{2}Y^{(4)}_{\bm{3^{0}},1} \\
	\end{pmatrix}\,.
\end{eqnarray}
The phases of the couplings $\alpha$, $\beta$, $\gamma$ and $g_{1}$ can be absorbed by lepton fields while $g_{2}$ is complex without gCP. This model can agree with the experimental data very well for certain values of the input parameters. We find $\chi^2_{min}=2.871$ and the best fit values of the free parameters are determined to be
\begin{eqnarray}
	\nonumber &&\Re\langle\tau\rangle=-0.329, \qquad  \Im\langle\tau\rangle=1.080, \qquad  \beta/\alpha=105.467, \qquad \gamma/\alpha=12.954, \\
	&&|g_2/g_1|=0.816, \qquad \text{arg}(g_2/g_1)=0.958\pi,\qquad  g_1v^2_u/\Lambda=29.321\,\text{meV}\,.
\end{eqnarray}
The observable quantities are predicted to be
\begin{eqnarray}
	\nonumber &&\sin^2\theta_{13}=0.02217, \qquad \sin^2\theta_{12}=0.304, \qquad \sin^2\theta_{23}=0.570, \qquad  \delta_{CP}=1.347\pi\,, \\
	\nonumber&&\alpha_{21}=1.942\pi,\qquad \alpha_{31}=0.953\pi, \qquad  m_1=37.424\,\text{meV} ,\qquad m_2=38.399\,\text{meV}\,, \\
	&&  m_3=62.588\,\text{meV}, \qquad \sum_i m_i=138.411\,\text{meV},\qquad m_{\beta\beta}=37.661\,\text{meV}\,,
\end{eqnarray}
If gCP is imposed on the model, the coupling $g_2$ would be real and it can be either positive or negative. We find that the minimum value of $\chi^2$ function is $\chi^2_{min}=2.914$ when the input parameters take the following values
\begin{eqnarray}
	\nonumber &&	\Re\langle\tau\rangle=-0.334, \qquad  \Im\langle\tau\rangle=1.092, \qquad  \beta/\alpha=105.717, \qquad \gamma/\alpha=12.696, \\
	&&g_2/g_1=-0.810,\qquad  g_1v^2_u/\Lambda=30.315\,\text{meV}\,.
\end{eqnarray}
The best fitting results for mixing parameters and neutrino masses are given by
\begin{eqnarray}
	\nonumber &&\sin^2\theta_{13}=0.02216, \qquad \sin^2\theta_{12}=0.304, \qquad \sin^2\theta_{23}=0.568, \qquad  \delta_{CP}=1.347\pi\,, \\
	\nonumber&&\alpha_{21}=1.954\pi,\qquad \alpha_{31}=0.952\pi, \qquad  m_1=38.275\,\text{meV} ,\qquad m_2=39.229\,\text{meV}\,, \\
	&&  m_3=63.101\,\text{meV}, \qquad \sum_i m_i=140.604\,\text{meV}, \qquad m_{\beta\beta}=38.557\,\text{meV}\,.
\end{eqnarray}
We numerically scan over the parameter space of the model, the lepton masses and mixing parameters are required to lie in the experimentally preferred $3\sigma$ regions. The allowed region of the modulus $\tau$ and the correlations among different observables are shown in figure~\ref{fig:M4}, where the red and green regions are the results with gCP and without gCP, respectively. We see that the model is very predictive and the neutrino mixing parameters are determined to vary in small regions. After including the gCP symmetry, the predictive power is improved further and the allowed regions shrink drastically.

\subsection{\label{sec:typ_5_model} Type V models }

\begin{figure}[t!]
\centering
\begin{tabular}{c}
\includegraphics[width=0.95\linewidth]{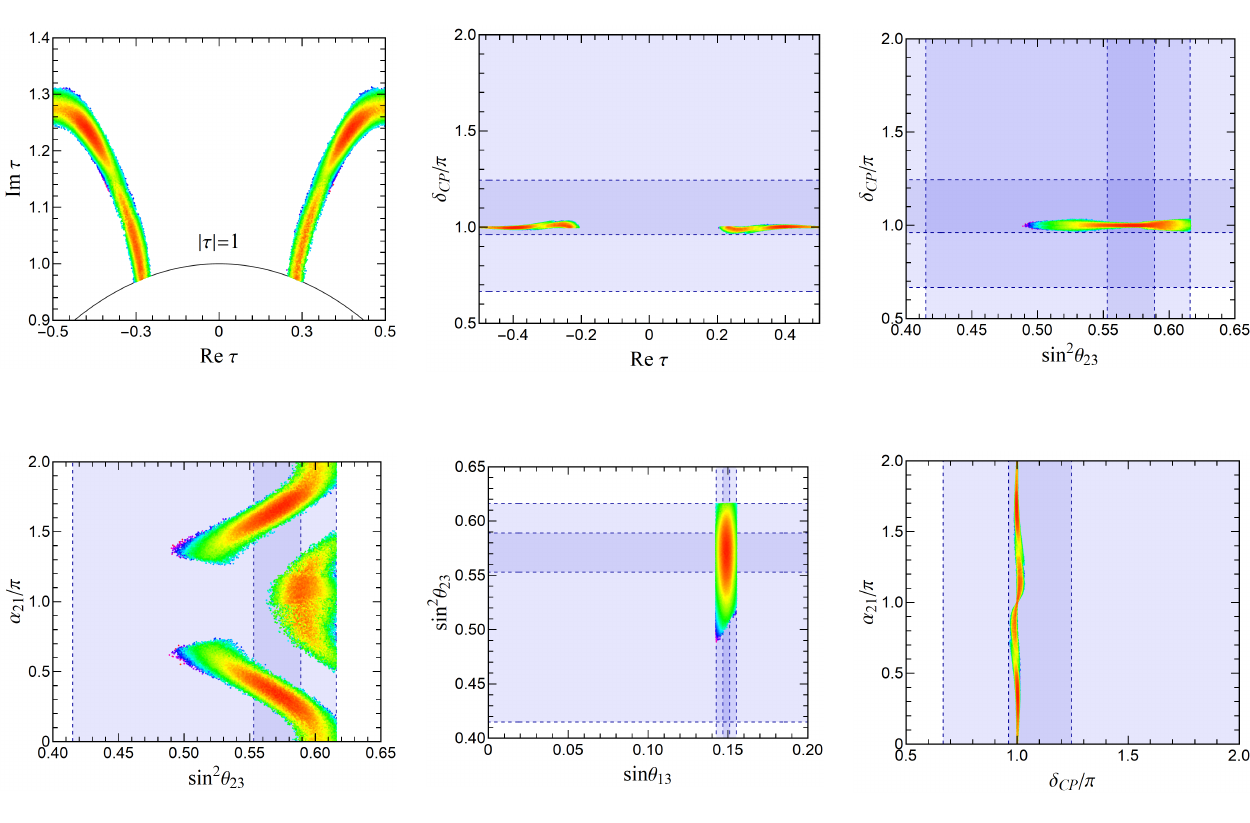}\\
\includegraphics[width=0.48\linewidth]{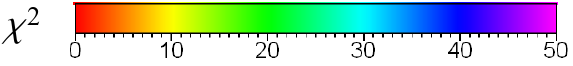}
\end{tabular}
\caption{\label{fig:corr_2RHN_CP}
The correlations between the neutrino parameters predicted in the example model of type V. The same convention as figure~\ref{fig:M2} is adopted here. }
\end{figure}

As the finite modular group $\Gamma^\prime_6$ has nine two-dimensional irreducible representations $\bm{2_k}$ and $\bm{2^{r}_k}$ with $r=0,1$, $k=0,1,2$, it is convenient to construct minimal seesaw models involving two right-handed neutrinos in $\Gamma^\prime_6$ modular symmetry. The two right-handed neutrinos $N^c=(N^c_1,N^c_2)^T$ are embedded into a doublet of $\Gamma^\prime_{6}$ and its modular weight is $k_{N^c}$. In order to take full advantage of the abundant two-dimensional representations of $\Gamma^\prime_6$, the three generations of left-handed lepton doublets are arranged into one singlet and one doublet of $\Gamma^\prime_{6}$. The three right-handed charged leptons are assumed to transform as a triplet under $\Gamma^\prime_{6}$ otherwise more modular invariant terms as well as more free parameters would be needed. There are many different possible representation assignments for the matter fields, and it is too lengthy to list them. For illustration, we give one example, the transformation properties and modular weights of the lepton fields are given by
\begin{eqnarray}
\nonumber &&L_{1}\sim\bm{1^0_2}, ~~~ L_{d}=(L_2, L_3)\sim \bm{2_2}, ~~~ E^c\sim\bm{3^0}, ~~~ N^c\sim\bm{2^1_2}, \\
&& k_{L_{1}}=1, ~~~ k_{L_{d}}=3, ~~~ k_{E^c}=3, ~~~ k_{N^c}=2\,,
\end{eqnarray}
which completely fixes the superpotential of the lepton masses as follows,
\begin{eqnarray}
\nonumber \mathcal{W}&=&\alpha  L_{1}\left(Y^{(4)}_{\bm{3^0}}E^c\right)_{\bm{1^0_1}}H_d+\beta \left(Y^{(6)}_{\bm{6i}}L_{d}E^c\right)_{\bm{1^0_0}}H_d+\gamma \left(Y^{(6)}_{\bm{6ii}}L_{d}E^c\right)_{\bm{1^0_0}}H_d+\delta \left(Y^{(6)}_{\bm{6iii}}L_{d}E^c\right)_{\bm{1^0_0}}H_d \\
	&& +h_1\left(Y^{(5)}_{\bm{4_2i}}L_{d}N^c\right)_{\bm{1^0_0}}H_u
	 +h_2\left(Y^{(5)}_{\bm{4_2ii}}L_{d}N^c\right)_{\bm{1^0_0}}H_u
	+\frac{g\Lambda}{2}\left(Y^{(4)}_{\bm{3^0}}N^cN^c\right)_{\bm{1^0_0}}\,.
\end{eqnarray}
We impose gCP symmetry such that all coupling constants are real. Notice that the neutrino Yukawa coupling between $L_1$ and $N^c$ is forbidden by modular invariance. The neutrino Dirac mass matrix and Majorana mass matrix are given by
\begin{eqnarray}
\nonumber&&m_{D}=\begin{pmatrix}
0 & ~-h_{1}Y^{(5)}_{\bm{4_2i},4}-h_{2}Y^{(5)}_{\bm{4_2ii},4} &~ h_{1}Y^{(5)}_{\bm{4_2i},3}+h_{2}Y^{(5)}_{\bm{4_2ii},3} \\
0 &~ h_{1}Y^{(5)}_{\bm{4_2i},2}+h_{2}Y^{(5)}_{\bm{4_2ii},2} &~-h_{1}Y^{(5)}_{\bm{4_2i},1}-h_{2}Y^{(5)}_{\bm{4_2ii},1} \\
\end{pmatrix}v_u, \\
&&m_{N}=
\frac{g\Lambda}{2}\begin{pmatrix}
\sqrt{2} Y^{(4)}_{\bm{3^{0}},1} &~ -Y^{(4)}_{\bm{3^{0}},3}\\
-Y^{(4)}_{\bm{3^{0}},3} &~ -\sqrt{2} Y^{(4)}_{\bm{3^{0}},2}\\
\end{pmatrix}\,.
\end{eqnarray}
The predictions for the lepton mixing parameters and lepton masses agree rather well with their measured values for certain values of the input parameters. The best fit point is found to be given by
\begin{eqnarray}
	\nonumber &&	\Re\langle\tau\rangle=-0.393, \qquad  \Im\langle\tau\rangle=1.237, \qquad  \beta/\alpha=0.0287, \qquad \gamma/\alpha=0.0293, \\
	&&\delta/\alpha=-0.875,  \qquad h_2/h_1=-1.347, \qquad  h^2_1v^2_u/(g\Lambda)=3.688\,\text{meV}\,,
\end{eqnarray}
with $\chi^2_{min}=0.464$. Accordingly the neutrino masses and mixing parameters are
\begin{eqnarray}
	\nonumber &&\sin^2\theta_{13}=0.02219, \qquad \sin^2\theta_{12}=0.304, \qquad \sin^2\theta_{23}=0.573, \qquad  \delta_{CP}=1.004\pi\,\\
	\nonumber&&\alpha_{21}=0.340\pi, \qquad  m_1=0\,\text{meV},\qquad m_2=8.597\,\text{meV}, \qquad m_3=50.167\,\text{meV}\,, \\
	&&   \sum_i m_i=58.763\,\text{meV},\qquad m_{\beta\beta}=3.224\,\text{meV}\,,
\end{eqnarray}
Similar to previous cases, a numerical analysis is performed, and the results are plotted in figure~\ref{fig:corr_2RHN_CP}. It is worth noting that the Dirac CP phase $\delta_{CP}$ is limited to a very narrow range.

\section{\label{Sec:model_Tp}Model building based on $\Gamma(2)$ modular symmetry with finite modular group $T^\prime$}

\begin{figure}[ht!]
	\centering
	\begin{tikzpicture}[scale=2.5]
		\draw[->] (-1.5,0) -- (1.5,0) ;
		\draw[->] (0,0) -- (0,1.88);
		\draw  (-1,0) -- (-1,1.7);
		\draw  (1,0) -- (1,1.7);
		\draw  (1,0) arc [start angle=0, end angle=180, radius=0.5cm];
		\draw  (0,0) arc [start angle=0, end angle=180, radius=0.5cm];
\node [black,below] at (0,0) {\normalsize  $0$};
		\node [black,below] at (-0.5,0) {\normalsize $-1/2$};
		\node [black,below] at (0.5,0) {\normalsize  $1/2$};
		\node [black,below] at (-1,0) {\normalsize  $-1$};
		\node [black,below] at (1,0) {\normalsize  $1$};
		\node [black] at (-0.7,1.55) {\normalsize  $\mathcal{D}_{\Gamma(2)}$};
		\node [black,below] at (1.5,0) {\normalsize  $\text{Re}(\tau)$};
		\node [black,right] at (0,1.88) {\normalsize  $\text{Im}(\tau)$};
		\begin{scope}	
	\clip (-1,0) rectangle (1,1.7);
	\clip   (1,1.7) -| (-1,0) arc (180:0:0.5) arc (180:0:0.5) -- cycle;			
\fill[red] (-1,-1) rectangle (1,7);
\draw[->] (0,0) -- (0,1.88);
		\end{scope}
	\end{tikzpicture}
	\caption{The fundamental domain $\mathcal{D}_{\Gamma(2)}\cong \mathcal{H}/\Gamma(2)$ of $\Gamma(2)$.}
	\label{fig:fundomain_Gamma2}
\end{figure}
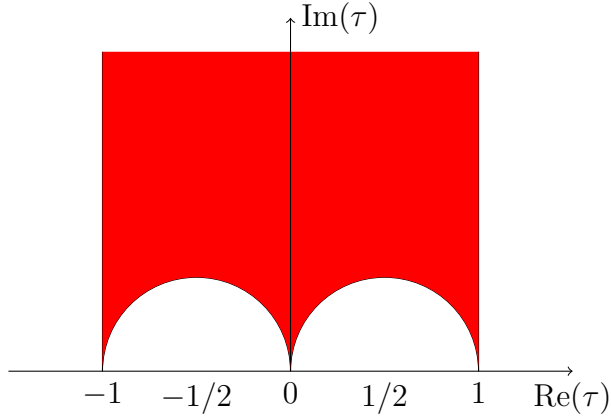

Applying the general formalism outlined in section~\ref{Sec:MS_MF}, we consider construction of modular models for $N'=2$ and $N''=6$ in this section, the full flavor symmetry is the $\Gamma(2)$ modular group. It is worth mentioning that the $\Gamma(2)$ modular symmetry was used to understand the quantum hall effect~\cite{georgelin20002}. In this context, the inequivalent vacuum of the modulus $\tau$  is given by the fundamental domain of $\Gamma(2)$:
\begin{equation}
\mathcal{D}_{\Gamma(2)}=\left\{\tau~\big|~ |\Re{\tau}|\leq1, \Im{\tau}>0, |\tau\pm 1/2|\geq1/2 \right\}\,,
\end{equation}
which is displayed in figure~\ref{fig:fundomain_Gamma2}.
The modular forms of level 6 can be arranged into irreducible multiplets of the finite modular group $\Gamma(2)/\Gamma(6) \cong T^\prime$, as is summarized in table~\ref{tab:MF_summary_Tp}. It can be seen that at a given weight, there are several modular multiplets in the same representation of $T'$. Comparing with the modular forms of level 3 and their decomposition under $\Gamma'_3\cong T^\prime$~\cite{Liu:2019khw}, we see there are more modular multiplets at a given weight. The $T'$ group has three one-dimensional representations $\bm{1^0_k}$ with $k=0, 1, 2$, three two-dimensional representations $\bm{2^0_k}$ and a three-dimensional representation $\bm{3^0}$. The Kronecker products between different representations of $T^\prime$ can be straightforwardly obtained from Eq.~\eqref{eq:KP-gamma6prime} as follows,
\begin{eqnarray}
\nonumber&&\bm{1^{0}_{i}}\otimes\bm{1^{0}_{j}}=\bm{1^{0}_{m}},\qquad
\bm{1^{0}_{i}}\otimes\bm{2^{0}_{j}}=\bm{2^{0}_{m}},\qquad \bm{1^{0}_{i}}\otimes\bm{3^{0}}=\bm{3^{0}},\qquad
\bm{2^{0}_{i}}\otimes\bm{2^{0}_{j}}=\bm{1^{0}_{m}}\oplus\bm{3^{0}},\\
&&
\bm{2^{0}_{i}}\otimes\bm{3^{0}}=\bm{2^{0}_{0}}\oplus
\bm{2^{0}_{1}}\oplus\bm{2^{0}_{2}}, \qquad
\bm{3^{0}}\otimes\bm{3^{0}}=\bm{1^{0}_{0}}\oplus
\bm{1^{0}_{1}}\oplus\bm{1^{0}_{2}}\oplus
\bm{3^{0}_{1}}\oplus\bm{3^{0}_{2}}\,,
\end{eqnarray}
Moreover, one can easily obtain the CG coefficients of $T^\prime$ group from those of $\Gamma^\prime_{6}$ in appendix~\ref{app:group-theory}.

In the same fashion as section~\ref{Sec:model_G6p}, one can also systematically analyze the possible modular models based on $\Gamma(2)$ modular symmetry with finite modular group $T^\prime$. It is too lengthy to present all possibilities here, for illustration we shall give an example model in the following. The left-handed lepton doublets $L$ are assigned to a triplet $\bm{3^{0}}$ of $T^\prime$ with the modular weight $k_L=1$. The three right-handed charged leptons transform as $E^c_1\sim \mathbf{1}^0_0$ and $E^c_{d}=(E^c_2,E^c_3)\sim \mathbf{2}^0_1$ under $T'$, and their modular weights are $k_{E^c_1}=1$ and $k_{E^c_{d}}=0$, respectively. We assume the neutrino masses are described by the effective Weinberg operator. From the Kronecker product $\bm{3^{0}}\otimes\bm{3^{0}}=\bm{1^{0}_{0}}\oplus
\bm{1^{0}_{1}}\oplus\bm{1^{0}_{2}}\oplus
\bm{3^{0}_{1}}\oplus\bm{3^{0}_{2}}$, we know that all the six weight 2 modular multiplets $Y^{\prime(2)}_{\bm{1^0_0i}}$, $Y^{\prime(2)}_{\bm{1^0_0ii}}$, $Y^{\prime(2)}_{\bm{1^0_2}}$, $Y^{\prime(2)}_{\bm{3^0i}}$, $Y^{\prime(2)}_{\bm{3^0ii}}$ and $Y^{\prime(2)}_{\bm{3^0iii}}$ can couple with the operator $LLH_{u}H_{u}$ to fulfill modular invariance. Thus the superpotential for neutrino masses is given by
\begin{eqnarray}
\nonumber \mathcal{W}_{\nu}&=&\frac{g_1}{2\Lambda}Y^{\prime(2)}_{\bm{1^0_0i}}\left(LL\right)_{\bm{1^0_0}}H_uH_u
+\frac{g_2}{2\Lambda}Y^{\prime(2)}_{\bm{1^0_0ii}}\left(LL\right)_{\bm{1^0_0}}H_uH_u
+\frac{g_3}{2\Lambda}Y^{\prime(2)}_{\bm{1^0_2}}\left(LL\right)_{\bm{1^0_1}}H_uH_u\\
\nonumber  &&\frac{g_4}{2\Lambda}\left(\left(LL\right)_{\bm{3^0_1}}Y^{\prime(2)}_{\bm{3^0i}}\right)_{\bm{1^0_0}}H_uH_u+
\frac{g_5}{2\Lambda}\left(\left(LL\right)_{\bm{3^0_1}}Y^{\prime(2)}_{\bm{3^0ii}}\right)_{\bm{1^0_0}}H_uH_u \\
\label{eq:Wnu-Gamm2}&& +
\frac{g_6}{2\Lambda}\left(\left(LL\right)_{\bm{3^0_1}}Y^{\prime(2)}_{\bm{3^0iii}}\right)_{\bm{1^0_0}}H_uH_u\,.
\end{eqnarray}
From the CG coefficients given in appendix~\ref{app:group-theory}, we can easily read off the light neutrino mass matrix. Furthermore, the Yukawa interactions for the charged leptons are of the following form
\begin{eqnarray}
\nonumber \hskip-0.19in \mathcal{W}_e&=&\alpha_1E^c_1\left(LY^{\prime(2)}_{\bm{3^0i}}\right)_{\bm{1^0_0}}H_d
+\alpha_2E^c_1\left(LY^{\prime(2)}_{\bm{3^0ii}}\right)_{\bm{1^0_0}}H_d
+\alpha_3E^c_1\left(LY^{\prime(2)}_{\bm{3^0iii}}\right)_{\bm{1^0_0}}H_d\\
\label{eq:We-Gamm2}&&\hskip-0.2in+\beta_1\left(\left(LE^c_{d}\right)_{\bm{2^0_1}}Y^{\prime(1)}_{\bm{2^0_2}}\right)_{\bm{1^0_0}}H_d+
\beta_2\left(\left(LE^c_{d}\right)_{\bm{2^0_2}}Y^{\prime(1)}_{\bm{2^0_1i}}\right)_{\bm{1^0_0}}H_d+
\beta_3\left(\left(LE^c_{d}\right)_{\bm{2^0_2}}Y^{\prime(1)}_{\bm{2^0_1ii}}\right)_{\bm{1^0_0}}H_d\,,
\end{eqnarray}
which leads to the charged lepton mass matrix
\begin{equation}
	m_{e}=\begin{pmatrix}
 \alpha_{1} Y^{\prime(2)}_{4}+\alpha_{2} Y^{\prime(2)}_{7}+\alpha_{3} Y^{\prime(2)}_{10} &~ \alpha_{1} Y^{\prime(2)}_{6}+\alpha_{2} Y^{\prime(2)}_{9}+\alpha_{3} Y^{\prime(2)}_{12} &~ \alpha_{1} Y^{\prime(2)}_{5}+\alpha_{2} Y^{\prime(2)}_{8}+\alpha_{3} Y^{\prime(2)}_{11} \\
 \beta_{1} Y_{2} &~ -\sqrt{2} \beta_{1} Y_{1}-\beta_{2} Y_{4}-\beta_{3} Y_{6} &~ \sqrt{2} (\beta_{2} Y_{3}+\beta_{3} Y_{5}) \\
 \beta_{1} Y_{1}-\sqrt{2} (\beta_{2} Y_{4}+\beta_{3} Y_{6}) &~ -\beta_{2} Y_{3}-\beta_{3} Y_{5} &~ \sqrt{2} \beta_{1} Y_{2}
	\end{pmatrix}v_d\,.
\end{equation}
All the coupling constants in Eqs.~(\ref{eq:Wnu-Gamm2},\ref{eq:We-Gamm2}) are real due to gCP invariance. This model contains enough free parameters to fit the experimental data, and we find the best fit values of all observables can be reproduced for the following values of the input parameters:
\begin{eqnarray}
\nonumber &&	\Re\langle\tau\rangle=-0.0263, \quad  \Im\langle\tau\rangle=1.789, \quad  \alpha_{2}/\alpha_{1}=285.994, \quad \alpha_{3}/\alpha_{1}=199.982,\\
\nonumber && \beta_{1}/\alpha_{1}=623.202, \quad \beta_{2}/\alpha_{1}=886.017, \quad  \beta_{3}/\alpha_{1}=3.243, \quad
g_2/g_1=972.843\,,   \\
&&g_3/g_1=469.822, \quad g_4/g_1=455.144, \quad g_5/g_1=789.301, \quad g_6/g_1=172.997 \,.
\end{eqnarray}
The resulting models usually contain a large number of free parameters if the modular flavor symmetry is some principal congruence subgroup $\Gamma(N')$ rather than the modular group $SL(2,\mathbb{Z})\equiv\Gamma(1)$, consequently the predictive power is reduced to certain extent. The top-down construction in string theory generally lead to both modular symmetry and traditional flavor symmetry, the full flavor symmetry named as eclectic flavor group is a nontrivial product of traditional
favor group and finite modular group~\cite{Nilles:2020nnc,Nilles:2020kgo,Nilles:2020tdp,Baur:2020jwc,Nilles:2020gvu}. The couplings in modular invariant models are subject to strong constraints from traditional flavor symmetry, and some terms allowed by the modular group could be forbidden by the traditional flavor group. Thus we expect that the modular invariant models based on $\Gamma(N')$ and finite modular group can become much more predictive by combining with certain traditional flavor group in the paradigm of eclectic flavor group.

\section{\label{sec:Con}Conclusion}

Modular symmetry is a promising framework to understand the hierarchical masses and flavor mixing patterns of quarks and leptons~\cite{Feruglio:2017spp}. In the present work, we generalize the modular invariance approach~\cite{Feruglio:2017spp} and a new route towards finite modular groups is proposed. The theory is assumed to be invariant under the principal congruence subgroup $\Gamma(N')$, the finite modular group is the quotient of two principal congruence subgroups $\Gamma(N')/\Gamma(N'')$. The known homogeneous finite modular groups can be reproduced, as shown in Eq.~\eqref{eq:finite-modGrp}, and the original modular invariant theory is the special case of $N'=1$.

For $N'=1$ and $N''=6$, the finite modular group is the homogeneous finite modular group $\Gamma(1)/\Gamma(6)=\Gamma'_6$ which is isomorphic to $S_{3}\times T^{\prime}$. The gCP symmetry consistent with the modular symmetry is considered. The representation matrices of both modular generators $S$ and $T$ are symmetric and unitary in all irreducible representations of $\Gamma'_6$, consequently the gCP transformation is a unit matrix up to modular transformations and all coupling constants are real if gCP invariance is imposed. We have constructed the weight 1 modular forms of level 6 in terms of the Dedekind eta function, and they can be arranged into a doublet and a quartet of $\Gamma'_6$. We also give the expressions of the  linearly independent higher weight modular forms which are the tensor products of weight 1 modular forms.

We have performed a comprehensive analysis of models for lepton masses and mixing with $\Gamma'_6$ modular symmetry, and five types of models summarized in table~\ref{tab:five_mod} have been studied. In the type I models, both left-handed leptons $L$ and right-handed neutrinos $N^c$ are assumed to be triplets of $\Gamma'_6$, the right-handed charged fermions are singlets of $\Gamma^\prime_{6}$. It turns out that the type I models with small number of free parameters coincide with the $A_4$ modular models discussed in~\cite{Yao:2020qyy}. Different from the type I models, the three right-handed neutrinos are assigned to the direct product of a singlet and a doublet of $\Gamma^\prime_{6}$ in the type II models. We find that five of the twenty modular models can accommodate the experimental data no matter whether gCP symmetry is included or not, as shown in table~\ref{tab:bf_N1100N2320}. The type III models differ from the type I models in the assignment of right-handed charged leptons which are assumed to be singlet plus doublet of $\Gamma'_6$. The light neutrino masses arise from the effective Weinberg operator and the minimal seesaw with two right-handed neutrinos in the type IV and type V models respectively. Moreover, we numerically scan over the parameter space of some example models, and the correlations among the mixing parameters are plotted.

Furthermore, we consider the modular invariant models for $N'=2$ and $N''=6$. Then the full modular flavor symmetry is $\Gamma(2)$ and the finite modular group is $\Gamma(2)/\Gamma(6)\cong T'$. The modular forms of level 6 can be decomposed into multiplets of $T'$.  From table~\ref{tab:MF_summary_Tp} we see that there are a number of modular forms in the same representation of $T'$ at a given weight. As a consequence, the resulting model has a plenty of free parameters. We expect that the independent free parameters could be reduced considerably if the modular symmetry is combined with the traditional flavor symmetry for instance in the context of eclectic flavor group~\cite{Nilles:2020nnc,Nilles:2020kgo,Nilles:2020tdp,Baur:2020jwc,Nilles:2020gvu}.

It is well-known that top quark is much heavier than other quark masses. Therefore the first and the second generations of quark fields are usually assigned to a doublet of certain flavor symmetry group while the third generation is assigned to a singlet. The modular group $\Gamma'_6$ has abundant one-dimensional and two-dimensional irreducible representations. It is interesting to apply the $\Gamma'_6$ modular symmetry to understand the hierarchical quark masses and mixing angles and construct a quark-lepton unification model which can describe the experimentally measured values of both leptons and quarks for a common modulus $\tau$. This is beyond the scope of present work, and the results of this work pave the way for such future studies.

\vskip0.4in

\section*{Acknowledgements}
CCL is grateful to Dr. Jun-Nan Lu for their kind help on numerical analysis. CCL is supported by the National Natural Science Foundation of China under Grant Nos. 12005167, 12047502 and the Anhui Province Natural Science Foundation Grant No. 1908085QA24. XGL and GJD are supported by the National Natural Science Foundation of China under Grant Nos. 11975224, 11835013 and 11947301.

\newpage

\section*{Appendix}

\begin{appendix}

\section{\label{app:group-theory}Group Theory of $\Gamma^\prime_{6}$}

The homogeneous finite modular group $\Gamma'_6$ can be generated by the modular generators $S$ and $T$ obeying the multiplication rules~\footnote{The defining relations of inhomogeneous finite modular group $\Gamma_6$ are $ S^{2}=T^{6}=(ST)^{3}=ST^2ST^3ST^4ST^3=1$, and $\Gamma_6$ is isomorphic to $S_{3}\times A_{4}$.}:
\begin{equation}
  S^{4}=T^{6}=(ST)^{3}=ST^2ST^3ST^4ST^3=1, \qquad S^2T=TS^2\,.
\end{equation}
From Eq.~\eqref{eq:factor-hom-FMG}, we know that $\Gamma'_{6}$ is isomorphic to $\Gamma'_{2}\times\Gamma'_{3}=S_{3}\times T^{\prime}$ where $\Gamma'_{3}\cong T^\prime$ is the double covering of the tetrahedron group $A_4$ and $\Gamma'_{2}\cong S_3$ is the permutation group of order 3.

The group $\Gamma^\prime_{6}\cong S_{3}\times T^{\prime}$ can also be generated by four generators $\tilde{a}$, $\tilde{b}$, $\tilde{c}$ and $\tilde{d}$ fulfilling the following relations
\begin{eqnarray}
\nonumber&& \tilde{a}^{4}=\tilde{b}^{3}=(\tilde{a}\tilde{b})^{3}=1, ~~ \tilde{a}^2\tilde{b}=\tilde{b}\tilde{a}^2,\\
\nonumber&&\tilde{c}^2=\tilde{d}^2=(\tilde{c}\tilde{d})^3=1\,,\\
&&\tilde{a}\tilde{c}=\tilde{c}\tilde{a}\,,~~ \tilde{a}\tilde{d}=\tilde{d}\tilde{a}, ~~ \tilde{b}\tilde{c}=\tilde{c}\tilde{b}, ~~ \tilde{b}\tilde{d}=\tilde{d}\tilde{b}\,.
\end{eqnarray}
where $\tilde{a}$ and $\tilde{b}$ are the generators of $T^\prime$, and $\tilde{c}$ and $\tilde{d}$ are the generators of $S_3$. It is easy to check that the multiplication rules of $\Gamma^\prime_{6}$ in above two bases are fulfilled by
\begin{equation}
\label{eq:gens_relation}
\tilde{a}=TST^4S^3T, \qquad  \tilde{b}=T^4, \qquad  \tilde{c}=S^3T^3S, \qquad  \tilde{d}=T^3\,,
\end{equation}
In the same fashion, the following relations can also be obtained
\begin{equation}
S=\tilde{a}\tilde{c}\tilde{d}\tilde{c}, \qquad  T=\tilde{b}\tilde{d}\,.
\end{equation}
The 144 elements of the modular group $\Gamma^\prime_{6}$ can be divided into 21 conjugacy classes and it has 104 non-trivial abelian subgroups which include seven $Z_{2}$ subgroups, thirteen $Z_{3}$ subgroups, twelve $Z_{4}$ subgroups, three $K_{4}$ subgroups, thirty-seven $Z_{6}$ subgroups,  nine $Z_2\times Z_4$, four $Z_3\times Z_3$ subgroups, three $Z_{12}$ subgroups, twelve $Z_2\times Z_6$ subgroups and four $Z_3\times Z_6$ subgroups. The $\Gamma^\prime_6$ group has twenty-one irreducible representations: six singlet representations which are labeled as $\bm{1^{r}_{k}}$, nine two-dimensional representations defined as $\bm{2_{k}}$ and $\bm{2^{r}_{k}}$,  two three-dimensional representations $\bm{3^r}$, three four-dimensional representations labeled as $\bm{4_{k}}$ and one six-dimensional representation $\bm{6}$, where the superscript $r$ and the subscript $k$ can take the values $r=0,1$ and $k=0,1,2$, respectively. In the present work, the explicit forms of the generators $S$ and $T$ ($\tilde{a}$, $\tilde{b}$, $\tilde{c}$ and $\tilde{d}$) in the twenty-one irreducible representations are summarized in table~\ref{tab:rep} (table~\ref{tab:rep2}), where the matrices $\bm{a_2}$, $\bm{b_2}$, $\bm{a_3}$ and $\bm{b_3}$ in the two tables are taken to be
\begin{equation}\label{eq:a3b3}
\hskip-0.1in \bm{a_2}=\frac{i}{\sqrt{3}}
\begin{pmatrix}
 1 &~ \sqrt{2} \\
 \sqrt{2} &~ -1 \\
\end{pmatrix},~
\bm{b_2}=
\begin{pmatrix}
 1 &~ 0 \\
 0 &~ \omega \\
\end{pmatrix},~
\bm{a_3}=\frac{1}{3}
\begin{pmatrix}
 -1 &~ 2 &~ 2 \\
 2 &~ -1 &~ 2 \\
 2 &~ 2 &~ -1 \\
\end{pmatrix},~
\bm{b_3}=\begin{pmatrix}
 1 &~ 0 &~ 0 \\
 0 &~ \omega  &~ 0 \\
 0 &~ 0 &~ \omega ^2 \\
\end{pmatrix}\,.
\end{equation}

\begin{table}[t!]
	\begin{center}
		\renewcommand{\tabcolsep}{1.5mm}
		\begin{tabular}{|c|c|c|c|c|c|c|}\hline\hline		
~~ & $\bm{1^{r}_k}$ & 	$\bm{2_k}$ & 	$\bm{2^{r}_k}$ & 	$\bm{3^{r}}$ & $\bm{4_k}$ & $\bm{6}$ \\  [0.01in]  \hline   &&&&&& \\ [-0.16in]
$S$  & $(-1)^{r}$   &  $\frac{1}{2}
\begin{pmatrix}
	-1 &~  \sqrt{3} \\
	\sqrt{3} &~ 1 \\
\end{pmatrix} $
&  $(-1)^{r}\bm{a_2} $
&  $(-1)^{r}\bm{a_3}$
& $\frac{1}{2}
\begin{pmatrix}
	-\bm{a_2} &~  \sqrt{3}\bm{a_2} \\
	\sqrt{3}\bm{a_2} &~ \bm{a_2} \\
\end{pmatrix}$
& $\frac{1}{2}
\begin{pmatrix}
	-\bm{a_3} &~  \sqrt{3}\bm{a_3} \\
	\sqrt{3}\bm{a_3} &~ \bm{a_3} \\
\end{pmatrix}$ \\  [0.08in]  \hline
&&&&&& \\ [-0.16in]
$T$   &  $(-1)^{r}\omega^k$
&   $\omega^k\begin{pmatrix}
	1 &~ 0 \\
	0 &~ -1 \\
\end{pmatrix} $
 &  $(-1)^{r}\omega^{k+1}\bm{b_2} $
 & $(-1)^{r}\bm{b_3} $
  &  $\omega^{k+1}\begin{pmatrix}
 	\bm{b_2} &~  \mathbb{0}_{2} \\
 	\mathbb{0}_{2} &~ -\bm{b_2} \\
 \end{pmatrix} $
& $\begin{pmatrix}
	\bm{b_3} &~  \mathbb{0}_{3} \\
	\mathbb{0}_{3} &~ -\bm{b_3} \\
\end{pmatrix}$   \\ [0.08in] \hline\hline
\end{tabular}
\caption{\label{tab:rep}The representation matrices of the generators $S$ and $T$ for the twenty-one irreducible representations of $\Gamma^\prime_6$ in the chosen basis, where $\omega=e^{2\pi i/3}$, $r=0,1$, $k=0,1,2$ and matrices $\bm{a_2}$, $\bm{b_2}$, $\bm{a_3}$ and $\bm{b_3}$ are shown in Eq.~\eqref{eq:a3b3}.}
\end{center}
\end{table}

\begin{table}[t!]
\begin{center}
\renewcommand{\tabcolsep}{3.0mm}
\begin{tabular}{|c|c|c||c|c|c|}\hline\hline
\multicolumn{3}{|c||}{$T^\prime$} & \multicolumn{3}{|c|}{$S_3$} \\ \hline \hline
&   &   & & &    \\ [-0.16in]
~~  &  $\tilde{a}$  &   $\tilde{b}$  &  ~~ &  $\tilde{c}$  &   $\tilde{d}$     \\ \hline
&   &   & & &    \\ [-0.16in]
$\bm{1^0_k}$ & $1$   &  $\omega^k$ & $\bm{1^{r}_{0}}$ & $(-1)^{r}$   &  $(-1)^{r}$    \\ \hline
&   &   & & &    \\ [-0.16in]
$\bm{2^0_k}$ &	$\bm{a_2} $	& $\omega^{k+1}\bm{b_2} $ &	$\bm{2_0}$ &	$-\frac{1}{2}
\begin{pmatrix}	1 &~  \sqrt{3} \\
\sqrt{3} &~ -1 \\
\end{pmatrix} $
&  $\begin{pmatrix}
1 &~ 0 \\
0 &~ -1 \\
\end{pmatrix}$\\ [0.12in]\hline
&   &  & & &     \\ [-0.16in]
$\bm{3^0}$ & $\bm{a_3}$& $\bm{b_3}$& -- & -- & --\\ \hline \hline
			
\multicolumn{6}{|c|}{}    \\ [-0.16in]
\multicolumn{6}{|c|}{$S_{3}\times T^{\prime} $} \\ \hline \hline
&   &   & \multicolumn{2}{|c|}{} &    \\ [-0.16in]
~~  &  $\tilde{a}$  &   $\tilde{b}$  &  \multicolumn{2}{|c|}{$\tilde{c}$}  &   $\tilde{d}$ \\ \hline
&   &  & \multicolumn{2}{|c|}{} &    \\ [-0.16in]
$\bm{1^{r}_k}=\bm{1^{r}_{0}}\times\bm{1^{0}_k}$ & $1$   &  $\omega^k$ & \multicolumn{2}{|c|}{$(-1)^{r}$}   &  $(-1)^{r}$    \\[0.02in] \hline
&   &   & \multicolumn{2}{|c|}{} &    \\ [-0.16in]
$\bm{2_k}=\bm{2_{0}}\times\bm{1^{0}_k}$ &
$\mathbb{1}_{2}$ & $\omega^k\mathbb{1}_{2} $
&  \multicolumn{2}{|c|}{$-\frac{1}{2}
\begin{pmatrix}
1 &~  \sqrt{3} \\
\sqrt{3} &~ -1 \\
\end{pmatrix} $}
&  $\begin{pmatrix}
1 &~ 0 \\
0 &~ -1 \\
\end{pmatrix}$	\\ [0.12in]\hline
&   &  & \multicolumn{2}{|c|}{} &   \\ [-0.16in]
			
$\bm{2^{r}_k}=\bm{1^{r}_{0}}\times\bm{2^{0}_k}$ &  $\bm{a_2} $ & $\omega^{k+1}\bm{b_2} $
&  \multicolumn{2}{|c|}{$(-1)^{r}\mathbb{1}_{2} $}
&  $(-1)^{r}\mathbb{1}_{2}$ \\ [0.02in]\hline
&   &  & \multicolumn{2}{|c|}{} &    \\ [-0.16in]
			
$\bm{3^{r}}=\bm{1^{r}_{0}}\times\bm{3^{0}}$ & $\bm{a_3}$
& $\bm{b_3}$
& \multicolumn{2}{|c|}{$(-1)^{r}\mathbb{1}_{3}$} & $(-1)^{r}\mathbb{1}_{3}$ \\ [0.02in]\hline
&   &  & \multicolumn{2}{|c|}{} &    \\ [-0.16in]
			
$\bm{4_k}=\bm{2_{0}}\times\bm{2^{0}_k}$ &
$\begin{pmatrix}
\bm{a_2} &~ \mathbb{0}_{2}  \\
\mathbb{0}_{2} &~ \bm{a_2}  \\
\end{pmatrix}$&
 $\omega^{k+1}\begin{pmatrix}
\bm{b_2} &~ \mathbb{0}_{2}  \\
\mathbb{0}_{2} &~ \bm{b_2}  \\
\end{pmatrix}$
&  \multicolumn{2}{|c|}{$-\frac{1}{2} \begin{pmatrix}
\mathbb{1}_{2} &~ \sqrt{3}\mathbb{1}_{2}  \\
\sqrt{3}\mathbb{1}_{2} &~ -\mathbb{1}_{2}  \\
\end{pmatrix}$} &
$\begin{pmatrix}
\mathbb{1}_{2} &~ \mathbb{0}_{2}  \\
\mathbb{0}_{2} &~ -\mathbb{1}_{2}  \\
\end{pmatrix}$\\ [0.14in]\hline
&   &  &  \multicolumn{2}{|c|}{} &    \\ [-0.16in]
			
$\bm{6}=\bm{2_{0}}\times\bm{3^{0}}$ & $
\begin{pmatrix}
\bm{a_3} &~ \mathbb{0}_{3}  \\
\mathbb{0}_{3} &~ \bm{a_3}  \\
\end{pmatrix}$
& $ \begin{pmatrix}
\bm{b_3} &~ \mathbb{0}_{3}  \\
\mathbb{0}_{3} &~ \bm{b_3}  \\
\end{pmatrix}$
& \multicolumn{2}{|c|}{$-\frac{1}{2} \begin{pmatrix}
\mathbb{1}_{3} &~ \sqrt{3}\mathbb{1}_{3}  \\
\sqrt{3}\mathbb{1}_{3} &~ -\mathbb{1}_{3}  \\
\end{pmatrix}$}
& $\begin{pmatrix}
\mathbb{1}_{3} &~ \mathbb{0}_{3}  \\
\mathbb{0}_{3} &~ -\mathbb{1}_{3}  \\
\end{pmatrix}$\\[0.16in] \hline\hline
\end{tabular}
\caption{\label{tab:rep2}The representation matrices of the generators $\tilde{a}$, $\tilde{b}$, $\tilde{c}$ and $\tilde{d}$ for the irreducible representations of groups $T^\prime$, $S_3$ and $\Gamma^\prime_6$ in the chosen basis, where $\omega=e^{2\pi i/3}$, $r=0,1$ and $k=0,1,2$.  In the present work, the representation matrices of $T^\prime$ are related to those of~\cite{Ding:2008rj} by the similarity transformation $u_2=\text{diag}(1,e^{\frac{5\pi i}{12}})P_{2}$ for two-dimensional representations $\bm{2^0_k}$ and $u_3=\text{diag}(1,\omega^2,\omega)$ for three-dimensional representation $\bm{3^0}$. The representation matrices of generators of $S_{3}$ can be found in~\cite{Ishimori:2010au}. The irreducible representations of $\Gamma^\prime_{6}=S_{3}\times T^{\prime}$ can be obtained from the direct products of those of $S_{3}$ and $T^\prime$~\cite{Tung:1985na}. For example, the two-dimensional representation $\bm{2_k}$ of $\Gamma^\prime_{6}$ is exactly the direct product $\bm{2_{0}}\times \bm{1^{0}_k}$. }
	\end{center}
\end{table}
From the representation matrices of generators in table~\ref{tab:rep} or table~\ref{tab:rep2}, one can easy to obtain the character table of $\Gamma^\prime_{6}$. Then the Kronecker products between different irreducible representations read as
\begin{eqnarray}
\nonumber \hskip-0.3in&&\bm{1^{r}_{i}}\otimes\bm{1^{s}_{j}}=\bm{1^{t}_{m}},\qquad
 \bm{1^{r}_{i}}\otimes\bm{2_{j}}=\bm{2_{m}},\qquad
\bm{1^{r}_{i}}\otimes\bm{2^{s}_{j}}=\bm{2^{t}_{m}},\qquad \bm{1^{r}_{i}}\otimes\bm{3^{s}}=\bm{3^{t}},\\
\nonumber \hskip-0.3in&&
\bm{1^{r}_{i}}\otimes\bm{4_{j}}=\bm{4_{m}},\qquad
\bm{1^{r}_{i}}\otimes\bm{6}=\bm{6}, \qquad
\bm{2_{i}}\otimes\bm{2_{j}}=\bm{1^{0}_{m}}\oplus
\bm{1^{1}_{m}}\oplus\bm{2_{m}},\qquad
\bm{2_{i}}\otimes\bm{2^{r}_{j}}=\bm{4_{m}}, \\
\nonumber \hskip-0.3in&&
\bm{2_{i}}\otimes\bm{3^{r}}=\bm{6},\qquad
\bm{2_{i}}\otimes\bm{4_{j}}=\bm{2^{0}_{m}}\oplus
\bm{2^{1}_{m}}\oplus\bm{4_{m}},\qquad
\bm{2_{i}}\otimes\bm{6}=\bm{3^{0}}\oplus\bm{3^{1}}
\oplus\bm{6}, \\
\nonumber \hskip-0.3in&&  \bm{2^{r}_{i}}\otimes\bm{2^{s}_{j}}
=\bm{1^{t}_{m}}\oplus\bm{3^{t}},
\qquad
\bm{2^{r}_{i}}\otimes\bm{3^{s}}=\bm{2^{t}_{0}}\oplus
\bm{2^{t}_{1}}\oplus\bm{2^{t}_{2}}, \qquad
\bm{2^{r}_{i}}\otimes\bm{4_{j}}=\bm{2_{m}}\oplus
\bm{6}, \\
\nonumber \hskip-0.3in&&
\bm{2^{r}_{i}}\otimes\bm{6}=\bm{4_{0}}\oplus
\bm{4_{1}}\oplus\bm{4_{2}}, \qquad
\bm{3^{r}}\otimes\bm{3^{s}}=\bm{1^{t}_{0}}\oplus
\bm{1^{t}_{1}}\oplus\bm{1^{t}_{2}}\oplus
\bm{3^{t}_{1}}\oplus\bm{3^{t}_{2}},
\\
\nonumber \hskip-0.3in&& \bm{3^{r}}\otimes\bm{4_{i}}
=\bm{4_{0}}\oplus\bm{4_{1}}\oplus\bm{4_{2}}, \qquad
\bm{3^{r}}\otimes\bm{6}
=\bm{2_{0}}\oplus\bm{2_{1}}\oplus\bm{2_{2}}
\oplus\bm{6_{1}}\oplus\bm{6_{2}},\\
\nonumber  \hskip-0.3in&& \bm{4_{i}}\otimes\bm{4_{j}}~
=\bm{1^{0}_{m}}\oplus\bm{1^{1}_{m}}\oplus\bm{2_{m}}\oplus
\bm{3^{0}}\oplus\bm{3^{1}}\oplus\bm{6}, \\
\nonumber \hskip-0.3in&& \bm{4_{i}}\otimes\bm{6}~
=\bm{2^{0}_{0}}\oplus\bm{2^{0}_{1}}\oplus\bm{2^{0}_{2}}
\oplus\bm{2^{1}_{0}}\oplus\bm{2^{1}_{1}}
\oplus\bm{2^{1}_{2}}
\oplus\bm{4_{0}}\oplus\bm{4_{1}}\oplus\bm{4_{2}}\,,\\
\nonumber  \hskip-0.3in&& \bm{6}\otimes\bm{6}~
=\bm{1^{0}_{0}}\oplus\bm{1^{0}_{1}}\oplus\bm{1^{0}_{2}}\oplus\bm{1^{1}_{0}}\oplus\bm{1^{1}_{1}}\oplus\bm{1^{1}_{2}} \oplus\bm{2_{0}}\oplus\bm{2_{1}}\oplus\bm{2_{2}}\oplus\bm{3^{0}_{S}}\oplus\bm{3^{0}_{A}}\oplus\bm{3^{1}_{S}}
\oplus\bm{3^{1}_{A}}\oplus\bm{6_{S}}\oplus\bm{6_{A}}\,, \\\label{eq:KP-gamma6prime}
\end{eqnarray}
where $i,j=0,1,2$ and $r,s=0,1$, and we have defined $m\equiv i+j~(\text{mod}~3)$ and $t\equiv r+s~(\text{mod}~2)$. The symbol $\bm{6_{1}}$ and $\bm{6_{2}}$ stand for two six-dimensional representations $\bm{6}$ that appear in the Kronecker products. The subscript ``$S$'' (``$A$'') refers to symmetric (antisymmetric) combinations. In the following, we list the CG coefficients in our basis, the irreducible representation $\bm{R_{\,i}}$ refers to $\bm{R_{\,i\, \text{mod}\,3}}$ in the case of the lower index $i>2$. We use the notation $\alpha_{i}$ ($\beta_{i}$) to denote the elements of the first (second) representation. Furthermore, we shall adopt the following notations facilitate the expressions of CG coefficients
\begin{equation}\label{eq:sim_mart}
 P_{2}=\begin{pmatrix}
0 &~ 1 \\ -1 &~0
\end{pmatrix},~
P_{3}=\begin{pmatrix}
0 &~0 &~1 \\ 1 &~0 &~0 \\ 0 &~1 &~0
\end{pmatrix},~
P_{4}=\begin{pmatrix}
 \mathbb{0}_{2} &~\mathbb{1}_{2}  \\
 -\mathbb{1}_{2} &~ \mathbb{0}_{2} \\
\end{pmatrix},~
 P_{6}(r,i)=\begin{pmatrix}
 \mathbb{0}_{3} &~\mathbb{1}_{3}  \\
 -\mathbb{1}_{3} &~ \mathbb{0}_{3} \\
\end{pmatrix}^r
\begin{pmatrix}
 P_{3} &~\mathbb{0}_{3}  \\
 \mathbb{0}_{3} &~ P_{3} \\
\end{pmatrix}^i\,,
\end{equation}

\begin{eqnarray*}
\begin{array}{|c|c|c|c|c|c|c|}\hline\hline
~\bm{1^{r}_{i}}\otimes\bm{2_j}=\bm{2_{m}}~  & ~ \bm{1^{r}_{i}}\otimes\bm{2^{s}_{j}}=\bm{2^{t}_{m}} ~ &  ~\bm{1^{r}_{i}}\otimes\bm{3^{s}}=\bm{3^{t}}~ & ~\bm{1^{r}_{i}}\otimes\bm{4_{j}}=\bm{4_{m}} & ~\bm{1^{r}_{i}}\otimes\bm{6}=\bm{6} \\ \hline
 &  & & &  \\[-0.18in]

~\bm{2_{m}}:
\alpha _1 P^{r}_{2} \begin{pmatrix}
\beta _1\\
 \beta _2
\end{pmatrix}
~ & \bm{2^{t}_{m}}:\alpha _1 \begin{pmatrix}
\beta _1\\
 \beta _2
\end{pmatrix}
~ &~\bm{3^{t}}: \alpha _1P^{i}_{3}\begin{pmatrix}
 \beta _1\\
\beta _2\\
 \beta _3
\end{pmatrix} ~
  & ~ \bm{4_{m}}:\alpha _1P^{r}_{4}\begin{pmatrix}
 \beta _1\\
\beta _2\\
 \beta _3 \\
 \beta _4
\end{pmatrix}  & ~ \bm{6}:\alpha _1P_{6}(r,i)\begin{pmatrix}
 \beta _1\\
\beta _2\\
 \beta _3 \\
 \beta _4 \\
 \beta _5 \\
 \beta _6
\end{pmatrix}   \\
 &  & & & \\[-0.18in] \hline \hline
\end{array}
\end{eqnarray*}

\begin{eqnarray*}
\begin{array}{|c|c|c|}\hline\hline
 ~\bm{2}_{i}\otimes\bm{2}_{j}=\bm{1}^{0}_{m}\oplus\bm{1}^{1}_{m}\oplus\bm{2}_{m}~
 & ~ \bm{2_{i}}\otimes\bm{2^{r}_{j}}=\bm{4_{m}} ~
 & ~ \bm{2_{i}}\otimes\bm{3^{r}}=\bm{6} ~\\ \hline
 &&    \\[-0.15in]

~\bm{1^{0}_m}:
\alpha _1 \beta _1+\alpha _2 \beta _2~ & ~ & ~  \\
 & &    \\ [-0.02in]
~\bm{1^{1}_{m}}:
\alpha _1 \beta _2-\alpha _2 \beta _1 &~  ~ & ~ ~ \\
& &   \\[-0.80in]
~ &~ \bm{4_{m}}:
P^{r}_{4}\begin{pmatrix}
\alpha _1 \beta _1\\ \alpha _1 \beta _2 \\ \alpha _2 \beta _1\\ \alpha _2 \beta _2\\
\end{pmatrix} ~ & ~
~\bm{6}: P_{6}(r,i)\begin{pmatrix}
\alpha _1 \beta _1\\ \alpha _1 \beta _2\\ \alpha _1 \beta _3\\ \alpha _2 \beta _1\\ \alpha _2 \beta _2\\ \alpha _2 \beta _3\\
\end{pmatrix} ~   \\
& &   \\[-0.60in]
~\bm{2_{m}}:
\begin{pmatrix}
\alpha _1 \beta _1-\alpha _2 \beta _2\\
-\alpha _1 \beta _2-\alpha _2 \beta _1
\end{pmatrix}~  & & \\
 & &  \\[-0.18in] \hline \hline
~\bm{2_{i}}\otimes\bm{4_{j}}=\bm{2^{0}_{m}}\oplus\bm{2^{1}_{m}}\oplus\bm{4_{m}}~
& \multicolumn{2}{|c|}{~\bm{2_{i}}\otimes\bm{6}=\bm{3^{0}}\oplus\bm{3^{1}\oplus\bm{6}}~} \\ \hline
~ \bm{2^{0}_{m}}: \begin{pmatrix}
\alpha _1 \beta _1+\alpha _2 \beta _3\\ \alpha _1 \beta _2+\alpha _2 \beta _4
\end{pmatrix} ~ &
\multicolumn{2}{|c|}{~ \bm{3^{0}}: P^{i}_{3} \begin{pmatrix}
\alpha _1 \beta _1+\alpha _2 \beta _4 \\
\alpha _1 \beta _2+\alpha _2 \beta _5\\
\alpha _1 \beta _3+\alpha _2 \beta _6
\end{pmatrix} ~} \\
~\bm{2^{1}_{m}}: \begin{pmatrix}
\alpha _1 \beta _3-\alpha _2 \beta _1\\ \alpha _1 \beta _4-\alpha _2 \beta _2
\end{pmatrix} ~ &
\multicolumn{2}{|c|}{~ \bm{3^{1}}:  P^{i}_{3} \begin{pmatrix}
\alpha _1 \beta _4-\alpha _2 \beta _1\\
\alpha _1 \beta _5-\alpha _2 \beta _2\\
\alpha _1 \beta _6-\alpha _2 \beta _3
\end{pmatrix}~} \\
 ~\bm{4_{m}}:\begin{pmatrix}
\alpha _1 \beta _1-\alpha _2 \beta _3\\
\alpha _1 \beta _2-\alpha _2 \beta _4\\
-\alpha _1 \beta _3-\alpha _2 \beta _1\\
-\alpha _1 \beta _4-\alpha _2 \beta _2
\end{pmatrix}~ &
 \multicolumn{2}{|c|}{~ \bm{6}: P_{6}(0,i)\begin{pmatrix}
\alpha _1 \beta _1-\alpha _2 \beta _4\\
\alpha _1 \beta _2-\alpha _2 \beta _5\\
\alpha _1 \beta _3-\alpha _2 \beta _6\\
-\alpha _1 \beta _4-\alpha _2 \beta _1\\
-\alpha _1 \beta _5-\alpha _2 \beta _2\\
-\alpha _1 \beta _6-\alpha _2 \beta _3
\end{pmatrix}~} \\ \hline \hline
\end{array}
\end{eqnarray*}

\begin{eqnarray*}
\begin{array}{|c|c|}\hline\hline
 ~\bm{2^{r}_{i}}\otimes\bm{2^{s}_{j}}=\bm{1^{t}_{m}}\oplus\bm{3^{t}}~  &
 ~\bm{2^{r}_{i}}\otimes\bm{4_{j}}=\bm{2_{m}}\oplus\bm{6} ~   \\ \hline
 &    \\[-0.15in]
 \bm{1^{t}_{m}}: \alpha _1 \beta _2-\alpha _2 \beta _1 ~ &
~\bm{2_{m}}: P^{r}_{2}\begin{pmatrix}
\alpha _1 \beta _2-\alpha _2 \beta _1\\
\alpha _1 \beta _4-\alpha _2 \beta _3
\end{pmatrix}    \\[0.13in]
~\bm{3^{t}}:
P^{m}_{3}\begin{pmatrix}
\alpha _1 \beta _2+\alpha _2 \beta _1\\
\sqrt{2} \alpha _2 \beta _2\\
-\sqrt{2} \alpha _1 \beta _1
\end{pmatrix}
~   &
~\bm{6}:
P_{6}(r,i)\begin{pmatrix}
\alpha _1 \beta _2+\alpha _2 \beta _1\\
\sqrt{2} \alpha _2 \beta _2\\
-\sqrt{2} \alpha _1 \beta _1\\
\alpha _1 \beta _4+\alpha _2 \beta _3\\
\sqrt{2} \alpha _2 \beta _4\\
-\sqrt{2} \alpha _1 \beta _3
\end{pmatrix}~ \\[0.05in] \hline \hline
 ~\bm{2^{r}_{i}}\otimes\bm{3^{s}}=\bm{2^{t}_{0}}\oplus
\bm{2^{t}_{1}}\oplus\bm{2^{t}_{2}}~& ~\bm{2^{r}_{i}}\otimes\bm{6}=\bm{4_{0}}\oplus\bm{4_{1}}\oplus\bm{4_{2}}~   \\ \hline
 &   \\[-0.15in]
 ~ \bm{2^{t}_{i}}: \begin{pmatrix}
\alpha _1 \beta _1+\sqrt{2} \alpha _2 \beta _3 \\
\sqrt{2} \alpha _1 \beta _2-\alpha _2 \beta _1
\end{pmatrix} ~ &
~ \bm{4_{i}}:P^{r}_{4}\begin{pmatrix}
\alpha _1 \beta _1+\sqrt{2} \alpha _2 \beta _3\\
\sqrt{2} \alpha _1 \beta _2-\alpha _2 \beta _1\\
\alpha _1 \beta _4+\sqrt{2} \alpha _2 \beta _6\\
\sqrt{2} \alpha _1 \beta _5-\alpha _2 \beta _4
\end{pmatrix}~ \\
 ~ \bm{2^{t}_{1+i}}: \begin{pmatrix}
\alpha _1 \beta _2+\sqrt{2} \alpha _2 \beta _1\\
\sqrt{2} \alpha _1 \beta _3-\alpha _2 \beta _2
\end{pmatrix} ~ &
~ \bm{4_{1+i}}:P^{r}_{4}\begin{pmatrix}
\alpha _1 \beta _2+\sqrt{2} \alpha _2 \beta _1\\
\sqrt{2} \alpha _1 \beta _3-\alpha _2 \beta _2\\
\alpha _1 \beta _5+\sqrt{2} \alpha _2 \beta _4\\
\sqrt{2} \alpha _1 \beta _6-\alpha _2 \beta _5
\end{pmatrix}~  \\[0.05in]
 ~ \bm{2^{t}_{2+i}}: \begin{pmatrix}
\alpha _1 \beta _3+\sqrt{2} \alpha _2 \beta _2\\
\sqrt{2} \alpha _1 \beta _1-\alpha _2 \beta _3
\end{pmatrix} ~
 & ~ \bm{4_{2+i}}:P^{r}_{4}\begin{pmatrix}
\alpha _1 \beta _3+\sqrt{2} \alpha _2 \beta _2\\
\sqrt{2} \alpha _1 \beta _1-\alpha _2 \beta _3\\
\alpha _1 \beta _6+\sqrt{2} \alpha _2 \beta _5\\
\sqrt{2} \alpha _1 \beta _4-\alpha _2 \beta _6
\end{pmatrix}~
 \\
 & \\[-0.15in] \hline \hline
\end{array}
\end{eqnarray*}

\begin{eqnarray*}
	\begin{array}{|c|c|c|}\hline\hline
		~\bm{3^{r}}\otimes\bm{3^{s}}=\bm{1^{t}_{0}}\oplus
		\bm{1^{t}_{1}}\oplus\bm{1^{t}_{2}}\oplus
		\bm{3^{t}_{1}}\oplus\bm{3^{t}_{2}} ~  &  ~\bm{3^{r}}\otimes\bm{4_{i}}
		=\bm{4_{0}}\oplus\bm{4_{1}}\oplus\bm{4_{2}} ~ &
		~ \bm{3^{r}}\otimes\bm{6}
		=\bm{2_{0}}\oplus\bm{2_{1}}\oplus\bm{2_{2}}
		\oplus\bm{6_{1}}\oplus\bm{6_{2}}\\ \hline
		& &   \\[-0.18in]
		~\bm{1^{t}_{0}}:\alpha _1 \beta _1+\alpha _2 \beta _3+\alpha _3 \beta _2~  & ~ &
		~\bm{2_{0}}:P^{r}_{2}\begin{pmatrix}\alpha _1 \beta _1+\alpha _2 \beta _3+\alpha _3 \beta _2\\
			\alpha _1 \beta _4+\alpha _2 \beta _6+\alpha _3 \beta _5\end{pmatrix} \\
		& &
		~  \\[-0.38in]
		~\bm{1^{t}_{1}}:\alpha _1 \beta _2+\alpha _2 \beta _1+\alpha _3 \beta _3~  & ~\bm{4_{i}}: P^{r}_{4}\begin{pmatrix}
			\alpha _1 \beta _1+\sqrt{2} \alpha _3 \beta _2\\
			-\alpha _1 \beta _2+\sqrt{2} \alpha _2 \beta _1\\
			\alpha _1 \beta _3+\sqrt{2} \alpha _3 \beta _4\\
			-\alpha _1 \beta _4+\sqrt{2} \alpha _2 \beta _3
		\end{pmatrix}~ &
		~\bm{2_{1}}:P^{r}_{2}\begin{pmatrix}\alpha _1 \beta _2+\alpha _2 \beta _1+\alpha _3 \beta _3\\
			\alpha _1 \beta _5+\alpha _2 \beta _4+\alpha _3 \beta _6\end{pmatrix}~ \\
		& & \\[-0.38in]
		~\bm{1^{t}_{2}}:\alpha _1 \beta _3+\alpha _2 \beta _2+\alpha _3 \beta _1~  & ~ &
		~\bm{2_{2}}:P^{r}_{2}\begin{pmatrix}\alpha _1 \beta _3+\alpha _2 \beta _2+\alpha _3 \beta _1\\
			\alpha _1 \beta _6+\alpha _2 \beta _5+\alpha _3 \beta _4\end{pmatrix}~ \\
		& &  \\[-0.18in]
		~\bm{3^{t}_{1}}:\begin{pmatrix}
			2 \alpha _1 \beta _1-\alpha _2 \beta _3-\alpha _3 \beta _2\\
			-\alpha _1 \beta _2-\alpha _2 \beta _1+2 \alpha _3 \beta _3\\
			-\alpha _1 \beta _3+2 \alpha _2 \beta _2-\alpha _3 \beta _1
		\end{pmatrix}~  & ~\bm{4_{1+i}}: P^{r}_{4}\begin{pmatrix}
			\sqrt{2} \alpha _1 \beta _2+\alpha _2 \beta _1\\
			-\alpha _2 \beta _2+\sqrt{2} \alpha _3 \beta _1\\
			\sqrt{2} \alpha _1 \beta _4+\alpha _2 \beta _3\\
			-\alpha _2 \beta _4+\sqrt{2} \alpha _3 \beta _3
		\end{pmatrix} &
		~\bm{6_{1}}:
		P_{6}(r,0)\begin{pmatrix}
			\alpha _1 \beta _1-\alpha _3 \beta _2\\
			-\alpha _2 \beta _1+\alpha _3 \beta _3\\
			-\alpha _1 \beta _3+\alpha _2 \beta _2\\
			\alpha _1 \beta _4-\alpha _3 \beta _5\\
			-\alpha _2 \beta _4+\alpha _3 \beta _6\\
			-\alpha _1 \beta _6+\alpha _2 \beta _5
		\end{pmatrix} ~\\
		& & \\[-0.18in]
		~\bm{3^{t}_{2}}:\begin{pmatrix}
			-\alpha _2 \beta _3+\alpha _3 \beta _2\\
			-\alpha _1 \beta _2+\alpha _2 \beta _1\\
			\alpha _1 \beta _3-\alpha _3 \beta _1
		\end{pmatrix}~  & ~\bm{4_{2+i}}: P^{r}_{4}\begin{pmatrix}
			\sqrt{2} \alpha _2 \beta _2+\alpha _3 \beta _1\\
			\sqrt{2} \alpha _1 \beta _1-\alpha _3 \beta _2\\
			\sqrt{2} \alpha _2 \beta _4+\alpha _3 \beta _3\\
			\sqrt{2} \alpha _1 \beta _3-\alpha _3 \beta _4
		\end{pmatrix} &
		~\bm{6_{2}}:
		P_{6}(r,0)\begin{pmatrix}
			\alpha _2 \beta _3-\alpha _3 \beta _2\\
			\alpha _1 \beta _2-\alpha _2 \beta _1\\
			-\alpha _1 \beta _3+\alpha _3 \beta _1\\
			\alpha _2 \beta _6-\alpha _3 \beta _5\\
			\alpha _1 \beta _5-\alpha _2 \beta _4\\
			-\alpha _1 \beta _6+\alpha _3 \beta _4
		\end{pmatrix}~\\
		& & \\[-0.18in]  \hline \hline
	\end{array}
\end{eqnarray*}

\begin{eqnarray*}
\begin{array}{|cc|}\hline\hline
\multicolumn{2}{|c|}{\bm{4_{i}}\otimes\bm{4_{j}}~
=\bm{1^{0}_{m}}\oplus\bm{1^{1}_{m}}\oplus\bm{2_{m}}\oplus
\bm{3^{0}}\oplus\bm{3^{1}}\oplus\bm{6}} \\ \hline
 &    \\[-0.08in]
\bm{1^{0}_{m}}:\alpha _1 \beta _2-\alpha _2 \beta _1+\alpha _3 \beta _4-\alpha _4 \beta _3~  & ~ \\
&  \\[-0.45in]
 & ~\bm{3^{1}}: P^{m}_{3}\begin{pmatrix}
 \alpha _1 \beta _4+\alpha _2 \beta _3-\alpha _3 \beta _2-\alpha _4 \beta _1\\
 \sqrt{2} \left(\alpha _2 \beta _4-\alpha _4 \beta _2\right)\\
 \sqrt{2} \left(-\alpha _1 \beta _3+\alpha _3 \beta _1\right)
\end{pmatrix}~   \\
&   \\[-0.45in]
~\bm{1^{1}_{m}}:\alpha _1 \beta _4-\alpha _2 \beta _3-\alpha _3 \beta _2+\alpha _4 \beta _1~ & \\
&   \\
\bm{2_{m}}:\begin{pmatrix}
 \alpha _1 \beta _2-\alpha _2 \beta _1-\alpha _3 \beta _4+\alpha _4 \beta _3\\
 -\alpha _1 \beta _4+\alpha _2 \beta _3-\alpha _3 \beta _2+\alpha _4 \beta _1
\end{pmatrix}~  &
~ ~\\
&   \\[-0.68in]
& ~\bm{6}:
P_{6}(0,m)\begin{pmatrix}
\alpha _1 \beta _2+\alpha _2 \beta _1-\alpha _3 \beta _4-\alpha _4 \beta _3\\
\sqrt{2} \left(\alpha _2 \beta _2-\alpha _4 \beta _4\right)\\
\sqrt{2} \left(-\alpha _1 \beta _1+\alpha _3 \beta _3\right)\\
-\alpha _1 \beta _4-\alpha _2 \beta _3-\alpha _3 \beta _2-\alpha _4 \beta _1\\
-\sqrt{2} \left(\alpha _2 \beta _4+\alpha _4 \beta _2\right)\\
\sqrt{2} \left(\alpha _1 \beta _3+\alpha _3 \beta _1\right)
\end{pmatrix}~ \\
&   \\[-0.78in]
~\bm{3^{0}}: P^{m}_{3}\begin{pmatrix}
 \alpha _1 \beta _2+\alpha _2 \beta _1+\alpha _3 \beta _4+\alpha _4 \beta _3\\
 \sqrt{2} \left(\alpha _2 \beta _2+\alpha _4 \beta _4\right)\\
 -\sqrt{2} \left(\alpha _1 \beta _1+\alpha _3 \beta _3\right)
\end{pmatrix}~ & \\
&  \\[-0.08in]  \hline \hline
\multicolumn{2}{|c|}{\bm{4_{i}}\otimes\bm{6}~
=\bm{2^{0}_{0}}\oplus\bm{2^{0}_{1}}\oplus\bm{2^{0}_{2}}
\oplus\bm{2^{1}_{0}}\oplus\bm{2^{1}_{1}}
\oplus\bm{2^{1}_{2}}
\oplus\bm{4_{0}}\oplus\bm{4_{1}}\oplus\bm{4_{2}}} \\ \hline
 &    \\[-0.18in]
~\bm{2^{0}_{i}}:\begin{pmatrix}
 \alpha _1 \beta _1+\sqrt{2} \alpha _2 \beta _3+\alpha _3 \beta _4+\sqrt{2} \alpha _4 \beta _6\\
 \sqrt{2} \alpha _1 \beta _2-\alpha _2 \beta _1+\sqrt{2} \alpha _3 \beta _5-\alpha _4 \beta _4
\end{pmatrix}~  &  \\
&  \\[-0.62in]
  &~ \bm{4_{i}}:\begin{pmatrix}
\alpha _1 \beta _1+\sqrt{2} \alpha _2 \beta _3-\alpha _3 \beta _4-\sqrt{2} \alpha _4 \beta _6\\
\sqrt{2} \alpha _1 \beta _2-\alpha _2 \beta _1-\sqrt{2} \alpha _3 \beta _5+\alpha _4 \beta _4\\
-\alpha _1 \beta _4-\sqrt{2} \alpha _2 \beta _6-\alpha _3 \beta _1-\sqrt{2} \alpha _4 \beta _3\\
-\sqrt{2} \alpha _1 \beta _5+\alpha _2 \beta _4-\sqrt{2} \alpha _3 \beta _2+\alpha _4 \beta _1
\end{pmatrix} ~ \\
&  \\[-0.62in]
~\bm{2^{0}_{1+i}}: \begin{pmatrix}
 \alpha _1 \beta _2+\sqrt{2} \alpha _2 \beta _1-\alpha _3 \beta _5+\sqrt{2} \alpha _4 \beta _4\\
 \sqrt{2} \alpha _1 \beta _3-\alpha _2 \beta _2+\sqrt{2} \alpha _3 \beta _6-\alpha _4 \beta _5
\end{pmatrix}~ &  \\
&   \\[-0.18in]
~\bm{2^{0}_{2+i}}:\begin{pmatrix}
\alpha _1 \beta _3+\sqrt{2} \alpha _2 \beta _2-\alpha _3 \beta _6+\sqrt{2} \alpha _4 \beta _5\\
\sqrt{2} \alpha _1 \beta _1-\alpha _2 \beta _3+\sqrt{2} \alpha _3 \beta _4-\alpha _4 \beta _6
\end{pmatrix} ~ & \\
&  \\[-0.62in]
  &~ \bm{4_{1+i}}:\begin{pmatrix}
\alpha _1 \beta _2+\sqrt{2} \alpha _2 \beta _1-\alpha _3 \beta _5-\sqrt{2} \alpha _4 \beta _4\\
\sqrt{2} \alpha _1 \beta _3-\alpha _2 \beta _2-\sqrt{2} \alpha _3 \beta _6+\alpha _4 \beta _5\\
-\alpha _1 \beta _5-\sqrt{2} \alpha _2 \beta _4-\alpha _3 \beta _2-\sqrt{2} \alpha _4 \beta _1\\
-\sqrt{2} \alpha _1 \beta _6+\alpha _2 \beta _5-\sqrt{2} \alpha _3 \beta _3+\alpha _4 \beta _2
\end{pmatrix}~  \\
&  \\[-0.62in]
~\bm{2^{1}_{i}}: \begin{pmatrix}
\alpha _1 \beta _4+\sqrt{2} \alpha _2 \beta _6-\alpha _3 \beta _1-\sqrt{2} \alpha _4 \beta _3\\
\sqrt{2} \alpha _1 \beta _5-\alpha _2 \beta _4-\sqrt{2} \alpha _3 \beta _2+\alpha _4 \beta _1
\end{pmatrix}~ &     \\
&   \\[-0.18in]
~\bm{2^{1}_{1+i}}:\begin{pmatrix}
 \alpha _1 \beta _5+\sqrt{2} \alpha _2 \beta _4-\alpha _3 \beta _2-\sqrt{2} \alpha _4 \beta _1\\
 \sqrt{2} \alpha _1 \beta _6-\alpha _2 \beta _5-\sqrt{2} \alpha _3 \beta _3+\alpha _4 \beta _2
\end{pmatrix}~  &
~ ~\\
&  \\[-0.62in]
  & ~\bm{4_{2+i}}:\begin{pmatrix}
\alpha _1 \beta _3+\sqrt{2} \alpha _2 \beta _2-\alpha _3 \beta _6-\sqrt{2} \alpha _4 \beta _5\\
\sqrt{2} \alpha _1 \beta _1-\alpha _2 \beta _3-\sqrt{2} \alpha _3 \beta _4+\alpha _4 \beta _6\\
-\alpha _1 \beta _6-\sqrt{2} \alpha _2 \beta _5-\alpha _3 \beta _3-\sqrt{2} \alpha _4 \beta _2\\
-\sqrt{2} \alpha _1 \beta _4+\alpha _2 \beta _6-\sqrt{2} \alpha _3 \beta _1+\alpha _4 \beta _3
\end{pmatrix}~  \\
&  \\[-0.62in]
~\bm{2^{1}_{2+i}}:\begin{pmatrix}
 \alpha _1 \beta _6+\sqrt{2} \alpha _2 \beta _5-\alpha _3 \beta _3-\sqrt{2} \alpha _4 \beta _2\\
 \sqrt{2} \alpha _1 \beta _4-\alpha _2 \beta _6-\sqrt{2} \alpha _3 \beta _1+\alpha _4 \beta _3
\end{pmatrix}~  &
~ ~\\
&  \\[-0.18in]  \hline \hline
\end{array}
\end{eqnarray*}

\begin{eqnarray*}
\begin{array}{|c|} \hline\hline
~\bm{6}\otimes\bm{6}=\bm{1^{0}_{0}}\oplus\bm{1^{0}_{1}}\oplus\bm{1^{0}_{2}}\oplus\bm{1^{1}_{0}}\oplus\bm{1^{1}_{1}}\oplus\bm{1^{1}_{2}} \oplus\bm{2_{0}}\oplus\bm{2_{1}}\oplus\bm{2_{2}}\oplus\bm{3^{0}_{S}}\oplus\bm{3^{0}_{A}}\oplus\bm{3^{1}_{S}}\oplus\bm{3^{1}_{A}}\oplus\bm{6_{S}}\oplus\bm{6_{A}}~ \\ \hline \\[-0.16in]
\bm{1^{0}_{0}}:\alpha _1 \beta _1+\alpha _2 \beta _3+\alpha _3 \beta _2+\alpha _4 \beta _4+\alpha _5 \beta _6+\alpha _6 \beta _5~ \\
\bm{1^{0}_{1}}:\alpha _1 \beta _2+\alpha _2 \beta _1+\alpha _3 \beta _3+\alpha _4 \beta _5+\alpha _5 \beta _4+\alpha _6 \beta _6~ \\
\bm{1^{0}_{2}}:\alpha _1 \beta _3+\alpha _2 \beta _2+\alpha _3 \beta _1+\alpha _4 \beta _6+\alpha _5 \beta _5+\alpha _6 \beta _4~ \\
\bm{1^{1}_{0}}:\alpha _1 \beta _4+\alpha _2 \beta _6+\alpha _3 \beta _5-\alpha _4 \beta _1-\alpha _5 \beta _3-\alpha _6 \beta _2~ \\
\bm{1^{1}_{1}}:\alpha _1 \beta _5+\alpha _2 \beta _4+\alpha _3 \beta _6-\alpha _4 \beta _2-\alpha _5 \beta _1-\alpha _6 \beta _3~ \\
\bm{1^{1}_{2}}:\alpha _1 \beta _6+\alpha _2 \beta _5+\alpha _3 \beta _4-\alpha _4 \beta _3-\alpha _5 \beta _2-\alpha _6 \beta _1~ \\
\bm{2_{0}}:\begin{pmatrix}
\alpha _1 \beta _1+\alpha _2 \beta _3+\alpha _3 \beta _2-\alpha _4 \beta _4-\alpha _5 \beta _6-\alpha _6 \beta _5\\
-(\alpha _1 \beta _4+\alpha _2 \beta _6+\alpha _3 \beta _5+\alpha _4 \beta _1+\alpha _5 \beta _3+\alpha _6 \beta _2)
\end{pmatrix} \\
\bm{2_{1}}:\begin{pmatrix}
\alpha _1 \beta _2+\alpha _2 \beta _1+\alpha _3 \beta _3-\alpha _4 \beta _5-\alpha _5 \beta _4-\alpha _6 \beta _6\\
-(\alpha _1 \beta _5+\alpha _2 \beta _4+\alpha _3 \beta _6+\alpha _4 \beta _2+\alpha _5 \beta _1+\alpha _6 \beta _3)
\end{pmatrix}~ \\
\bm{2_{2}}:\begin{pmatrix}
\alpha _1 \beta _3+\alpha _2 \beta _2+\alpha _3 \beta _1-\alpha _4 \beta _6-\alpha _5 \beta _5-\alpha _6 \beta _4\\
-(\alpha _1 \beta _6+\alpha _2 \beta _5+\alpha _3 \beta _4+\alpha _4 \beta _3+\alpha _5 \beta _2+\alpha _6 \beta _1)
\end{pmatrix}~ \\
~\bm{3^{0}_{S}}:
\begin{pmatrix}
2 \alpha _1 \beta _1-\alpha _2 \beta _3-\alpha _3 \beta _2+2 \alpha _4 \beta _4-\alpha _5 \beta _6-\alpha _6 \beta _5\\
2 \alpha _3 \beta _3-\alpha _1 \beta _2-\alpha _2 \beta _1+2 \alpha _6 \beta _6-\alpha _4 \beta _5-\alpha _5 \beta _4\\
2 \alpha _2 \beta _2-\alpha _1 \beta _3-\alpha _3 \beta _1+2 \alpha _5 \beta _5-\alpha _4 \beta _6-\alpha _6 \beta _4
\end{pmatrix}
~ \\
~\bm{3^{0}_{A}}:
\begin{pmatrix}
 \alpha _2 \beta _3-\alpha _3 \beta _2+\alpha _5 \beta _6-\alpha _6 \beta _5\\
\alpha _1 \beta _2 -\alpha _2 \beta _1+\alpha _4 \beta _5-\alpha _5 \beta _4\\
 -\alpha _1 \beta _3+\alpha _3 \beta _1-\alpha _4 \beta _6+\alpha _6 \beta _4
\end{pmatrix}
~ \\
~\bm{3^{1}_{S}}:
\begin{pmatrix}
\alpha _2 \beta _6-\alpha _3 \beta _5-\alpha _5 \beta _3+\alpha _6 \beta _2\\
\alpha _1 \beta _5-\alpha _2 \beta _4-\alpha _4 \beta _2+\alpha _5 \beta _1\\
-\alpha _1 \beta _6+\alpha _3 \beta _4+\alpha _4 \beta _3-\alpha _6 \beta _1
\end{pmatrix}
~ \\
~\bm{3^{1}_{A}}:
\begin{pmatrix}
 2 \alpha _1 \beta _4-\alpha _2 \beta _6-\alpha _3 \beta _5-2 \alpha _4 \beta _1+\alpha _5 \beta _3+\alpha _6 \beta _2\\
 -\alpha _1 \beta _5-\alpha _2 \beta _4+2 \alpha _3 \beta _6+\alpha _4 \beta _2+\alpha _5 \beta _1-2 \alpha _6 \beta _3\\
 -\alpha _1 \beta _6+2 \alpha _2 \beta _5-\alpha _3 \beta _4+\alpha _4 \beta _3-2 \alpha _5 \beta _2+\alpha _6 \beta _1
\end{pmatrix}
~ \\
~\bm{6_{S}}:
\begin{pmatrix}
2 \alpha _1 \beta _1-\alpha _2 \beta _3-\alpha _3 \beta _2-2 \alpha _4 \beta _4+\alpha _5 \beta _6+\alpha _6 \beta _5\\
-\alpha _1 \beta _2-\alpha _2 \beta _1+2 \alpha _3 \beta _3+\alpha _4 \beta _5+\alpha _5 \beta _4-2 \alpha _6 \beta _6\\
-\alpha _1 \beta _3+2 \alpha _2 \beta _2-\alpha _3 \beta _1+\alpha _4 \beta _6-2 \alpha _5 \beta _5+\alpha _6 \beta _4\\
-2 \alpha _1 \beta _4+\alpha _2 \beta _6+\alpha _3 \beta _5-2 \alpha _4 \beta _1+\alpha _5 \beta _3+\alpha _6 \beta _2\\
\alpha _1 \beta _5+\alpha _2 \beta _4-2 \alpha _3 \beta _6+\alpha _4 \beta _2+\alpha _5 \beta _1-2 \alpha _6 \beta _3\\
\alpha _1 \beta _6-2 \alpha _2 \beta _5+\alpha _3 \beta _4+\alpha _4 \beta _3-2 \alpha _5 \beta _2+\alpha _6 \beta _1
\end{pmatrix}
~ \\
~\bm{6_{A}}:
\begin{pmatrix}
\alpha _2 \beta _3-\alpha _3 \beta _2-\alpha _5 \beta _6+\alpha _6 \beta _5\\
\alpha _1 \beta _2-\alpha _2 \beta _1-\alpha _4 \beta _5+\alpha _5 \beta _4\\
-\alpha _1 \beta _3+\alpha _3 \beta _1+\alpha _4 \beta _6-\alpha _6 \beta _4\\
-\alpha _2 \beta _6+\alpha _3 \beta _5-\alpha _5 \beta _3+\alpha _6 \beta _2\\
-\alpha _1 \beta _5+\alpha _2 \beta _4-\alpha _4 \beta _2+\alpha _5 \beta _1\\
\alpha _1 \beta _6-\alpha _3 \beta _4+\alpha _4 \beta _3-\alpha _6 \beta _1
 \end{pmatrix}.~ \\ [0.28in]
  \hline\hline
\end{array}
\end{eqnarray*}

\section{\label{app:hig_wei}Higher weight modular forms of $\Gamma^{\prime}_6$}

The space of modular forms of weight 2 has dimension $6\times2=12$ and they can be calculated by the contractions of $Y_{\bm{2^0_2}}^{(1)}$ and  $Y^{(1)}_{\bm{4_1}}$ which are the modular multiplets of weight 1 discussed in section~\ref{Sec:MF_N=6}. Hence there are 9 constraints between the 21 products $Y_i Y_j$. The 12 linearly independent modular forms of weight 2 can be arranged into the following multiplets of $\Gamma^{\prime}_6$:
\begin{equation}\label{eq:Yw2l}
\begin{array}{ll}
Y^{(2)}_{\bm{1^{1}_{2}}}(\tau)
=\left(Y^{(1)}_{\bm{4_1}}Y^{(1)}_{\bm{4_1}}\right)_{\bm{1^{1}_{2}}}\,,  & \qquad
 Y^{(2)}_{\bm{2_0}}(\tau)
 =\left(Y^{(1)}_{\bm{2^{0}_2}}Y^{(1)}_{\bm{4_1}}\right)_{\bm{2_0}}\,, \\
 Y^{(2)}_{\bm{3^0}}(\tau)
 =\left(Y^{(1)}_{\bm{2^{0}_2}}Y^{(1)}_{\bm{2^{0}_2}}\right)_{\bm{3^0}}\,,   &\qquad
 Y^{(2)}_{\bm{6}}(\tau)
  =\left(Y^{(1)}_{\bm{2^{0}_2}}Y^{(1)}_{\bm{4_1}}\right)_{\bm{6}}\,,
 \end{array}
\end{equation}
where $(\ldots)_{\bm{r}}$ denotes a contraction into the $\Gamma^\prime_{6}$ irreducible representation $\bm{r}$ according to the CG coefficients listed in appendix~\ref{app:group-theory}.
\begin{table}[t!]
\centering
\begin{tabular}{|c|c|}
\hline  \hline

& Modular form $Y^{(k)}_{\bm{r}}$ \\  \hline
&  \\[-0.18in]

$k=1$ &  $Y_{\bm{2^0_2}}^{(1)}$,  $Y^{(1)}_{\bm{4_1}}$ \\
&  \\[-0.18in] \hline

&  \\[-0.18in]
$k=2$ & $Y^{(2)}_{\bm{1^1_2}}$, $Y^{(2)}_{\bm{2_0}}$, $Y^{(2)}_{\bm{3^0}}$, $Y^{(2)}_{\bm{6}}$  \\
&  \\[-0.18in] \hline

&  \\[-0.18in]
$k=3$ &   $Y^{(3)}_{\bm{2^0_1}}$, $Y^{(3)}_{\bm{2^0_2}}$, $Y^{(3)}_{\bm{2^1_1}}$, $Y^{(3)}_{\bm{4_0}}$, $Y^{(3)}_{\bm{4_1}}$, $Y^{(3)}_{\bm{4_2}}$  \\
&  \\[-0.18in]\hline

&  \\[-0.18in]
$k=4$ & $Y^{(4)}_{\bm{1^{0}_{0}}} $, $Y^{(4)}_{\bm{1^{0}_{1}}} $, $Y^{(4)}_{\bm{2_{0}}} $, $Y^{(4)}_{\bm{2_{2}}} $, $Y^{(4)}_{\bm{3^{0}}}$, $Y^{(4)}_{\bm{3^{1}}}$, $Y^{(4)}_{\bm{6i}}$, $Y^{(4)}_{\bm{6ii}}$ \\
&  \\[-0.18in]\hline

&  \\[-0.18in]
$k=5$ & $Y^{(5)}_{\bm{2^0_0}}$, $Y^{(5)}_{\bm{2^0_1}}$, $Y^{(5)}_{\bm{2^0_2}}$, $Y^{(5)}_{\bm{2^1_0}}$, $Y^{(5)}_{\bm{2^1_1}}$, $Y^{(5)}_{\bm{4_0}}$, $Y^{(5)}_{\bm{4_1i}}$, $Y^{(5)}_{\bm{4_1ii}}$, $Y^{(5)}_{\bm{4_2i}}$, $Y^{(5)}_{\bm{4_2ii}}$ \\
&  \\[-0.18in] \hline

&  \\[-0.18in]
$k=6$ & $Y^{(6)}_{\bm{1^{0}_{0}}} $, $Y^{(6)}_{\bm{1^{1}_{0}}} $, $Y^{(6)}_{\bm{1^{1}_{2}}} $, $Y^{(6)}_{\bm{2_{0}}}$, $Y^{(6)}_{\bm{2_{1}}}$,  $Y^{(6)}_{\bm{2_{2}}}$, $Y^{(6)}_{\bm{3^{0}i}}$, $Y^{(6)}_{\bm{3^{0}ii}}$, $Y^{(6)}_{\bm{3^{1}}}$, $Y^{(6)}_{\bm{6i}}$, $Y^{(6)}_{\bm{6ii}}$, $Y^{(6)}_{\bm{6iii}}$   \\
&  \\[-0.18in] \hline \hline
\end{tabular}
\caption{\label{tab:MF_summary}Integral weight modular multiplets of level $6$ up to weight 6, the subscript $\bm{r}$ denotes the transformation property under  homogeneous finite group $\Gamma^{\prime}_6$. Here $Y^{(4)}_{\bm{6i}}$ and $Y^{(4)}_{\bm{6ii}}$ stand for two six-dimensional modular multiplets of $\Gamma_6$ of weight 4. The same convention is adopted for other modular forms. }
\end{table}
We summarize the modular forms of level $N=6$ up to weight 6 in table~\ref{tab:MF_summary}, and their explicit forms are given in the following. For weight 3, the modular forms take the following form
\begin{equation}\label{eq:Yw3l}
\begin{array}{lll}
 Y^{(3)}_{\bm{2^{0}_{1}}}(\tau)
 =\left(Y^{(1)}_{\bm{2^{0}_2}}Y^{(2)}_{\bm{3}}\right)_{\bm{2^{0}_{1}}}, &\quad
Y^{(3)}_{\bm{2^{0}_{2}}}(\tau)
 =\left(Y^{(1)}_{\bm{2^{0}_2}}Y^{(2)}_{\bm{3}}\right)_{\bm{2^{0}_{2}}}, & \quad
Y^{(3)}_{\bm{2^{1}_{1}}}(\tau)
 =\left(Y^{(1)}_{\bm{2^{0}_2}}Y^{(2)}_{\bm{1^{1}_{2}}}\right)_{\bm{2^{1}_{1}}}\,, \\
Y^{(3)}_{\bm{4_{0}}}(\tau)
 =\left(Y^{(1)}_{\bm{2^{0}_2}}Y^{(2)}_{\bm{6}}\right)_{\bm{4_{0}}}, & \quad
Y^{(3)}_{\bm{4_{1}}}(\tau)
 =\left(Y^{(1)}_{\bm{2^{0}_2}}Y^{(2)}_{\bm{6}}\right)_{\bm{4_{1}}}, & \quad
Y^{(3)}_{\bm{4_{2}}}(\tau)
 =\left(Y^{(1)}_{\bm{2^{0}_2}}Y^{(2)}_{\bm{2_0}}\right)_{\bm{4_{2}}}\,.
 \end{array}
\end{equation}
The modular multiplets of weight 4 are given by
\begin{equation}\label{eq:Yw4l}
\begin{array}{lll}
 Y^{(4)}_{\bm{1^{0}_{0}}}(\tau)    =\left(Y^{(2)}_{\bm{2_0}}Y^{(2)}_{\bm{2_0}}\right)_{\bm{1^{0}_{0}}} \,,
 & \quad Y^{(4)}_{\bm{1^{0}_{1}}}(\tau)  =\left(Y^{(2)}_{\bm{1^{1}_2}}Y^{(2)}_{\bm{1^{1}_2}}\right)_{\bm{1^{0}_{1}}} \,,  & \quad  Y^{(4)}_{\bm{2_0}}(\tau)  =\left(Y^{(2)}_{\bm{2_0}}Y^{(2)}_{\bm{2_0}}\right)_{\bm{2_0}}  \, ,\\
 Y^{(4)}_{\bm{2_2}}(\tau)  =\left(Y^{(2)}_{\bm{1^{1}_2}}Y^{(2)}_{\bm{2_0}}\right)_{\bm{2_2}}  \,, & \quad
Y^{(4)}_{\bm{3^{0}}}(\tau)  =\left(Y^{(2)}_{\bm{2_0}}Y^{(2)}_{\bm{6}}\right)_{\bm{3^{0}}}  \,,
&\quad Y^{(4)}_{\bm{3^{1}}}(\tau)  =\left(Y^{(2)}_{\bm{1^{1}_2}}Y^{(2)}_{\bm{3^{0}}}\right)_{\bm{3^{1}}}  \,,\\
Y^{(4)}_{\bm{6i}}(\tau)=\left(Y^{(2)}_{\bm{1^{1}_2}}Y^{(2)}_{\bm{6}}\right)_{\bm{6}} \,,
& \quad  Y^{(4)}_{\bm{6ii}}(\tau) =
\left(Y^{(2)}_{\bm{2_0}}Y^{(2)}_{\bm{3^{0}}}\right)_{\bm{6}}  \,.
 \end{array}
\end{equation}
From the tensor product of the modular multiplets of weight 2 and weight 3, we find that the modular forms of weight 5 are
\begin{equation}\label{eq:Yw5l}
\begin{array}{lll}
 Y^{(5)}_{\bm{2^{0}_{0}}}(\tau)
 =\left(Y^{(2)}_{\bm{1^{1}_2}}Y^{(3)}_{\bm{2^{1}_{1}}}\right)_{\bm{2^{0}_{0}}}\,, & \quad
 Y^{(5)}_{\bm{2^{0}_{1}}}(\tau)
 =\left(Y^{(2)}_{\bm{2_0}}Y^{(3)}_{\bm{4_{1}}}\right)_{\bm{2^{0}_{1}}}\,, & \quad
Y^{(5)}_{\bm{2^{0}_{2}}}(\tau)
 =\left(Y^{(2)}_{\bm{2_0}}Y^{(3)}_{\bm{4_{2}}}\right)_{\bm{2^{0}_{2}}}\,, \\
 Y^{(5)}_{\bm{2^{1}_{0}}}(\tau)
 =\left(Y^{(2)}_{\bm{1^{1}_2}}Y^{(3)}_{\bm{2^{0}_{1}}}\right)_{\bm{2^{1}_{0}}}\,,  & \quad
Y^{(5)}_{\bm{2^{1}_{1}}}(\tau)
 =\left(Y^{(2)}_{\bm{1^{1}_2}}Y^{(3)}_{\bm{2^{0}_{2}}}\right)_{\bm{2^{1}_{1}}}\,, & \quad
 Y^{(5)}_{\bm{4_{0}}}(\tau)
 =\left(Y^{(2)}_{\bm{1^{1}_2}}Y^{(3)}_{\bm{4_{1}}}\right)_{\bm{4_{0}}}\,, \\
Y^{(5)}_{\bm{4_{1}i}}(\tau)
 =\left(Y^{(2)}_{\bm{1^{1}_2}}Y^{(3)}_{\bm{4_{2}}}\right)_{\bm{4_{1}}}\,, & \quad
 Y^{(5)}_{\bm{4_{1}ii}}(\tau)
 =\left(Y^{(2)}_{\bm{2_0}}Y^{(3)}_{\bm{2^{0}_{1}}}\right)_{\bm{4_{1}}}\,, & \quad
Y^{(5)}_{\bm{4_{2}i}}(\tau)
 =\left(Y^{(2)}_{\bm{1^{1}_2}}Y^{(3)}_{\bm{4_{0}}}\right)_{\bm{4_{2}}}\,, \\
Y^{(5)}_{\bm{4_{2}ii}}(\tau)
 =\left(Y^{(2)}_{\bm{2_0}}Y^{(3)}_{\bm{2^{0}_{2}}}\right)_{\bm{4_{2}}}\,.
 \end{array}
\end{equation}

There are 36 linearly independent modular forms arising at weight 6 and level $N=6$ which may be necessary in model construction, we give them in the following,
\begin{equation}\label{eq:Yw6l}
\begin{array}{lll}
Y^{(6)}_{\bm{1^{0}_{0}}}(\tau) =\left(Y^{(3)}_{\bm{2^{0}_{1}}}Y^{(3)}_{\bm{2^{0}_{2}}}\right)_{\bm{1^{0}_{0}}}\,, & \quad
Y^{(6)}_{\bm{1^{1}_{0}}}(\tau) =\left(Y^{(3)}_{\bm{2^{0}_{2}}}Y^{(3)}_{\bm{2^{1}_{1}}}\right)_{\bm{1^{1}_{0}}}\,, & \quad
Y^{(6)}_{\bm{1^{1}_{2}}}(\tau) =\left(Y^{(3)}_{\bm{2^{0}_{1}}}Y^{(3)}_{\bm{2^{1}_{1}}}\right)_{\bm{1^{1}_{2}}} \,,\\
Y^{(6)}_{\bm{2_0}}(\tau) =\left(Y^{(3)}_{\bm{2^{0}_{1}}}Y^{(3)}_{\bm{4_{2}}}\right)_{\bm{2_0}} \,,  & \quad
Y^{(6)}_{\bm{2_1}}(\tau) =\left(Y^{(3)}_{\bm{2^{0}_{1}}}Y^{(3)}_{\bm{4_{0}}}\right)_{\bm{2_1}}  \,,  & \quad
Y^{(6)}_{\bm{2_2}}(\tau) =\left(Y^{(3)}_{\bm{2^{0}_{1}}}Y^{(3)}_{\bm{4_{1}}}\right)_{\bm{2_2}} \,,   \\
Y^{(6)}_{\bm{3^{0}i}}(\tau)  =\left(Y^{(3)}_{\bm{2^{0}_{1}}}Y^{(3)}_{\bm{2^{0}_{1}}}\right)_{\bm{3^{0}}}  \,, & \quad
Y^{(6)}_{\bm{3^{0}ii}}(\tau)  =\left(Y^{(3)}_{\bm{2^{0}_{1}}}Y^{(3)}_{\bm{2^{0}_{2}}}\right)_{\bm{3^{0}}}  \,,  & \quad
Y^{(6)}_{\bm{3^{1}}}(\tau)=
\left(Y^{(3)}_{\bm{2^{0}_{1}}}Y^{(3)}_{\bm{2^{1}_{1}}}\right)_{\bm{3^{1}}}  \,,  \\
Y^{(6)}_{\bm{6i}}(\tau)  =\left(Y^{(3)}_{\bm{2^{0}_{1}}}Y^{(3)}_{\bm{4_{0}}}\right)_{\bm{6}}  \,, & \quad
Y^{(6)}_{\bm{6ii}}(\tau)  =\left(Y^{(3)}_{\bm{2^{0}_{1}}}Y^{(3)}_{\bm{4_{1}}}\right)_{\bm{6}} \,,  & \quad
Y^{(6)}_{\bm{6iii}} (\tau) =\left(Y^{(3)}_{\bm{2^{0}_{1}}}Y^{(3)}_{\bm{4_{2}}}\right)_{\bm{6}} \,.
\end{array}
\end{equation}

\section{\label{app:hig_wei_Tp}Higher weight modular forms of level $N=6$ under $ T'$}

\begin{table}[t!]
\centering
\begin{tabular}{|c|c|}
\hline  \hline

& Modular form $Y^{(k)}_{\bm{r}}$ \\  \hline
&  \\[-0.18in]

$k=1$ &  $ Y^{\prime(1)}_{\bm{2^0_2}}(\tau)$, $Y^{\prime(1)}_{\bm{2^0_1i}}(\tau) $, $Y^{\prime(1)}_{\bm{2^0_1ii}}(\tau)$ \\
&  \\[-0.18in] \hline

&  \\[-0.18in]
$k=2$ & $Y^{\prime(2)}_{\bm{1^0_0i}}$, $Y^{\prime(2)}_{\bm{1^0_0ii}}$, $Y^{\prime(2)}_{\bm{1^0_2}}$, $Y^{\prime(2)}_{\bm{3^0i}}$, $Y^{\prime(2)}_{\bm{3^0ii}}$, $Y^{\prime(2)}_{\bm{3^0iii}}$   \\
&  \\[-0.18in] \hline

&  \\[-0.18in]
$k=3$ &  $Y^{\prime(3)}_{\bm{2^0_0i}}$, $Y^{\prime(3)}_{\bm{2^0_0ii}}$, $Y^{\prime(3)}_{\bm{2^0_1i}}$, $Y^{\prime(3)}_{\bm{2^0_1ii}}$, $Y^{\prime(3)}_{\bm{2^0_1iii}}$, $Y^{\prime(3)}_{\bm{2^0_1iv}}$, $Y^{\prime(3)}_{\bm{2^0_2i}}$, $Y^{\prime(3)}_{\bm{2^0_2ii}}$, $Y^{\prime(3)}_{\bm{2^0_2iii}}$  \\
&  \\[-0.18in]\hline

&  \\[-0.18in]
$k=4$ & $Y^{\prime(4)}_{\bm{1^0_0i}} $, $Y^{\prime(4)}_{\bm{1^0_0ii}} $, $Y^{\prime(4)}_{\bm{1^0_0iii}} $, $Y^{\prime(4)}_{\bm{1^{0}_{1}}} $, $Y^{\prime(4)}_{\bm{1^{0}_{2}i}} $, $Y^{\prime(4)}_{\bm{1^{0}_{2}ii}} $, $Y^{\prime(4)}_{\bm{3^0i}}$, $Y^{\prime(4)}_{\bm{3^0ii}}$, $Y^{\prime(4)}_{\bm{3^0iii}}$, $Y^{\prime(4)}_{\bm{3^0iv}}$, $Y^{\prime(4)}_{\bm{3^0v}}$, $Y^{\prime(4)}_{\bm{3^0vi}}$  \\
&  \\[-0.18in]\hline

&  \\[-0.18in]
$k=5$ & $Y^{\prime(5)}_{\bm{2^0_0i}}$, $Y^{\prime(5)}_{\bm{2^0_0ii}}$, $Y^{\prime(5)}_{\bm{2^0_0iii}}$, $Y^{\prime(5)}_{\bm{2^0_0iv}}$, $Y^{\prime(5)}_{\bm{2^0_1i}}$, $Y^{\prime(5)}_{\bm{2^0_1ii}}$, $Y^{\prime(5)}_{\bm{2^0_1iii}}$, $Y^{\prime(5)}_{\bm{2^0_1iv}}$, $Y^{\prime(5)}_{\bm{2^0_1v}}$, $Y^{\prime(5)}_{\bm{2^0_1vi}}$,  $Y^{\prime(5)}_{\bm{2^0_2i}}$, $Y^{\prime(5)}_{\bm{2^0_2ii}}$, $Y^{\prime(5)}_{\bm{2^0_2iii}}$, $Y^{\prime(5)}_{\bm{2^0_2iv}}$, $Y^{\prime(5)}_{\bm{2^0_2v}}$ \\
&  \\[-0.18in] \hline

&  \\[-0.18in]
\multirow{2}{*}{$k=6$} & $Y^{\prime(6)}_{\bm{1^0_0i}} $, $Y^{\prime(6)}_{\bm{1^0_0ii}} $, $Y^{\prime(6)}_{\bm{1^0_0iii}} $, $Y^{\prime(6)}_{\bm{1^0_0iv}} $, $Y^{\prime(6)}_{\bm{1^{0}_{1}i}} $, $Y^{\prime(6)}_{\bm{1^{0}_{1}ii}} $, $Y^{\prime(6)}_{\bm{1^{0}_{2}i}} $, $Y^{\prime(6)}_{\bm{1^{0}_{2}ii}} $, $Y^{\prime(6)}_{\bm{1^{0}_{2}iii}} $, $Y^{\prime(6)}_{\bm{3^{0}i}}$, $Y^{\prime(6)}_{\bm{3^{0}ii}}$, $Y^{\prime(6)}_{\bm{3^0iii}}$, $Y^{\prime(6)}_{\bm{3^0iv}}$, $Y^{\prime(6)}_{\bm{3^0v}}$,$Y^{\prime(6)}_{\bm{3^0vi}}$, \\
 &  \\[-0.18in]
&  $Y^{\prime(6)}_{\bm{3^0vii}}$, $Y^{\prime(6)}_{\bm{3^{0}viii}}$, $Y^{\prime(6)}_{\bm{3^{0}ix}}$  \\
&  \\[-0.18in] \hline \hline
\end{tabular}
\caption{\label{tab:MF_summary_Tp}Summary of modular forms of level $N=6$ up to weight 6 in the irreducible multiplets of finite group $T^\prime$, where the subscripts denote the transformation property under the $T^{\prime}$ modular symmetry. }
\end{table}

In the case of $N'=2$ and $N''=6$, the modular forms of $\Gamma(6)$ can be decomposed into the irreducible multiplets of finite modular group $T'$, as listed in table~\ref{tab:MF_summary_Tp}, and their explicit forms up to weight 6 are given in the following. Through the tensor products of the modular forms $ Y^{\prime(1)}_{\bm{2^0_2}}(\tau)$, $
Y^{\prime(1)}_{\bm{2^0_1i}}(\tau) $ and $Y^{\prime(1)}_{\bm{2^0_1ii}}(\tau)$ in section~\ref{sub:MF_Tp}  one can find, at weight 2, the following linearly independent modular multiplets:
\begin{equation}\label{eq:Yw22}
\begin{array}{lll}
Y^{\prime(2)}_{\bm{1^0_0i}}(\tau)
=\left(Y^{\prime(1)}_{\bm{2^0_2}}Y^{\prime(1)}_{\bm{2^0_1i}}\right)_{\bm{1^{0}_{0}}}\,, & ~
Y^{\prime(2)}_{\bm{1^0_0ii}}(\tau)
=\left(Y^{\prime(1)}_{\bm{2^0_2}}Y^{\prime(1)}_{\bm{2^0_1ii}}\right)_{\bm{1^{0}_{0}}} \,,& ~
Y^{\prime(2)}_{\bm{1^{0}_{2}}}(\tau)
=\left(Y^{\prime(1)}_{\bm{2^0_1i}}Y^{\prime(1)}_{\bm{2^0_1ii}}\right)_{\bm{1^{0}_{2}}} \,,\\
 Y^{\prime(2)}_{\bm{3^0i}}(\tau)
 =\left(Y^{\prime(1)}_{\bm{2^{0}_2}}Y^{\prime(1)}_{\bm{2^{0}_2}}\right)_{\bm{3^0}} \,, & ~
 Y^{\prime(2)}_{\bm{3^0ii}}(\tau)
 =\left(Y^{\prime(1)}_{\bm{2^{0}_2}}Y^{\prime(1)}_{\bm{2^{0}_1i}}\right)_{\bm{3^0}} \,, & ~
 Y^{\prime(2)}_{\bm{3^0iii}}(\tau)
 =\left(Y^{\prime(1)}_{\bm{2^{0}_2}}Y^{\prime(1)}_{\bm{2^{0}_1ii}}\right)_{\bm{3^0}}\,.
 \end{array}
\end{equation}
The weight 3 modular forms decompose as
\begin{equation}\label{eq:Yw32}
\begin{array}{lll}
 Y^{\prime(3)}_{\bm{2^{0}_{0}i}}(\tau)
 =\left(Y^{\prime(1)}_{\bm{2^{0}_2}}Y^{\prime(2)}_{\bm{3^0ii}}\right)_{\bm{2^{0}_{0}}}\,, & ~
 Y^{\prime(3)}_{\bm{2^{0}_{0}ii}}(\tau)
 =\left(Y^{\prime(1)}_{\bm{2^{0}_2}}Y^{\prime(2)}_{\bm{3^0iii}}\right)_{\bm{2^{0}_{0}}}\,, &~
 Y^{\prime(3)}_{\bm{2^{0}_{1}i}}(\tau)
 =\left(Y^{\prime(1)}_{\bm{2^{0}_2}}Y^{\prime(2)}_{\bm{1^{0}_2}}\right)_{\bm{2^{0}_{1}}} \,, \\
Y^{\prime(3)}_{\bm{2^{0}_{1}ii}}(\tau)
 =\left(Y^{\prime(1)}_{\bm{2^{0}_2}}Y^{\prime(2)}_{\bm{3^0i}}\right)_{\bm{2^{0}_{1}}}\,, & ~
Y^{\prime(3)}_{\bm{2^{0}_{1}iii}}(\tau)
 =\left(Y^{\prime(1)}_{\bm{2^{0}_2}}Y^{\prime(2)}_{\bm{3^0ii}}\right)_{\bm{2^{0}_{1}}}\,, & ~
Y^{\prime(3)}_{\bm{2^{0}_{1}iv}}(\tau)
 =\left(Y^{\prime(1)}_{\bm{2^{0}_2}}Y^{\prime(2)}_{\bm{3^0iii}}\right)_{\bm{2^{0}_{1}}}\,, \\
Y^{\prime(3)}_{\bm{2^{0}_{2}i}}(\tau)
 =\left(Y^{\prime(1)}_{\bm{2^{0}_2}}Y^{\prime(2)}_{\bm{1^{0}_0i}}\right)_{\bm{2^{0}_{2}}}\,, & ~
Y^{\prime(3)}_{\bm{2^{0}_{2}ii}}(\tau)
 =\left(Y^{\prime(1)}_{\bm{2^{0}_2}}Y^{\prime(2)}_{\bm{1^{0}_0ii}}\right)_{\bm{2^{0}_{2}}}\,, & ~
Y^{\prime(3)}_{\bm{2^{0}_{2}iii}}(\tau)
 =\left(Y^{\prime(1)}_{\bm{2^{0}_2}}Y^{\prime(2)}_{\bm{3^0i}}\right)_{\bm{2^{0}_{2}}}\,.
 \end{array}
\end{equation}
The linear space of modular forms of level 6 and weight 4 under $T^\prime$ can be decomposed into
\begin{equation}\label{eq:Yw42}
\begin{array}{lll}
 Y^{\prime(4)}_{\bm{1^0_0i}}(\tau)    =\left(Y^{\prime(2)}_{\bm{1^0_0i}}Y^{\prime(2)}_{\bm{1^0_0i}}\right)_{\bm{1^{0}_{0}}}  \,, & ~ Y^{\prime(4)}_{\bm{1^0_0ii}}(\tau)    =\left(Y^{\prime(2)}_{\bm{1^0_0i}}Y^{\prime(2)}_{\bm{1^0_0ii}}\right)_{\bm{1^{0}_{0}}}  \,, & ~
Y^{\prime(4)}_{\bm{1^0_0iii}}(\tau)    =\left(Y^{\prime(2)}_{\bm{1^0_0i}}Y^{\prime(2)}_{\bm{1^0_0ii}}\right)_{\bm{1^{0}_{0}}} \,, \\
Y^{\prime(4)}_{\bm{1^{0}_{1}}}(\tau)    =\left(Y^{\prime(2)}_{\bm{1^0_2}}Y^{\prime(2)}_{\bm{1^0_2}}\right)_{\bm{1^{0}_{1}}} \,, & ~
Y^{\prime(4)}_{\bm{1^{0}_{2}i}} (\tau)   =\left(Y^{\prime(2)}_{\bm{1^0_0i}}Y^{\prime(2)}_{\bm{1^0_2}}\right)_{\bm{1^{0}_{2}}} \,, & ~
 Y^{\prime(4)}_{\bm{1^{0}_{2}ii}}(\tau)    =\left(Y^{\prime(2)}_{\bm{1^0_0ii}}Y^{\prime(2)}_{\bm{1^0_2}}\right)_{\bm{1^{0}_{2}}} \,, \\
Y^{\prime(4)}_{\bm{3^{0}i}}(\tau)=\left(Y^{\prime(2)}_{\bm{1^0_0i}}Y^{(2)}_{\bm{3^0i}}\right)_{\bm{3^{0}}}  \,, & ~
 Y^{\prime(4)}_{\bm{3^{0}ii}}(\tau)  =\left(Y^{(2)}_{\bm{1^0_0i}}Y^{\prime(2)}_{\bm{3^0ii}}\right)_{\bm{3^{0}}}  \,, & ~
Y^{\prime(4)}_{\bm{3^0iii}}(\tau)  =\left(Y^{(2)}_{\bm{1^0_0i}}Y^{\prime(2)}_{\bm{3^0iii}}\right)_{\bm{3^{0}}}  \,, \\
 Y^{\prime(4)}_{\bm{3^0iv}}(\tau)  =\left(Y^{(2)}_{\bm{1^0_0ii}}Y^{\prime(2)}_{\bm{3^0i}}\right)_{\bm{3^{0}}}  \,, & ~
Y^{\prime(4)}_{\bm{3^0v}}(\tau)  =\left(Y^{(2)}_{\bm{1^0_0ii}}Y^{\prime(2)}_{\bm{3^0ii}}\right)_{\bm{3^{0}}}  \,, & ~
 Y^{\prime(4)}_{\bm{3^0vi}}(\tau)  =\left(Y^{(2)}_{\bm{1^0_0ii}}Y^{\prime(2)}_{\bm{3^0iii}}\right)_{\bm{3^{0}}}  \,.
 \end{array}
\end{equation}
The modular forms of weight 5 under the finite modular group $T^\prime$ take the following form
\begin{equation}\label{eq:Yw52}
\hskip-0.1in\begin{array}{lll}
 Y^{\prime(5)}_{\bm{2^{0}_{0}i}}(\tau)
 =\left(Y^{\prime(2)}_{\bm{1^0_0i}}Y^{\prime(3)}_{\bm{2^{0}_{0}i}}\right)_{\bm{2^{0}_{0}}}
 \,, & ~
 Y^{\prime(5)}_{\bm{2^{0}_{0}ii}}(\tau)
 =\left(Y^{\prime(2)}_{\bm{1^0_0i}}Y^{\prime(3)}_{\bm{2^{0}_{0}ii}}\right)_{\bm{2^{0}_{0}}}\,, & ~
 Y^{\prime(5)}_{\bm{2^{0}_{0}iii}}(\tau)
 =\left(Y^{\prime(2)}_{\bm{1^0_0ii}}Y^{\prime(3)}_{\bm{2^{0}_{0}i}}\right)_{\bm{2^{0}_{0}}}\,, \\
 Y^{\prime(5)}_{\bm{2^{0}_{0}iv}}(\tau)
 =\left(Y^{\prime(2)}_{\bm{1^0_0ii}}Y^{\prime(3)}_{\bm{2^{0}_{0}ii}}\right)_{\bm{2^{0}_{0}}}\,, & ~
Y^{\prime(5)}_{\bm{2^{0}_{1}i}}(\tau)
 =\left(Y^{\prime(2)}_{\bm{1^0_0i}}Y^{\prime(3)}_{\bm{2^{0}_{1}i}}\right)_{\bm{2^{0}_{1}}}\,, & ~ Y^{\prime(5)}_{\bm{2^{0}_{1}ii}}(\tau)
 =\left(Y^{\prime(2)}_{\bm{1^0_0i}}Y^{\prime(3)}_{\bm{2^{0}_{1}ii}}\right)_{\bm{2^{0}_{1}}}\,, \\
Y^{\prime(5)}_{\bm{2^{0}_{1}iii}}(\tau)
 =\left(Y^{\prime(2)}_{\bm{1^0_0i}}Y^{\prime(3)}_{\bm{2^{0}_{1}iii}}\right)_{\bm{2^{0}_{1}}}\,, & ~ Y^{\prime(5)}_{\bm{2^{0}_{1}iv}}(\tau)
 =\left(Y^{\prime(2)}_{\bm{1^0_0i}}Y^{\prime(3)}_{\bm{2^{0}_{1}iv}}\right)_{\bm{2^{0}_{1}}}
\,, & ~
Y^{\prime(5)}_{\bm{2^{0}_{1}v}}(\tau)
 =\left(Y^{\prime(2)}_{\bm{1^0_0ii}}Y^{\prime(3)}_{\bm{2^{0}_{1}i}}\right)_{\bm{2^{0}_{1}}}\,, \\
Y^{\prime(5)}_{\bm{2^{0}_{1}vi}}(\tau)
 =\left(Y^{\prime(2)}_{\bm{1^0_0ii}}Y^{\prime(3)}_{\bm{2^{0}_{1}ii}}\right)_{\bm{2^{0}_{1}}}\,, & ~
Y^{\prime(5)}_{\bm{2^{0}_{2}i}}(\tau)
 =\left(Y^{\prime(2)}_{\bm{1^0_0i}}Y^{\prime(3)}_{\bm{2^{0}_{2}i}}\right)_{\bm{2^{0}_{2}}}\,, & ~ Y^{\prime(5)}_{\bm{2^{0}_{2}ii}}(\tau)
 =\left(Y^{\prime(2)}_{\bm{1^0_0i}}Y^{\prime(3)}_{\bm{2^{0}_{2}ii}}\right)_{\bm{2^{0}_{2}}}\,, \\
Y^{\prime(5)}_{\bm{2^{0}_{2}iii}}(\tau)
 =\left(Y^{\prime(2)}_{\bm{1^0_0i}}Y^{\prime(3)}_{\bm{2^{0}_{2}iii}}\right)_{\bm{2^{0}_{2}}}
\,, & ~
 Y^{\prime(5)}_{\bm{2^{0}_{2}iv}}(\tau)
 =\left(Y^{\prime(2)}_{\bm{1^0_0ii}}Y^{\prime(3)}_{\bm{2^{0}_{2}ii}}\right)_{\bm{2^{0}_{2}}}\,, & ~
Y^{\prime(5)}_{\bm{2^{0}_{2}v}}(\tau)
 =\left(Y^{\prime(2)}_{\bm{1^0_0ii}}Y^{\prime(3)}_{\bm{2^{0}_{2}iii}}\right)_{\bm{2^{0}_{2}}}\,.
 \end{array}
\end{equation}
Finally there are 36 independent weight 6 modular forms of level 6 under the finite modular group $T^\prime$:
\begin{equation}\label{eq:Yw62}
\hskip-0.19in\begin{array}{lll}
Y^{\prime(6)}_{\bm{1^0_0i}}(\tau) =\left(Y^{\prime(3)}_{\bm{2^{0}_{0}i}}Y^{\prime(3)}_{\bm{2^{0}_{0}ii}}\right)_{\bm{1^{0}_{0}}} \,, & ~ Y^{\prime(6)}_{\bm{1^0_0ii}}(\tau) =\left(Y^{\prime(3)}_{\bm{2^{0}_{1}ii}}Y^{\prime(3)}_{\bm{2^{0}_{2}i}}\right)_{\bm{1^{0}_{0}}} \,, & ~
Y^{\prime(6)}_{\bm{1^0_0iii}}(\tau) =\left(Y^{\prime(3)}_{\bm{2^{0}_{1}ii}}Y^{\prime(3)}_{\bm{2^{0}_{2}ii}}\right)_{\bm{1^{0}_{0}}} \,, \\
 Y^{\prime(6)}_{\bm{1^0_0iv}}(\tau) =\left(Y^{\prime(3)}_{\bm{2^{0}_{1}ii}}Y^{\prime(3)}_{\bm{2^{0}_{2}iii}}\right)_{\bm{1^{0}_{0}}} \,, & ~
Y^{\prime(6)}_{\bm{1^{0}_{1}i}}(\tau) =\left(Y^{\prime(3)}_{\bm{2^{0}_{0}i}}Y^{\prime(3)}_{\bm{2^{0}_{1}i}}\right)_{\bm{1^{0}_{1}}} \,, & ~ Y^{\prime(6)}_{\bm{1^{0}_{1}ii}}(\tau) =\left(Y^{\prime(3)}_{\bm{2^{0}_{1}i}}Y^{\prime(3)}_{\bm{2^{0}_{1}ii}}\right)_{\bm{1^{0}_{1}}} \,, \\
Y^{\prime(6)}_{\bm{1^{0}_{2}i}}(\tau) =\left(Y^{\prime(3)}_{\bm{2^{0}_{0}i}}Y^{\prime(3)}_{\bm{2^{0}_{2}i}}\right)_{\bm{1^{0}_{2}}} \,, & ~ Y^{\prime(6)}_{\bm{1^{0}_{2}ii}}(\tau) =\left(Y^{\prime(3)}_{\bm{2^{0}_{1}i}}Y^{\prime(3)}_{\bm{2^{0}_{2}ii}}\right)_{\bm{1^{0}_{2}}} \,, & ~
Y^{\prime(6)}_{\bm{1^{0}_{2}iii}}(\tau) =\left(Y^{\prime(3)}_{\bm{2^{0}_{0}ii}}Y^{\prime(3)}_{\bm{2^{0}_{2}i}}\right)_{\bm{1^{0}_{2}}} \,,\\
 Y^{\prime(6)}_{\bm{3^{0}i}}(\tau)  =\left(Y^{\prime(3)}_{\bm{2^{0}_{0}i}}Y^{\prime(3)}_{\bm{2^{0}_{0}i}}\right)_{\bm{3^{0}}} \,, & ~
Y^{\prime(6)}_{\bm{3^{0}ii}}(\tau)  =\left(Y^{\prime(3)}_{\bm{2^{0}_{0}i}}Y^{\prime(3)}_{\bm{2^{0}_{0}ii}}\right)_{\bm{3^{0}}}  \,, & ~
Y^{\prime(6)}_{\bm{3^0iii}}(\tau)  =\left(Y^{\prime(3)}_{\bm{2^{0}_{0}i}}Y^{\prime(3)}_{\bm{2^{0}_{1}ii}}\right)_{\bm{3^{0}}}  \,, \\
Y^{\prime(6)}_{\bm{3^0iv}}(\tau)  =\left(Y^{\prime(3)}_{\bm{2^{0}_{0}i}}Y^{\prime(3)}_{\bm{2^{0}_{1}iii}}\right)_{\bm{3^{0}}}  \,, & ~
 Y^{\prime(6)}_{\bm{3^0v}}(\tau)  =\left(Y^{\prime(3)}_{\bm{2^{0}_{0}i}}Y^{\prime(3)}_{\bm{2^{0}_{1}iv}}\right)_{\bm{3^{0}}}  \,, & ~
Y^{\prime(6)}_{\bm{3^0vi}}(\tau)  =\left(Y^{\prime(3)}_{\bm{2^{0}_{0}ii}}Y^{\prime(3)}_{\bm{2^{0}_{1}ii}}\right)_{\bm{3^{0}}}  \,, \\
 Y^{\prime(6)}_{\bm{3^{0}vii}}(\tau)  =\left(Y^{\prime(3)}_{\bm{2^{0}_{1}ii}}Y^{\prime(3)}_{\bm{2^{0}_{1}ii}}\right)_{\bm{3^{0}}} \,, & ~
Y^{\prime(6)}_{\bm{3^{0}viii}}(\tau)  =\left(Y^{\prime(3)}_{\bm{2^{0}_{1}ii}}Y^{\prime(3)}_{\bm{2^{0}_{1}iii}}\right)_{\bm{3^{0}}}  \,, & ~
 Y^{\prime(6)}_{\bm{3^{0}ix}}(\tau)  =\left(Y^{\prime(3)}_{\bm{2^{0}_{1}ii}}Y^{\prime(3)}_{\bm{2^{0}_{1}iv}}\right)_{\bm{3^{0}}}  \,.
\end{array}
\end{equation}

\end{appendix}



\begin{thebibliography}{10}
	
	\bibitem{Esteban:2020cvm}
	I.~Esteban, M.~Gonzalez-Garcia, M.~Maltoni, T.~Schwetz, and A.~Zhou, ``{The
		fate of hints: updated global analysis of three-flavor neutrino
		oscillations},'' \href{http://dx.doi.org/10.1007/JHEP09(2020)178}{{\em JHEP}
		{\bfseries 2009} (2020) 178},
	\href{http://arxiv.org/abs/2007.14792}{{\ttfamily arXiv:2007.14792
			[hep-ph]}}.
	
	\bibitem{ParticleDataGroup:2020ssz}
	{\bfseries Particle Data Group} Collaboration, P.~A. Zyla {\em et~al.},
	``{Review of Particle Physics},''
	\href{http://dx.doi.org/10.1093/ptep/ptaa104}{{\em PTEP} {\bfseries 2020}
		no.~8, (2020) 083C01}.
	
	\bibitem{Altarelli:2010gt}
	G.~Altarelli and F.~Feruglio, ``{Discrete Flavor Symmetries and Models of
		Neutrino Mixing},'' \href{http://dx.doi.org/10.1103/RevModPhys.82.2701}{{\em
			Rev.Mod.Phys.} {\bfseries 82} (2010) 2701--2729},
	\href{http://arxiv.org/abs/1002.0211}{{\ttfamily arXiv:1002.0211 [hep-ph]}}.
	
	\bibitem{Ishimori:2010au}
	H.~Ishimori, T.~Kobayashi, H.~Ohki, Y.~Shimizu, H.~Okada, and M.~Tanimoto,
	``{Non-Abelian Discrete Symmetries in Particle Physics},''
	\href{http://dx.doi.org/10.1143/PTPS.183.1}{{\em Prog.Theor.Phys.Suppl.}
		{\bfseries 183} (2010) 1--163},
	\href{http://arxiv.org/abs/1003.3552}{{\ttfamily arXiv:1003.3552 [hep-th]}}.
	
	\bibitem{King:2013eh}
	S.~F. King and C.~Luhn, ``{Neutrino Mass and Mixing with Discrete Symmetry},''
	\href{http://dx.doi.org/10.1088/0034-4885/76/5/056201}{{\em Rept.Prog.Phys.}
		{\bfseries 76} (2013) 056201},
	\href{http://arxiv.org/abs/1301.1340}{{\ttfamily arXiv:1301.1340 [hep-ph]}}.
	
	\bibitem{King:2014nza}
	S.~F. King, A.~Merle, S.~Morisi, Y.~Shimizu, and M.~Tanimoto, ``{Neutrino Mass
		and Mixing: from Theory to Experiment},''
	\href{http://dx.doi.org/10.1088/1367-2630/16/4/045018}{{\em New J.Phys.}
		{\bfseries 16} (2014) 045018},
	\href{http://arxiv.org/abs/1402.4271}{{\ttfamily arXiv:1402.4271 [hep-ph]}}.
	
	\bibitem{King:2015aea}
	S.~F. King, ``{Models of Neutrino Mass, Mixing and CP Violation},''
	\href{http://dx.doi.org/10.1088/0954-3899/42/12/123001}{{\em J.Phys.}
		{\bfseries G42} (2015) 123001},
	\href{http://arxiv.org/abs/1510.02091}{{\ttfamily arXiv:1510.02091
			[hep-ph]}}.
	
	\bibitem{King:2017guk}
	S.~King, ``{Unified Models of Neutrinos, Flavour and CP Violation},''
	\href{http://dx.doi.org/10.1016/j.ppnp.2017.01.003}{{\em
			Prog.Part.Nucl.Phys.} {\bfseries 94} (2017) 217--256},
	\href{http://arxiv.org/abs/1701.04413}{{\ttfamily arXiv:1701.04413
			[hep-ph]}}.
	
	\bibitem{Feruglio:2019ybq}
	F.~Feruglio and A.~Romanino, ``{Lepton flavor symmetries},''
	\href{http://dx.doi.org/10.1103/RevModPhys.93.015007}{{\em Rev. Mod. Phys.}
		{\bfseries 93} no.~1, (2021) 015007},
	\href{http://arxiv.org/abs/1912.06028}{{\ttfamily arXiv:1912.06028
			[hep-ph]}}.
	
\bibitem{Feruglio:2017spp}
F.~Feruglio, \href{http://dx.doi.org/10.1142/9789813238053_0012}{``{Are
		neutrino masses modular forms?},''} in {\em From My Vast Repertoire ...:
	Guido Altarelli's Legacy}, A.~Levy, S.~Forte, and G.~Ridolfi, eds.,
pp.~227--266.
\newblock 2019.
\newblock
\href{http://arxiv.org/abs/1706.08749}{{\ttfamily arXiv:1706.08749 [hep-ph]}}.
\newblock
	
	\bibitem{Kobayashi:2018vbk}
	T.~Kobayashi, K.~Tanaka, and T.~H. Tatsuishi, ``{Neutrino mixing from finite
		modular groups},'' \href{http://dx.doi.org/10.1103/PhysRevD.98.016004}{{\em
			Phys.Rev.} {\bfseries D98} (2018) 016004},
	\href{http://arxiv.org/abs/1803.10391}{{\ttfamily arXiv:1803.10391
			[hep-ph]}}.
	
	\bibitem{Kobayashi:2018wkl}
	T.~Kobayashi, Y.~Shimizu, K.~Takagi, M.~Tanimoto, T.~H. Tatsuishi, and
	H.~Uchida, ``{Finite modular subgroups for fermion mass matrices and
		baryon/lepton number violation},''
	\href{http://dx.doi.org/10.1016/j.physletb.2019.05.034}{{\em Phys.Lett.}
		{\bfseries B794} (2019) 114--121},
	\href{http://arxiv.org/abs/1812.11072}{{\ttfamily arXiv:1812.11072
			[hep-ph]}}.
	
	\bibitem{Kobayashi:2019rzp}
	T.~Kobayashi, Y.~Shimizu, K.~Takagi, M.~Tanimoto, and T.~H. Tatsuishi,
	``{Modular $S_3$-invariant flavor model in SU(5) grand unified theory},''
	\href{http://dx.doi.org/10.1093/ptep/ptaa055}{{\em PTEP} {\bfseries 2020}
		(2020) 053B05}, \href{http://arxiv.org/abs/1906.10341}{{\ttfamily
			arXiv:1906.10341 [hep-ph]}}.
	
	\bibitem{Okada:2019xqk}
	H.~Okada and Y.~Orikasa, ``{Modular $S_3$ symmetric radiative seesaw model},''
	\href{http://dx.doi.org/10.1103/PhysRevD.100.115037}{{\em Phys.Rev.}
		{\bfseries D100} (2019) 115037},
	\href{http://arxiv.org/abs/1907.04716}{{\ttfamily arXiv:1907.04716
			[hep-ph]}}.
	
	\bibitem{Novichkov:2021evw}
	P.~Novichkov, J.~Penedo, and S.~Petcov, ``{Fermion Mass Hierarchies, Large
		Lepton Mixing and Residual Modular Symmetries},''
	\href{http://arxiv.org/abs/2102.07488}{{\ttfamily arXiv:2102.07488
			[hep-ph]}}.
	
	\bibitem{Criado:2018thu}
	J.~C. Criado and F.~Feruglio, ``{Modular Invariance Faces Precision Neutrino
		Data},'' \href{http://dx.doi.org/10.21468/SciPostPhys.5.5.042}{{\em SciPost
			Phys.} {\bfseries 5} (2018) 042},
	\href{http://arxiv.org/abs/1807.01125}{{\ttfamily arXiv:1807.01125
			[hep-ph]}}.
	
	\bibitem{Kobayashi:2018scp}
	T.~Kobayashi, N.~Omoto, Y.~Shimizu, K.~Takagi, M.~Tanimoto, and T.~H.
	Tatsuishi, ``{Modular A$_{4}$ invariance and neutrino mixing},''
	\href{http://dx.doi.org/10.1007/JHEP11(2018)196}{{\em JHEP} {\bfseries 1811}
		(2018) 196}, \href{http://arxiv.org/abs/1808.03012}{{\ttfamily
			arXiv:1808.03012 [hep-ph]}}.
	
	\bibitem{deAnda:2018ecu}
	F.~J. de~Anda, S.~F. King, and E.~Perdomo, ``{$SU(5)$ grand unified theory with
		$A_4$ modular symmetry},''
	\href{http://dx.doi.org/10.1103/PhysRevD.101.015028}{{\em Phys.Rev.}
		{\bfseries D101} (2020) 015028},
	\href{http://arxiv.org/abs/1812.05620}{{\ttfamily arXiv:1812.05620
			[hep-ph]}}.
	
	\bibitem{Okada:2018yrn}
	H.~Okada and M.~Tanimoto, ``{CP violation of quarks in $A_4$ modular
		invariance},'' \href{http://dx.doi.org/10.1016/j.physletb.2019.02.028}{{\em
			Phys.Lett.} {\bfseries B791} (2019) 54--61},
	\href{http://arxiv.org/abs/1812.09677}{{\ttfamily arXiv:1812.09677
			[hep-ph]}}.
	
	\bibitem{Novichkov:2018yse}
	P.~Novichkov, S.~Petcov, and M.~Tanimoto, ``{Trimaximal Neutrino Mixing from
		Modular A4 Invariance with Residual Symmetries},''
	\href{http://dx.doi.org/10.1016/j.physletb.2019.04.043}{{\em Phys.Lett.}
		{\bfseries B793} (2019) 247--258},
	\href{http://arxiv.org/abs/1812.11289}{{\ttfamily arXiv:1812.11289
			[hep-ph]}}.
	
	\bibitem{Nomura:2019jxj}
	T.~Nomura and H.~Okada, ``{A modular $A_4$ symmetric model of dark matter and
		neutrino},'' \href{http://dx.doi.org/10.1016/j.physletb.2019.134799}{{\em
			Phys.Lett.} {\bfseries B797} (2019) 134799},
	\href{http://arxiv.org/abs/1904.03937}{{\ttfamily arXiv:1904.03937
			[hep-ph]}}.
	
	\bibitem{Okada:2019uoy}
	H.~Okada and M.~Tanimoto, ``{Towards unification of quark and lepton flavors in
		$A_4$ modular invariance},''
	\href{http://dx.doi.org/10.1140/epjc/s10052-021-08845-y}{{\em Eur. Phys. J.
			C} {\bfseries 81} no.~1, (2021) 52},
	\href{http://arxiv.org/abs/1905.13421}{{\ttfamily arXiv:1905.13421
			[hep-ph]}}.
	
\bibitem{Nomura:2019yft}
T.~Nomura and H.~Okada, ``{A two loop induced neutrino mass model with modular
	$A_4$ symmetry},''
\href{http://dx.doi.org/10.1016/j.nuclphysb.2021.115372}{{\em Nucl. Phys. B}
	{\bfseries 966} (2021) 115372},
\href{http://arxiv.org/abs/1906.03927}{{\ttfamily arXiv:1906.03927
		[hep-ph]}}.
	
	\bibitem{Ding:2019zxk}
	G.-J. Ding, S.~F. King, and X.-G. Liu, ``{Modular A$_{4}$ symmetry models of
		neutrinos and charged leptons},''
	\href{http://dx.doi.org/10.1007/JHEP09(2019)074}{{\em JHEP} {\bfseries 1909}
		(2019) 074}, \href{http://arxiv.org/abs/1907.11714}{{\ttfamily
			arXiv:1907.11714 [hep-ph]}}.
	
	\bibitem{Okada:2019mjf}
	H.~Okada and Y.~Orikasa, ``{A radiative seesaw model in modular $A_4$
		symmetry},'' \href{http://arxiv.org/abs/1907.13520}{{\ttfamily
			arXiv:1907.13520 [hep-ph]}}.
	
	\bibitem{Nomura:2019lnr}
	T.~Nomura, H.~Okada, and O.~Popov, ``{A modular $A_4$ symmetric scotogenic
		model},'' \href{http://dx.doi.org/10.1016/j.physletb.2020.135294}{{\em
			Phys.Lett.} {\bfseries B803} (2020) 135294},
	\href{http://arxiv.org/abs/1908.07457}{{\ttfamily arXiv:1908.07457
			[hep-ph]}}.
	
	\bibitem{Kobayashi:2019xvz}
	T.~Kobayashi, Y.~Shimizu, K.~Takagi, M.~Tanimoto, and T.~H. Tatsuishi, ``{$A_4$
		lepton flavor model and modulus stabilization from $S_4$ modular symmetry},''
	\href{http://dx.doi.org/10.1103/PhysRevD.101.039904}{{\em Phys.Rev.}
		{\bfseries D100} (2019) 115045},
	\href{http://arxiv.org/abs/1909.05139}{{\ttfamily arXiv:1909.05139
			[hep-ph]}}.
	
	\bibitem{Asaka:2019vev}
	T.~Asaka, Y.~Heo, T.~H. Tatsuishi, and T.~Yoshida, ``{Modular $A_4$ invariance
		and leptogenesis},'' \href{http://dx.doi.org/10.1007/JHEP01(2020)144}{{\em
			JHEP} {\bfseries 2001} (2020) 144},
	\href{http://arxiv.org/abs/1909.06520}{{\ttfamily arXiv:1909.06520
			[hep-ph]}}.
	
	\bibitem{Gui-JunDing:2019wap}
	G.-J. Ding, S.~F. King, X.-G. Liu, and J.-N. Lu, ``{Modular S$_{4}$ and A$_{4}$
		symmetries and their fixed points: new predictive examples of lepton
		mixing},'' \href{http://dx.doi.org/10.1007/JHEP12(2019)030}{{\em JHEP}
		{\bfseries 1912} (2019) 030},
	\href{http://arxiv.org/abs/1910.03460}{{\ttfamily arXiv:1910.03460
			[hep-ph]}}.
	
	\bibitem{Zhang:2019ngf}
	D.~Zhang, ``{A modular $A_4$ symmetry realization of two-zero textures of the
		Majorana neutrino mass matrix},''
	\href{http://dx.doi.org/10.1016/j.nuclphysb.2020.114935}{{\em Nucl.Phys.}
		{\bfseries B952} (2020) 114935},
	\href{http://arxiv.org/abs/1910.07869}{{\ttfamily arXiv:1910.07869
			[hep-ph]}}.
	
	\bibitem{Nomura:2019xsb}
	T.~Nomura, H.~Okada, and S.~Patra, ``{An inverse seesaw model with $A_4$
		-modular symmetry},''
	\href{http://dx.doi.org/10.1016/j.nuclphysb.2021.115395}{{\em Nucl. Phys. B}
		{\bfseries 967} (2021) 115395},
	\href{http://arxiv.org/abs/1912.00379}{{\ttfamily arXiv:1912.00379
			[hep-ph]}}.
	
	\bibitem{Wang:2019xbo}
	X.~Wang, ``{Lepton flavor mixing and CP violation in the minimal type-(I+II)
		seesaw model with a modular $A_4$ symmetry},''
	\href{http://dx.doi.org/10.1016/j.nuclphysb.2020.115105}{{\em Nucl.Phys.}
		{\bfseries B957} (2020) 115105},
	\href{http://arxiv.org/abs/1912.13284}{{\ttfamily arXiv:1912.13284
			[hep-ph]}}.
	
	\bibitem{Kobayashi:2019gtp}
	T.~Kobayashi, T.~Nomura, and T.~Shimomura, ``{Type II seesaw models with
		modular $A_4$ symmetry},''
	\href{http://dx.doi.org/10.1103/PhysRevD.102.035019}{{\em Phys.Rev.}
		{\bfseries D102} (2020) 035019},
	\href{http://arxiv.org/abs/1912.00637}{{\ttfamily arXiv:1912.00637
			[hep-ph]}}.
	
	\bibitem{King:2020qaj}
	S.~J. King and S.~F. King, ``{Fermion mass hierarchies from modular
		symmetry},'' \href{http://dx.doi.org/10.1007/JHEP09(2020)043}{{\em JHEP}
		{\bfseries 2009} (2020) 043},
	\href{http://arxiv.org/abs/2002.00969}{{\ttfamily arXiv:2002.00969
			[hep-ph]}}.
	
	\bibitem{Ding:2020yen}
	G.-J. Ding and F.~Feruglio, ``{Testing Moduli and Flavon Dynamics with Neutrino
		Oscillations},'' \href{http://dx.doi.org/10.1007/JHEP06(2020)134}{{\em JHEP}
		{\bfseries 2006} (2020) 134},
	\href{http://arxiv.org/abs/2003.13448}{{\ttfamily arXiv:2003.13448
			[hep-ph]}}.
	
	\bibitem{Okada:2020rjb}
	H.~Okada and M.~Tanimoto, ``{Quark and lepton flavors with common modulus
		$\tau$ in $A_4$ modular symmetry},''
	\href{http://arxiv.org/abs/2005.00775}{{\ttfamily arXiv:2005.00775
			[hep-ph]}}.
	
	\bibitem{Nomura:2020opk}
	T.~Nomura and H.~Okada, ``{A linear seesaw model with $A_4$-modular flavor and
		local $U(1)_{B-L}$ symmetries},''
	\href{http://arxiv.org/abs/2007.04801}{{\ttfamily arXiv:2007.04801
			[hep-ph]}}.
	
	\bibitem{Asaka:2020tmo}
	T.~Asaka, Y.~Heo, and T.~Yoshida, ``{Lepton flavor model with modular $A_4$
		symmetry in large volume limit},''
	\href{http://dx.doi.org/10.1016/j.physletb.2020.135956}{{\em Phys.Lett.}
		{\bfseries B811} (2020) 135956},
	\href{http://arxiv.org/abs/2009.12120}{{\ttfamily arXiv:2009.12120
			[hep-ph]}}.
	
	\bibitem{Okada:2020brs}
	H.~Okada and M.~Tanimoto, ``{Spontaneous CP violation by modulus $\tau$ in
		$A_4$ model of lepton flavors},''
	\href{http://dx.doi.org/10.1007/JHEP03(2021)010}{{\em JHEP} {\bfseries 03}
		(2021) 010}, \href{http://arxiv.org/abs/2012.01688}{{\ttfamily
			arXiv:2012.01688 [hep-ph]}}.
	
	\bibitem{Yao:2020qyy}
	C.-Y. Yao, J.-N. Lu, and G.-J. Ding, ``{Modular Invariant $A_{4}$ Models for
		Quarks and Leptons with Generalized CP Symmetry},''
	\href{http://dx.doi.org/10.1007/JHEP05(2021)102}{{\em JHEP} {\bfseries 05}
		(2021) 102}, \href{http://arxiv.org/abs/2012.13390}{{\ttfamily
			arXiv:2012.13390 [hep-ph]}}.
	
	\bibitem{Feruglio:2021dte}
	F.~Feruglio, V.~Gherardi, A.~Romanino, and A.~Titov, ``{Modular invariant
		dynamics and fermion mass hierarchies around $\tau = i$},''
	\href{http://dx.doi.org/10.1007/JHEP05(2021)242}{{\em JHEP} {\bfseries 05}
		(2021) 242}, \href{http://arxiv.org/abs/2101.08718}{{\ttfamily
			arXiv:2101.08718 [hep-ph]}}.
	
	\bibitem{Chen:2021zty}
	P.~Chen, G.-J. Ding, and S.~F. King, ``{SU(5) GUTs with A$_{4}$ modular
		symmetry},'' \href{http://dx.doi.org/10.1007/JHEP04(2021)239}{{\em JHEP}
		{\bfseries 04} (2021) 239}, \href{http://arxiv.org/abs/2101.12724}{{\ttfamily
			arXiv:2101.12724 [hep-ph]}}.
	
	\bibitem{Okada:2021qdf}
	H.~Okada, Y.~Shimizu, M.~Tanimoto, and T.~Yoshida, ``{Modulus \ensuremath{\tau}
		linking leptonic CP violation to baryon asymmetry in A$_{4}$ modular
		invariant flavor model},''
	\href{http://dx.doi.org/10.1007/JHEP07(2021)184}{{\em JHEP} {\bfseries 07}
		(2021) 184}, \href{http://arxiv.org/abs/2105.14292}{{\ttfamily
			arXiv:2105.14292 [hep-ph]}}.
	
	\bibitem{Penedo:2018nmg}
	J.~Penedo and S.~Petcov, ``{Lepton Masses and Mixing from Modular $S_4$
		Symmetry},'' \href{http://dx.doi.org/10.1016/j.nuclphysb.2018.12.016}{{\em
			Nucl.Phys.} {\bfseries B939} (2019) 292--307},
	\href{http://arxiv.org/abs/1806.11040}{{\ttfamily arXiv:1806.11040
			[hep-ph]}}.
	
	\bibitem{Novichkov:2018ovf}
	P.~Novichkov, J.~Penedo, S.~Petcov, and A.~Titov, ``{Modular S$_{4}$ models of
		lepton masses and mixing},''
	\href{http://dx.doi.org/10.1007/JHEP04(2019)005}{{\em JHEP} {\bfseries 1904}
		(2019) 005}, \href{http://arxiv.org/abs/1811.04933}{{\ttfamily
			arXiv:1811.04933 [hep-ph]}}.
	
	\bibitem{deMedeirosVarzielas:2019cyj}
	I.~de~Medeiros~Varzielas, S.~F. King, and Y.-L. Zhou, ``{Multiple modular
		symmetries as the origin of flavor},''
	\href{http://dx.doi.org/10.1103/PhysRevD.101.055033}{{\em Phys.Rev.}
		{\bfseries D101} (2020) 055033},
	\href{http://arxiv.org/abs/1906.02208}{{\ttfamily arXiv:1906.02208
			[hep-ph]}}.
	
	\bibitem{Kobayashi:2019mna}
	T.~Kobayashi, Y.~Shimizu, K.~Takagi, M.~Tanimoto, and T.~H. Tatsuishi, ``{New
		$A_4$ lepton flavor model from $S_4$ modular symmetry},''
	\href{http://dx.doi.org/10.1007/JHEP02(2020)097}{{\em JHEP} {\bfseries 2002}
		(2020) 097}, \href{http://arxiv.org/abs/1907.09141}{{\ttfamily
			arXiv:1907.09141 [hep-ph]}}.
	
	\bibitem{King:2019vhv}
	S.~F. King and Y.-L. Zhou, ``{Trimaximal TM$_1$ mixing with two modular $S_4$
		groups},'' \href{http://dx.doi.org/10.1103/PhysRevD.101.015001}{{\em
			Phys.Rev.} {\bfseries D101} (2020) 015001},
	\href{http://arxiv.org/abs/1908.02770}{{\ttfamily arXiv:1908.02770
			[hep-ph]}}.
	
	\bibitem{Criado:2019tzk}
	J.~C. Criado, F.~Feruglio, and S.~J. King, ``{Modular Invariant Models of
		Lepton Masses at Levels 4 and 5},''
	\href{http://dx.doi.org/10.1007/JHEP02(2020)001}{{\em JHEP} {\bfseries 2002}
		(2020) 001}, \href{http://arxiv.org/abs/1908.11867}{{\ttfamily
			arXiv:1908.11867 [hep-ph]}}.
	
	\bibitem{Wang:2019ovr}
	X.~Wang and S.~Zhou, ``{The minimal seesaw model with a modular S$_{4}$
		symmetry},'' \href{http://dx.doi.org/10.1007/JHEP05(2020)017}{{\em JHEP}
		{\bfseries 2005} (2020) 017},
	\href{http://arxiv.org/abs/1910.09473}{{\ttfamily arXiv:1910.09473
			[hep-ph]}}.
	
	\bibitem{Wang:2020dbp}
	X.~Wang, ``{Dirac neutrino mass models with a modular $S_4$ symmetry},''
	\href{http://dx.doi.org/10.1016/j.nuclphysb.2020.115247}{{\em Nucl.Phys.}
		{\bfseries B962} (2021) 115247},
	\href{http://arxiv.org/abs/2007.05913}{{\ttfamily arXiv:2007.05913
			[hep-ph]}}.
	
	\bibitem{King:2021fhl}
	S.~F. King and Y.-L. Zhou, ``{Twin modular S$_{4}$ with SU(5) GUT},''
	\href{http://dx.doi.org/10.1007/JHEP04(2021)291}{{\em JHEP} {\bfseries 04}
		(2021) 291}, \href{http://arxiv.org/abs/2103.02633}{{\ttfamily
			arXiv:2103.02633 [hep-ph]}}.
	
	\bibitem{Ding:2021zbg}
	G.-J. Ding, S.~F. King, and C.-Y. Yao, ``{Modular $S_4\times SU(5)$ GUT},''
	\href{http://arxiv.org/abs/2103.16311}{{\ttfamily arXiv:2103.16311
			[hep-ph]}}.
	
	\bibitem{Qu:2021jdy}
	B.-Y. Qu, X.-G. Liu, P.-T. Chen, and G.-J. Ding, ``{Flavor mixing and CP
		violation from the interplay of $S_4$ modular group and gCP},''
	\href{http://arxiv.org/abs/2106.11659}{{\ttfamily arXiv:2106.11659
			[hep-ph]}}.
	
	\bibitem{Novichkov:2018nkm}
	P.~Novichkov, J.~Penedo, S.~Petcov, and A.~Titov, ``{Modular A$_{5}$ symmetry
		for flavour model building},''
	\href{http://dx.doi.org/10.1007/JHEP04(2019)174}{{\em JHEP} {\bfseries 1904}
		(2019) 174}, \href{http://arxiv.org/abs/1812.02158}{{\ttfamily
			arXiv:1812.02158 [hep-ph]}}.
	
	\bibitem{Ding:2019xna}
	G.-J. Ding, S.~F. King, and X.-G. Liu, ``{Neutrino mass and mixing with $A_5$
		modular symmetry},''
	\href{http://dx.doi.org/10.1103/PhysRevD.100.115005}{{\em Phys.Rev.}
		{\bfseries D100} (2019) 115005},
	\href{http://arxiv.org/abs/1903.12588}{{\ttfamily arXiv:1903.12588
			[hep-ph]}}.
	
	\bibitem{Ding:2020msi}
	G.-J. Ding, S.~F. King, C.-C. Li, and Y.-L. Zhou, ``{Modular Invariant Models
		of Leptons at Level 7},''
	\href{http://dx.doi.org/10.1007/JHEP08(2020)164}{{\em JHEP} {\bfseries 2008}
		(2020) 164}, \href{http://arxiv.org/abs/2004.12662}{{\ttfamily
			arXiv:2004.12662 [hep-ph]}}.
	
	\bibitem{Liu:2019khw}
	X.-G. Liu and G.-J. Ding, ``{Neutrino Masses and Mixing from Double Covering of
		Finite Modular Groups},''
	\href{http://dx.doi.org/10.1007/JHEP08(2019)134}{{\em JHEP} {\bfseries 1908}
		(2019) 134}, \href{http://arxiv.org/abs/1907.01488}{{\ttfamily
			arXiv:1907.01488 [hep-ph]}}.
	
	\bibitem{Lu:2019vgm}
	J.-N. Lu, X.-G. Liu, and G.-J. Ding, ``{Modular symmetry origin of texture
		zeros and quark lepton unification},''
	\href{http://dx.doi.org/10.1103/PhysRevD.101.115020}{{\em Phys.Rev.}
		{\bfseries D101} (2020) 115020},
	\href{http://arxiv.org/abs/1912.07573}{{\ttfamily arXiv:1912.07573
			[hep-ph]}}.
	
	\bibitem{Novichkov:2020eep}
	P.~P. Novichkov, J.~T. Penedo, and S.~T. Petcov, ``{Double cover of modular
		$S_4$ for flavour model building},''
	\href{http://dx.doi.org/10.1016/j.nuclphysb.2020.115301}{{\em Nucl. Phys. B}
		{\bfseries 963} (2021) 115301},
	\href{http://arxiv.org/abs/2006.03058}{{\ttfamily arXiv:2006.03058
			[hep-ph]}}.
	
	\bibitem{Liu:2020akv}
	X.-G. Liu, C.-Y. Yao, and G.-J. Ding, ``{Modular invariant quark and lepton
		models in double covering of $S_4$ modular group},''
	\href{http://dx.doi.org/10.1103/PhysRevD.103.056013}{{\em Phys. Rev. D}
		{\bfseries 103} no.~5, (2021) 056013},
	\href{http://arxiv.org/abs/2006.10722}{{\ttfamily arXiv:2006.10722
			[hep-ph]}}.
	
	\bibitem{Wang:2020lxk}
	X.~Wang, B.~Yu, and S.~Zhou, ``{Double covering of the modular $A_5$ group and
		lepton flavor mixing in the minimal seesaw model},''
	\href{http://dx.doi.org/10.1103/PhysRevD.103.076005}{{\em Phys. Rev. D}
		{\bfseries 103} no.~7, (2021) 076005},
	\href{http://arxiv.org/abs/2010.10159}{{\ttfamily arXiv:2010.10159
			[hep-ph]}}.
	
	\bibitem{Yao:2020zml}
	C.-Y. Yao, X.-G. Liu, and G.-J. Ding, ``{Fermion masses and mixing from the
		double cover and metaplectic cover of the $A_5$ modular group},''
	\href{http://dx.doi.org/10.1103/PhysRevD.103.095013}{{\em Phys. Rev. D}
		{\bfseries 103} no.~9, (2021) 095013},
	\href{http://arxiv.org/abs/2011.03501}{{\ttfamily arXiv:2011.03501
			[hep-ph]}}.
	
	\bibitem{Liu:2020msy}
	X.-G. Liu, C.-Y. Yao, B.-Y. Qu, and G.-J. Ding, ``{Half-integral weight modular
		forms and application to neutrino mass models},''
	\href{http://dx.doi.org/10.1103/PhysRevD.102.115035}{{\em Phys. Rev. D}
		{\bfseries 102} no.~11, (2020) 115035},
	\href{http://arxiv.org/abs/2007.13706}{{\ttfamily arXiv:2007.13706
			[hep-ph]}}.
	
	\bibitem{Okada:2020ukr}
	H.~Okada and M.~Tanimoto, ``{Modular invariant flavor model of $A_4$ and
		hierarchical structures at nearby fixed points},''
	\href{http://dx.doi.org/10.1103/PhysRevD.103.015005}{{\em Phys. Rev. D}
		{\bfseries 103} no.~1, (2021) 015005},
	\href{http://arxiv.org/abs/2009.14242}{{\ttfamily arXiv:2009.14242
			[hep-ph]}}.
	
	\bibitem{Novichkov:2019sqv}
	P.~Novichkov, J.~Penedo, S.~Petcov, and A.~Titov, ``{Generalised CP Symmetry in
		Modular-Invariant Models of Flavour},''
	\href{http://dx.doi.org/10.1007/JHEP07(2019)165}{{\em JHEP} {\bfseries 1907}
		(2019) 165}, \href{http://arxiv.org/abs/1905.11970}{{\ttfamily
			arXiv:1905.11970 [hep-ph]}}.
	
	\bibitem{Ding:2021iqp}
	G.-J. Ding, F.~Feruglio, and X.-G. Liu, ``{CP Symmetry and Symplectic Modular
		Invariance},'' \href{http://dx.doi.org/10.21468/SciPostPhys.10.6.133}{{\em
			SciPost Phys.} {\bfseries 10} (2021) 133},
	\href{http://arxiv.org/abs/2102.06716}{{\ttfamily arXiv:2102.06716
			[hep-ph]}}.
	
	\bibitem{Acharya:1995ag}
	B.~S. Acharya, D.~Bailin, A.~Love, W.~Sabra, and S.~Thomas, ``{Spontaneous
		breaking of CP symmetry by orbifold moduli},''
	\href{http://dx.doi.org/10.1016/S0370-2693(97)00912-X}{{\em Phys.Lett.}
		{\bfseries B357} (1995) 387--396}.
	
	\bibitem{Dent:2001cc}
	T.~Dent, ``{CP violation and modular symmetries},''
	\href{http://dx.doi.org/10.1103/PhysRevD.64.056005}{{\em Phys.Rev.}
		{\bfseries D64} (2001) 056005}.
	
	\bibitem{Giedt:2002ns}
	J.~Giedt, ``{CP violation and moduli stabilization in heterotic models},''
	\href{http://dx.doi.org/10.1142/S0217732302007879}{{\em Mod.Phys.Lett.}
		{\bfseries A17} (2002) 1465--1473}.
	
	\bibitem{Baur:2019kwi}
	A.~Baur, H.~P. Nilles, A.~Trautner, and P.~K. Vaudrevange, ``{Unification of
		Flavor, CP, and Modular Symmetries},''
	\href{http://dx.doi.org/10.1016/j.physletb.2019.03.066}{{\em Phys.Lett.}
		{\bfseries B795} (2019) 7--14},
	\href{http://arxiv.org/abs/1901.03251}{{\ttfamily arXiv:1901.03251
			[hep-th]}}.
	
	\bibitem{Ding:2020zxw}
	G.-J. Ding, F.~Feruglio, and X.-G. Liu, ``{Automorphic Forms and Fermion
		Masses},'' \href{http://dx.doi.org/10.1007/JHEP01(2021)037}{{\em JHEP}
		{\bfseries 01} (2021) 037}, \href{http://arxiv.org/abs/2010.07952}{{\ttfamily
			arXiv:2010.07952 [hep-th]}}.
	
	\bibitem{deAdelhartToorop:2011re}
	R.~de~Adelhart~Toorop, F.~Feruglio, and C.~Hagedorn, ``{Finite Modular Groups
		and Lepton Mixing},''
	\href{http://dx.doi.org/10.1016/j.nuclphysb.2012.01.017}{{\em Nucl.Phys.}
		{\bfseries B858} (2012) 437--467},
	\href{http://arxiv.org/abs/1112.1340}{{\ttfamily arXiv:1112.1340 [hep-ph]}}.
	
	\bibitem{Serre-1973A}
	J.~P. Serre, ``A course in arithmetic,'' {\em Springer-Verlag} (1973) .
	
	\bibitem{miyake2006modular}
	T.~Miyake, {\em Modular forms}.
	\newblock Springer Science \& Business Media, 2006.
	
	\bibitem{diamond2005first}
	F.~Diamond and J.~M. Shurman, {\em A first course in modular forms}, vol.~228
	of {\em Graduate Texts in Mathematics}.
	\newblock Springer, 2005.
	
	\bibitem{schultz2015notes}
	D.~Schultz, ``{Notes on Modular Forms}.''
	\url{https://faculty.math.illinois.edu/~schult25/ModFormNotes.pdf}, 2015.
	
	\bibitem{SageMath:2018}
	{The Sage Developers}, \emph{{S}ageMath, the {S}age {M}athematics {S}oftware
		{S}ystem ({V}ersion 8.4)}, 2018.
	
	\bibitem{KamLAND-Zen:2016pfg}
	{\bfseries KamLAND-Zen} Collaboration, A.~Gando {\em et~al.}, ``{Search for
		Majorana Neutrinos near the Inverted Mass Hierarchy Region with
		KamLAND-Zen},'' \href{http://dx.doi.org/10.1103/PhysRevLett.117.082503}{{\em
			Phys. Rev. Lett.} {\bfseries 117} no.~8, (2016) 082503},
	\href{http://arxiv.org/abs/1605.02889}{{\ttfamily arXiv:1605.02889
			[hep-ex]}}. {[Addendum: Phys.Rev.Lett. 117, 109903 (2016)]}.
	
	\bibitem{Planck:2018vyg}
	{\bfseries Planck} Collaboration, N.~Aghanim {\em et~al.}, ``{Planck 2018
		results. VI. Cosmological parameters},''
	\href{http://dx.doi.org/10.1051/0004-6361/201833910}{{\em Astron. Astrophys.}
		{\bfseries 641} (2020) A6}, \href{http://arxiv.org/abs/1807.06209}{{\ttfamily
			arXiv:1807.06209 [astro-ph.CO]}}.
	
	\bibitem{Aghanim:2018eyx}
	{\bfseries Planck} Collaboration, N.~Aghanim {\em et~al.}, ``{Planck 2018
		results. VI. Cosmological parameters},''
	\href{http://dx.doi.org/10.1051/0004-6361/201833910}{{\em Astron.Astrophys.}
		{\bfseries 641} (2020) A6}, \href{http://arxiv.org/abs/1807.06209}{{\ttfamily
			arXiv:1807.06209 [astro-ph.CO]}}.
	
	\bibitem{georgelin20002}
	Y.~Georgelin, T.~Masson, and J.-C. Wallet, ``$\gamma(2)$ modular symmetry,
	renormalization group flow and the quantum hall effect,'' {\em Journal of
		Physics A: Mathematical and General} {\bfseries 33} no.~1, (2000) 39.
	
	\bibitem{Nilles:2020nnc}
	H.~P. Nilles, S.~Ramos-S{\'a}nchez, and P.~K. Vaudrevange, ``{Eclectic Flavor
		Groups},'' \href{http://dx.doi.org/10.1007/JHEP02(2020)045}{{\em JHEP}
		{\bfseries 2002} (2020) 045},
	\href{http://arxiv.org/abs/2001.01736}{{\ttfamily arXiv:2001.01736
			[hep-ph]}}.
	
	\bibitem{Nilles:2020kgo}
	H.~P. Nilles, S.~Ramos-Sanchez, and P.~K. Vaudrevange, ``{Lessons from eclectic
		flavor symmetries},''
	\href{http://dx.doi.org/10.1016/j.nuclphysb.2020.115098}{{\em Nucl.Phys.}
		{\bfseries B957} (2020) 115098},
	\href{http://arxiv.org/abs/2004.05200}{{\ttfamily arXiv:2004.05200
			[hep-ph]}}.
	
	\bibitem{Nilles:2020tdp}
	H.~P. Nilles, S.~Ramos{\textendash}S{\'a}nchez, and P.~K. Vaudrevange,
	``{Eclectic flavor scheme from ten-dimensional string theory {\textendash} I.
		Basic results},''
	\href{http://dx.doi.org/10.1016/j.physletb.2020.135615}{{\em Phys.Lett.}
		{\bfseries B808} (2020) 135615},
	\href{http://arxiv.org/abs/2006.03059}{{\ttfamily arXiv:2006.03059
			[hep-th]}}.
	
	\bibitem{Baur:2020jwc}
	A.~Baur, M.~Kade, H.~P. Nilles, S.~Ramos-Sanchez, and P.~K. Vaudrevange, ``{The
		eclectic flavor symmetry of the $\boldsymbol{\mathbb{Z}_2}$ orbifold},''
	\href{http://dx.doi.org/10.1007/JHEP02(2021)018}{{\em JHEP} {\bfseries 2102}
		(2021) 018}, \href{http://arxiv.org/abs/2008.07534}{{\ttfamily
			arXiv:2008.07534 [hep-th]}}.
	
	\bibitem{Nilles:2020gvu}
	H.~P. Nilles, S.~Ramos{\textendash}S{\'a}nchez, and P.~K. Vaudrevange,
	``{Eclectic flavor scheme from ten-dimensional string theory - II detailed
		technical analysis},''
	\href{http://dx.doi.org/10.1016/j.nuclphysb.2021.115367}{{\em Nucl.Phys.}
		{\bfseries B966} (2021) 115367},
	\href{http://arxiv.org/abs/2010.13798}{{\ttfamily arXiv:2010.13798
			[hep-th]}}.
	
	\bibitem{Ding:2008rj}
	G.-J. Ding, ``{Fermion Mass Hierarchies and Flavor Mixing from T-prime
		Symmetry},'' \href{http://dx.doi.org/10.1103/PhysRevD.78.036011}{{\em Phys.
			Rev. D} {\bfseries 78} (2008) 036011},
	\href{http://arxiv.org/abs/0803.2278}{{\ttfamily arXiv:0803.2278 [hep-ph]}}.
	
	\bibitem{Tung:1985na}
	W.~K. Tung, {\em {GROUP THEORY IN PHYSICS}}.
	\newblock 1985.
	
\end{thebibliography}

\providecommand{\href}[2]{#2}\begingroup\raggedright\endgroup

\end{document}